\documentclass[twocolumn,epjc3]{svjour3}

\RequirePackage{fix-cm}
\RequirePackage{amsmath}
\RequirePackage{amssymb}
\RequirePackage{graphicx}

\journalname{Eur. Phys. J. C}

\usepackage{txfonts}

\newcommand{\be}{\begin{equation}}
\newcommand{\ee}{\end{equation}}
\newcommand{\ba}{\begin{eqnarray}}
\newcommand{\ea}{\end{eqnarray}}
\newcommand{\nn}{\nonumber\\}
\newcommand{\beq}{\begin{equation}}
\newcommand{\eeq}{\end{equation}}  
\newcommand{\beqa}{\begin{eqnarray}}
\newcommand{\eeqa}{\end{eqnarray}}
\newcommand{\bseq}{\begin{subequations}}
\newcommand{\eseq}{\end{subequations}}

\newcommand{\pt}{$p_{\mathrm{T}}$}
\def\p{{\boldsymbol p}}
\def\v{{\boldsymbol v}}
\def\u{{\boldsymbol u}}

\def\cal{\mathcal}
\def\r{{\boldsymbol r}}
\def\b{{\boldsymbol b}}
\def\x{{\boldsymbol x}}

\def\0{{\boldsymbol 0}}

\def\1d{{$(1\!+\!1)$-D}}
\def\2d{{$(2\!+\!1)$-D}}
\def\3d{{$(3\!+\!1)$-D}}
\DeclareMathOperator\erf{erf}

\begin{document}
 
\title{Relativistic viscous hydrodynamics for heavy-ion collisions with ECHO-QGP} 

\author{
L.~Del~Zanna\thanksref{email1,addr1,addr2,addr3}
\and 
V.~Chandra\thanksref{addr2}
\and
G.~Inghirami\thanksref{addr1,addr2}
\and
V.~Rolando\thanksref{addr4,addr5}
\and
A.~Beraudo\thanksref{addr6}
\and
A.~De~Pace\thanksref{addr7}
\and
G.~Pagliara\thanksref{addr4,addr5}
\and
A.~Drago\thanksref{addr4,addr5}
\and
F.~Becattini\thanksref{addr1,addr2,addr8}
}

\thankstext{email1}{e-mail: luca.delzanna@unifi.it}

\institute{
Dipartimento di Fisica e Astronomia, Universit\`a di Firenze,
Via G. Sansone 1, I-50019 Sesto F.no (Firenze), Italy\label{addr1}
\and
INFN - Sezione di Firenze, Via G. Sansone 1, I-50019 Sesto F.no (Firenze), Italy\label{addr2}
\and
INAF - Osservatorio Astrofisico di Arcetri, L.go E. Fermi 5, I-50125 Firenze, Italy\label{addr3}
\and
Dipartimento di Fisica e Scienze della Terra, Universit\`a di Ferrara, 
Via Saragat 1, I-44100 Ferrara, Italy\label{addr4}
\and
INFN - Sezione di Ferrara, Via Saragat 1, I-44100 Ferrara, Italy\label{addr5}
\and
Physics Department, Theory Unit, CERN, CH-1211 Gen\`eve 23, Switzerland\label{addr6}
\and
INFN - Sezione di Torino, Via P. Giuria 1, I-10125 Torino, Italy\label{addr7}
\and
Frankfurt Institute for Advanced Studies and University of Frankfurt, 
D-60438 Frankfurt am Main, Germany\label{addr8}
}

\date{Received / Accepted}

\maketitle

\begin{abstract}
We present ECHO-QGP, a numerical code for $(3+1)$-dimensional
relativistic viscous hydrodynamics designed for the modeling of the
space-time evolution of the matter created in high energy nuclear collisions. 
The code has been built on top of the \emph{Eulerian Conservative High-Order} 
astrophysical code for general relativistic magneto-hydrodynamics 
[\emph{Del Zanna et al., Astron. Astrophys. 473, 11, 2007}] 
and here it has been upgraded to handle the physics of the Quark-Gluon Plasma. 
ECHO-QGP features second-order treatment of causal relativistic
viscosity effects both in Minkowskian and in Bjorken coordinates; partial or
complete chemical equilibrium of hadronic species before kinetic
freeze-out; initial conditions based on the Glauber model,
including a Monte-Carlo routine for event-by-event fluctuating initial
conditions; a freeze-out procedure based on the Cooper-Frye prescription. 
The code is extensively validated against several test problems
and results always appear accurate, as guaranteed by the combination
of the conservative (shock-capturing) approach and the high-order methods employed.
ECHO-QGP can be extended to include evolution of the electromagnetic 
fields coupled to the plasma.
\end{abstract}

\section{Introduction}

Physics of strong interactions, described by quantum chromodynamics (QCD), 
has many fascinating aspects. 
Among them color confinement (absence of isolated quarks and gluons in nature)
and asymptotic freedom (quarks and gluons behave as if they were free at large 
energies/temperatures)  are the most celebrated ones. 
Based on these properties, QCD predicts a deconfined phase of matter 
at high temperature where quarks and gluons are effectively free beyond the nucleonic volume.
This state of matter, which one tries to reproduce in relativistic heavy-ion collision 
experiments, is commonly known as Quark-Gluon Plasma (QGP).

Experimental observations strongly suggest that the QGP, near the critical
temperature,  behaves more like a nearly perfect fluid than like a free gas of 
quarks and gluons~\cite{expt_rhic,alice}. To interpret the experimental signatures 
of deconfinement in heavy-ion collisions one would like to know the entire 
evolution of the produced matter. Relativistic hydrodynamic (RHD) modeling of the 
QGP evolution has been fairly successful in understanding particle spectra, flow 
and correlations (\emph{e.g.}~\cite{roma09}). 
The current state of the art is represented by second order viscous RHD calculations in \3d.
In particular, the RHD models were able to reproduce the transverse momentum 
spectra of hadrons in central and semi-central collisions, including the 
anisotropy in non-central collisions {-- quantified in terms of the elliptic flow coefficient 
$v_2$ --} in the range of the transverse momentum up to about $1.5-2.0$~GeV/c ~\cite{hirano,heinz},
which covers more than 99~\% of the emitted particles.
These outcomes from ideal RHD, found to be in agreement with the experimental data, 
were suggestive of an early thermalization of the QGP in ultrarelativistic heavy
ion collisions and of its strongly interacting (non-perturbative) nature, giving rise 
to the notion of a strongly-coupled QGP that behaves like an almost perfect fluid, 
{\it i.e.} with nearly vanishing shear viscosity, bulk viscosity, and thermal conductivity.

Initial formulations of relativistic viscous fluid dynamics were based on the 
extension of Navier-Stokes equations by Eckart~\cite{ek1940} and Landau~\cite{landau}. 
These descriptions ran into difficulties due to the acausal behavior of the 
propagation of short wavelength modes. To cure this problem, second order dissipative 
RHD was formulated almost four decades ago by Israel and Stewart~\cite{IS} and 
M\"uller~\cite{mueller_1,mueller_2}. 
In this formalism, the dissipative flows become independent dynamical 
entities whose kinetic equation of motions are coupled and need to be solved
simultaneously with the hydrodynamic evolution equations.
Viscous RHD numerical studies were initially limited to 
{\1d~\cite{mura,bair1,bair2,dum,moln1} and} to \2d assuming longitudinal 
boost-invariance~\cite{mura1,arch,rom,heinz1,heinz2,luzum,song},
both with averaged as well as well as with fluctuating initial conditions.
More recently, also \3d simulations of heavy-ion collisions 
have been performed, both with {and} without dissipative effects~\cite{hirano,schenke,bozek,pratt}, 
and they {turned out to be} quite successful in understanding the bulk and flow 
properties of the QGP~\cite{csernai,bozek_app1,bozek_app2,schenke_app1,schenke_app2}.

In this paper, we present a new viscous RHD code for heavy-ion collisions: ECHO-QGP.
It is based on the \emph{Eulerian Conservative High-Order} (ECHO) code developed by 
Del~Zanna {\it et al.}~\cite{luca_1} for astrophysical applications and currently adapted to 
face the physics of relativistic heavy-ion collisions. 
The original ECHO code can handle non-vanishing conserved-number currents as well 
as electromagnetic fields, which are essential for the astrophysical computations, in any \3d
metric of General Relativity. Recent developments 
include the coupling to elliptic solvers for Einstein equations, for situations where self-gravity 
is important~\cite{bucciantini1}, and modifications to the Ohm law in the presence of 
turbulent mean-field dynamo and dissipative effects~\cite{bucciantini2}. 
Here the conservative approach and the shock-capturing properties of the original code 
(needed to handle strongly non-linear effects),  as well as the high-order numerical 
procedures (to achieve accurate resolution of small scale features), are fully preserved. 
However, in view of the fact that the QGP produced at RHIC and LHC energies is almost baryon-free, 
we leave the inclusion of finite baryon density for the future (a sort of number density will be 
evolved as a tracer just for numerical reasons) and we neglect all possible electromagnetic 
effects at this stage.

Actually, in high-energy nuclear collisions a large magnetic field (up to $\sim 10^{15}$ Tesla) 
may be present, but its evolution in the plasma in a full electromagneto-RHD
calculation has not been tackled so far. There is a growing interest on its possible role, 
particularly in the presence of non-trivial topological configurations of the 
color-field~\cite{kharzeev_CMFI}. ECHO-QGP, developed starting from a relativistic 
magneto-hydrodynamic (MHD) code, could naturally allow the inclusion of a non-vanishing 
electromagnetic field: this will represent our next topic of research, first within an 
ideal MHD setup, then for the case of a resistive plasma.   
 
In the following, we shall present the formalism implemented into ECHO-QGP. 
Then we will describe the various equations of state, initial conditions
and diagnostics procedures employed, and we will solve the evolution equations 
(both in Minkowski and in Bjorken coordinates). Finally we will display the outcomes of a rich 
sample of validation tests and of simulations appropriate to compare with RHIC-type data.

The paper is organized as follows. In Sec.~II, the mathematical formalism and the
physics assumptions are described;
Sec.~III deals with the description of the algorithms and the numerical tests of the code. 
In Sec.~IV, the freeze-out procedure adopted in our analysis is described. 
Results as appropriate to RHIC-type data are presented and discussed in Sec.~V. 
Sec.~VI offers conclusions, plausible extensions and outlook of the work. Finally,
some of the formalism and computational details are presented in the Appendices.

\subsection{Notations}
We adopt the following notations in this article. 
Natural units are used throughout, that is $\hbar\!=\!c\!=\!k_B\!=\!1$.
The signature of the metric tensor is chosen to be 
$(-1,+1,+1,+1)$, as often employed in numerical relativity,
especially in the astrophysical community (and in the original ECHO code). 
Greek indices used in four-vectors range from 0 to 3, while Latin indices,
ranging from 1 to 3, are used for spatial components.
In the numerical tests and applications, we will consider two systems, 
{\it viz.} Minkowski coordinates $(t,x,y,z)$ and Bjorken coordinates $(\tau, x, y, \eta_s)$, 
which are the most appropriate for heavy ion collisions (see the appendices).
The covariant derivative is denoted as $d_\mu$, then $d_\lambda g_{\mu\nu}=0$,
and we adopt $\partial_\mu$ for the ordinary derivative and $\Gamma^\lambda_{\alpha\beta}$ 
for the connection. The action of the covariant derivative on any
(mixed, rank 2) tensor $T^{\alpha}_{\,\beta}$ is then as follows
$$
d_\mu T^{\alpha}_{\,\beta}  = \partial_\mu T^{\alpha}_{\,\beta} + 
\Gamma^{\alpha}_{\mu\lambda} T^{\lambda}_{\,\beta}-
\Gamma^{\lambda}_{\mu\beta} T^{\alpha}_{\,\lambda}, 
$$
where the Christoffel symbol is defined as
$$
\Gamma^{\lambda}_{\alpha\beta}  =\tfrac{1}{2} g^{\lambda\gamma} 
(\partial_{\alpha} g_{\gamma\beta}+ \partial_{\beta} g_{\alpha\gamma}
-\partial_{\gamma} g_{\alpha\beta}).
$$
The fluid four velocity is denoted as $u^\mu \equiv \gamma (1, v^i)$, 
with the normalizing condition $u_\mu u^\mu = -1$ providing the definition of
the fluid Lorentz factor $\gamma = (1 - g_{ij}v^iv^j)^{-1/2}$,
where $v^i\equiv u^i/\gamma$ and $\gamma\equiv u^0$.
The orthogonal projector is 
$$
\Delta^{\mu\nu}\equiv g^{\mu\nu}+u^{\mu}u^{\nu},
$$
and the quantities will be split according to it. The covariant derivative can also
be always written as
$$
d_\mu = -u_\mu D + \nabla_\mu,
$$
where the comoving derivative in the temporal direction has been defined as 
$D\!\equiv\! u^\alpha d_\alpha$, whereas the spatial comoving derivative as
$\nabla_{\mu}\equiv \Delta_\mu^{\,\alpha} d_\alpha$. 

\section{Formalism and physical assumptions}

\subsection{Relativistic viscous hydrodynamics}

The basic quantities needed to describe the dynamics of any relativistic fluid are the current $N^\mu$ 
(associated to any conserved charge) and the energy-momentum tensor $T^{\mu\nu}$, 
both conserved quantities satisfying the evolution laws
\begin{align}
& d_\mu N^\mu = 0, \\
& d_\mu T^{\mu\nu} = 0. \label{eqh}
\end{align}
In the presence of dissipation these quantities can be decomposed in general 
(see for instance~\cite{molnar}) as
\begin{align}
\label{eq1}
 N^{\mu} & = n u^\mu + V^\mu, \\
 T^{\mu\nu} & = e u^\mu u^\nu + (P+\Pi) \Delta^{\mu\nu}+\pi^{\mu\nu}+w^{\mu}u^\nu+w^\nu u^\mu,
\end{align}
where $n=-u_\mu N^\mu$ is the charge density, $V^\mu = \Delta^\mu_{\,\alpha}N^\alpha$ is the
particle diffusion flux, $e=u_\mu T^{\mu\nu} u_\nu$ is the energy density, 
$P + \Pi=\frac{1}{3} \Delta_{\mu\nu} T^{\mu\nu}$ is the isotropic pressure, and
$w^\mu=-\Delta^\mu_{\,\alpha} T^{\alpha\beta} u_\beta$ is the energy-momentum 
flow orthogonal to $u^\mu$. 
The quantities $\pi^{\mu\nu}$ and $\Pi$ denote the shear and bulk part of the 
viscous stress tensor, respectively. The shear viscous tensor is defined as
$\pi^{\mu\nu} = [\textstyle{\frac{1}{2}}(\Delta^\mu_{\,\alpha}\Delta^\nu_{\,\beta}+
\Delta^\mu_{\,\beta}\Delta^\nu_{\,\alpha})
-\textstyle{\frac{1}{3}}\Delta^{\mu\nu}\Delta_{\alpha\beta}]T^{\alpha\beta}$, and
it satisfies the orthogonality ($\pi^{\mu\nu}u_\nu=0$) and traceless ($\pi^\mu_{\,\mu}=0$) conditions.
In the absence of the dissipative quantities ($V^\mu=w^\mu=\pi^{\mu\nu}=\Pi=0$), we obtain 
the ideal decompositions $N^\mu_\mathrm{eq}=n u^\mu$ and
 $T^{\mu\nu}_\mathrm{eq}= e u^\mu u^\nu + P \Delta^{\mu\nu}$. 
In the local rest frame of the fluid (LRF), the quantities $n$ and $e$ are fixed to their 
equilibrium values by utilizing the Landau matching conditions 
($n=n_\mathrm{eq}$, $e=e_\mathrm{eq}$), and the 
pressure is obtained using an appropriate equation of state (EOS) as 
$P=\mathcal{P}(e,n)=\frac{1}{3} \Delta_{\mu\nu} T^{\mu\nu}_\mathrm{eq}$.

At this stage, there are two possibilities as far as the selection of the frame is concerned.
One can either choose the Landau frame in which $w^\mu=0$ (no net energy-momentum dissipative flow) 
or the Eckart frame in which $V^\mu=0$ (no charge dissipative flow). 
Since the QGP in experiments is created at vanishingly small baryon density
and the equation of state can be assumed in the form  $P=\mathcal{P}(e)$,
the former choice of frame is more convenient for us, so  we shall choose the Landau frame where 
$w^\mu=0$. Therefore, we have only one quantity left to describe the dynamics of the fluid under 
consideration, {\it viz.} $T^{\mu\nu}$, and the equation for $N^\mu$ becomes redundant. 

It is now convenient to decompose the conservation law in Eq.~(\ref{eqh}) 
along the directions parallel and orthogonal to $u^\mu$, in order to derive the 
energy and momentum equations, respectively. 
Before doing so, let us introduce some useful kinematic quantities.
The covariant derivative of the fluid velocity can be decomposed in its 
\emph{irreducible tensorial parts} as~\cite{muro07}
\begin{equation}
d_\mu u_\nu = \sigma_{\mu\nu} + \omega_{\mu\nu} - u_\mu D u_\nu + 
\tfrac{1}{3}\Delta_{\mu\nu} \theta,
\end{equation}
where we define the (transverse, traceless, and symmetric) \emph{shear tensor},
the (transverse, traceless, and antisymmetric) \emph{vorticity tensor}, 
and the \emph{expansion scalar} respectively as
\begin{align}
\sigma_{\mu\nu} & = 
\tfrac{1}{2}(\nabla_\mu u_\nu + \nabla_\nu u_\mu) - \tfrac{1}{3}\Delta_{\mu\nu}\theta, 
 \nonumber \\
& = \tfrac{1}{2} (d_\mu u_\nu + d_\nu u_\mu) + 
\tfrac{1}{2} (u_\mu D u_\nu + u_\nu D u_\mu)
- \tfrac{1}{3}\Delta_{\mu\nu}\theta , \\
\omega_{\mu\nu} & =
\tfrac{1}{2}(\nabla_\mu u_\nu - \nabla_\nu u_\mu) \nonumber \\
& = \tfrac{1}{2} (d_\mu u_\nu - d_\nu u_\mu) + 
\tfrac{1}{2} (u_\mu D u_\nu - u_\nu D u_\mu), \\
\theta & = \nabla_\mu u^\mu = d_\mu u^\mu.
\end{align}
With the above definitions, the relativistic energy and momentum equations
can be written as
\begin{align}
& De+(e+P+\Pi)\theta + \pi^{\mu\nu}\sigma_{\mu\nu} = 0, \\
& (e+P+\Pi) Du_\nu + \nabla_\nu (P+\Pi) + \Delta_\nu^{\,\beta} \nabla_\alpha \pi^\alpha_{\,\beta} + 
D u^\mu\,\pi_{\mu\nu} = 0,
\end{align}
and the latter is clearly orthogonal to $u_\nu$.

The bulk and shear viscous parts of stress tensor, including terms up to second-order 
in the velocity gradients, satisfies the following evolution equations:
\begin{align}
&  D\Pi  = - \tfrac{1}{\tau_\Pi}(\Pi+\zeta\theta)-\tfrac{4}{3} \Pi \theta, \\
 & \Delta^\mu_{\,\alpha} \Delta^\nu_{\,\beta} D \pi^{\alpha\beta}  \! = \!
-\tfrac{1}{\tau_{\pi}} (\pi^{\mu\nu} \! + \! 2 \eta \sigma^{\mu\nu})
 \! - \! \tfrac{4}{3} \pi^{\mu\nu} \theta \! - \!
 \lambda (\pi^{\mu\lambda} \omega^\nu_{\,\lambda}+
\pi^{\nu\lambda} \omega^\mu_{\,\lambda})
\label{pimn}
\end{align}
For their derivation and for the most general structure of the evolution equations
for $\Pi$ and $\pi^{\mu\nu}$ we refer the reader to \cite{minwalla,roma_pi}.
In our analysis, we ignore terms that are quadratic in $\Pi$, $\pi^{\mu\nu}$ and $\omega^{\mu\nu}$.
To obtain the solution of the above evolution equations we shall need 
to specify $\eta$, $\zeta$, the shear and bulk relaxation times, $\tau_\pi$ and $\tau_\Pi$,
and the other second-order transport parameter $\lambda\equiv \lambda_2/\eta$~\cite{roma_pi}. 
The parameter $\lambda_2$ is known in $\cal{N}=4$ Super-symmetric Yang Mills theory, 
but not in QCD in a non-perturbative domain. The vorticity contribution in Eq.~(\ref{pimn}) 
in the present analysis will be mostly ignored by letting $\lambda=0$, 
whereas in specific runs it will be chosen to be 1 as in~\cite{molnar}.

Now, after employing the definition of the orthogonal projector and the 
orthogonality condition $u_\mu \pi^{\mu\nu}=0$,  we can rewrite Eq. (\ref{pimn}) as
\begin{equation}
 D \pi^{\mu\nu} =  -\tfrac{1}{\tau_\pi}(\pi^{\mu\nu}+2\eta\sigma^{\mu\nu})-\tfrac{4}{3} \pi^{\mu\nu}
 \theta  +\mathcal{I}_1^{\mu\nu}+\mathcal{I}_2^{\mu\nu},
\end{equation}
where the source term $\mathcal{I}_{1}^{\mu\nu}$ comes from the orthogonal projection, while
$\mathcal{I}_2^{\mu\nu}$ is the vorticity contribution term. Their expression is the following
\begin{align}
\mathcal{I}_1^{\mu\nu}&=(\pi^{\lambda\mu}u^\nu+\pi^{\lambda\nu}u^\mu) D u_\lambda, \\
\mathcal{I}_2^{\mu\nu}&=  - \lambda (\pi^{\mu\lambda} \omega^\nu_{\,\lambda}
+\pi^{\nu\lambda} \omega^\mu_{\,\lambda}).
\end{align}

\subsection{Implementation in ECHO-QGP}

Let us look at the hydrodynamic equations that are obtained by invoking the 
conservation of $T^{\mu\nu}$ and the evolution equations 
for $\Pi$ and $\pi^{\mu\nu}$ in the context of the ECHO code. 
The ECHO-QGP equations must be written as conservative balance laws and
here we will make an effort to remain as close as possible to the formalism originally
employed in~\cite{luca_1}. 
Even when working in the Landau frame, it is convenient to evolve the continuity
equation (in the limit $V^\mu=0$) for numerical reasons. This becomes
\begin{equation}
Dn + n\theta = 0,
\end{equation}
and basically the number density $n$ must be interpreted just as a tracer
responding to the evolution of the fluid velocity through the expansion scalar $\theta$.
In conservation form this is
\be
d_\mu N^\mu = \vert g \vert^{-\frac{1}{2}}\partial_\mu (\vert g \vert^{\frac{1}{2}} N^\mu) = 0
\ee
or also
\be
\partial_0( \vert g \vert^{\frac{1}{2}} N^0)+
\partial_k( \vert g \vert^{\frac{1}{2}} N^k)=0,
\ee
where $N^0=n\gamma$ and $N^k=n\gamma v^k$.
The energy-momentum conservation equation can be expressed as
\be
d_\mu T^\mu_{\,\nu} = \vert g \vert^{-\frac{1}{2}}\partial_\mu (\vert g \vert^{\frac{1}{2}}
T^\mu_{\,\nu})-\Gamma^\mu_{\nu\lambda} T^\lambda_{\,\mu}=0,
\ee
where the relation
$\Gamma^{\mu}_{\mu\lambda}=\vert g \vert^{-\frac{1}{2}} \partial_\lambda \vert g \vert^{\frac{1}{2}}$ 
has been employed, $g$ being the determinant of the metric tensor.
We can further rewrite this equation as
\be
\label{eqb}
\partial_0 (\vert g\vert^{\frac{1}{2}} T^0_{\,\nu})+\partial_k (\vert g \vert^{\frac{1}{2}} T^k_{\,\nu})=
\vert g\vert^{\frac{1}{2}} \Gamma^\mu_{\nu\lambda} T^\lambda_{\,\,\mu} =
\vert g\vert^{\frac{1}{2}} \tfrac{1}{2} T^{\lambda\mu}\partial_\nu g_{\lambda\mu},
\ee
where the symmetry properties of the energy-momentum tensor have been now exploited.

Since our aim is to write all the equations in a conservative form, so that the same numerical
techniques can be conveniently used for the whole system, the evolution equation of 
the dissipative fluxes $\pi^{\mu\nu}$ and $\Pi$ can also be cast in this above form
of balance laws. We then use the $d_\mu N^\mu = 0$ relation in order to rewrite
the $D\equiv u^\mu d_\mu$ derivatives. If one multiplies the viscous evolution 
equations by the tracer $n$ one has
\begin{equation}
\label{eqd}
\partial_0( \vert g \vert^{\frac{1}{2}} N^0 \Pi) + \partial_k( \vert g \vert^{\frac{1}{2}} N^k \Pi)
= \vert g \vert^{\frac{1}{2}} n[ -\tfrac{1}{\tau_\Pi}(\Pi+\zeta \theta)-\tfrac{4}{3}\Pi \theta ]
\end{equation}
and
\begin{align}
\label{eqc}
& \partial_0( \vert g \vert^{\frac{1}{2}} N^0 \pi^{\mu\nu})+
\partial_k( \vert g \vert^{\frac{1}{2}} N^k \pi^{\mu\nu})= \nonumber \\
& \vert g \vert^{\frac{1}{2}} n[ -\tfrac{1}{\tau_\pi}(\pi^{\mu\nu}+2\eta\sigma^{\mu\nu})
 -\tfrac{4}{3} \pi^{\mu\nu} \theta +\mathcal{I}_0^{\mu\nu}+\mathcal{I}_1^{\mu\nu}+\mathcal{I}_2^{\mu\nu} ],
\end{align} 
where we have singled out
\begin{equation}
\mathcal{I}_0^{\mu\nu} = -u^\alpha (\Gamma^\mu_{\lambda\alpha}\pi^{\lambda\nu}+
\Gamma^\nu_{\lambda\alpha}\pi^{\mu\lambda})
\end{equation}
as a separated source term, which clearly vanishes in the Minkowski metric. 
The technique of introducing a conserved-number current as a tracer 
is exploited in a similar way within a recent code for \2d Lagrangian 
hydrodynamics~\cite{nhostler} to solve the evolution equation of the 
bulk viscous pressure $\Pi$.

Due to the orthogonality condition, only 6 out of 10 components of the viscous stress tensor
are independent (here we decide not to impose the additional traceless condition). 
Our choice for ECHO-QGP is to evolve only the 6 spatial components $\pi^{ij}$.
Then we can combine  Eqs. (\ref{eqb}-\ref{eqd}) in matrix form as a system of
$1+3+1+1+6=12$ balance laws  
\begin{equation}
\label{echoeq}
\partial_0 {\bf U}+ \partial_k {\bf F}^k ={\bf S}, 
\end{equation}
where
\begin{equation}
\label{echoeq1}
{\bf U}= \vert g \vert^{\frac{1}{2}}\left(\begin{array}{c}
N\equiv N^0 \\
S_i \equiv T^0_{\,i}\\
 E\equiv - T^0_{\,0} \\
 N\Pi \\
N\pi^{ij}
\end{array}\right), \quad
{\bf F}^k = \vert g \vert^{\frac{1}{2}}\left(\begin{array}{c}
N^k \\
 T^k_{\,i}\\
 - T^k_{\, 0} \\
 N^k \Pi \\
N^k \pi^{ij}
\end{array}\right)
\end{equation}
are respectively the set of \emph{conservative variables} and \emph{fluxes}, 
while the source terms are given by
\begin{equation}
\label{echoeq2}
{\bf S}=\vert g \vert^{\frac{1}{2}}\left(\begin{array}{c}
0 \\
\tfrac{1}{2} T^{\mu\nu}\partial_i g_{\mu\nu} \\
- \tfrac{1}{2} T^{\mu\nu}\partial_0 g_{\mu\nu} \\
n [ -\frac{1}{\tau_\pi}(\Pi+\zeta \theta)-\frac{4}{3}\Pi \theta] \\
\!\! n [-\frac{1}{\tau_\pi}(\pi^{ij}+2\eta\sigma^{ij})
 -\frac{4}{3} \pi^{ij} \theta + \mathcal{I}_0^{ij} +  \mathcal{I}_1^{ij}+\mathcal{I}_2^{ij}] \!\!
\end{array}\right).
\end{equation}
The above Eqs.~(\ref{echoeq}-\ref{echoeq2}) represent the set of ECHO-QGP 
equations in the most general form (we recall that ECHO can work in any kind
of GR metric).
We shall specify them in Minkowski and Bjorken coordinates in full \3d
in Appendix~B. 

Let us now proceed to the estimation of the \emph{primitive variables} like the fluid velocity and the 
local energy density, which are needed at every timestep of the
evolution to calculate the above fluxes and source terms, as well as all the quantities for diagnostics. 
In order to do so, we first rewrite the energy-momentum
components as needed in our evolution equations above, assuming for simplicity a metric
in which $g_{00}=-1$ and $g_{0i}=0$ (both conditions are met either by flat space
and Bjorken coordinates). 
First, the orthogonality conditions $\pi^{\mu\nu}u_\nu=0$ yield the relations
\begin{equation}
\pi^{0i} = \pi^{ij}v_j, \quad \pi^{00} = \pi^{0i} v_i = \pi^{ij}v_i v_j,
\end{equation}
where $v^i=u^i/\gamma$, $v_i=g_{ij}v^j$, and $\gamma = (1 - v_i v^i)^{-1/2}$.
Then, we rewrite the conservative variables as
\begin{align}
& N = n \gamma, \\
& S^i = (e + P + \Pi)\gamma^2 v^i + \pi^{0i}, \\
& E = (e + P + \Pi)\gamma^2 - (P + \Pi) + \pi^{00},
\end{align}
where we have substituted $S^i=g^{ij}S_j$ and $\pi^{00}=-\pi^0_{\,0}$.
Now, provided that $\pi^{ij}$ are also conserved variables (since
$N$ is a conserved variable in turn), if the $v^i$ components were known then the LRF
charge and energy densities would be given by
\begin{equation}
e = E - g_{ij}S^iv^j, \quad n = N/\gamma,
\end{equation}
then also the pressure $P=\mathcal{P}(e)$ or even $P=\mathcal{P}(e,n)$ can be
worked out. 
However, the procedure that we have found more stable is reported below, 
which is basically an iteration of the one for ideal relativistic hydrodynamics, 
assuming that corrections due to viscous terms are small (therefore a few iterations of it will be required).
\begin{itemize}
\item An external cycle on $v^i$ components is performed, starting from the
values at the previous time-step. Then we define the quantities
$$
\tilde{E} = E - \pi^{00}, \quad \tilde{S}^i = S^i - \pi^{0i};
$$
\item An inner cycle on $P$ is performed, then we define
$$
\tilde{P}(P)=P+\Pi, \quad v^2 (P)= \tilde{S}^2/(\tilde{E} + \tilde{P})^2
$$
and
$$
e(P)=(\tilde{E}+\tilde{P})(1-v^2)-\tilde{P}, \quad n(P)=N\sqrt{1-v^2}.
$$
The cycle can be iterated via a Newton-Raphson procedure, trying to
minimize the quantity
$$
f(P) = \mathcal{P}[e(P),n(P)] - P,
$$
and then
$$
P_\mathrm{new} = P - f(P)/f^\prime (P),
$$
where
$$
f^\prime (P) = \frac{\partial\mathcal{P}}{\partial e}
\frac{de}{dP} + \frac{\partial\mathcal{P}}{\partial n}\frac{dn}{dP} - 1,
$$
until $|P_\mathrm{new} - P|\to 0$ to a given tolerance.
\item
Once the pressure for the old choice of $v^i$ components has been found, the
new choice is provided by
$$
v^i_\mathrm{new} = \tilde{S}^i/(\tilde{E}+\tilde{P}),
$$
and the external loop is closed when a given tolerance in $v^i_\mathrm{new}-v^i$
terms is reached.
\end{itemize}
Notice that, in the ideal case, there is no need of the external iteration since
$\tilde{E}\equiv E$, $\tilde{S}^i\equiv S^i$, $\tilde{P}\equiv P$. The same would hold if
also the $\pi^{00}$ and $\pi^{0i}$ viscous terms were evolved. Thanks to
the conservative nature of $N$, they would be in fact conservative variables in turn,
without the need to retrieve them via the orthogonality condition (which implies the
use of unknown velocity terms).

Before concluding the section, we anticipate another ingredient required by the 
numerical evolution scheme. In the evaluation of \emph{numerical fluxes}
(and also of the maximum time-step allowed) the local fastest characteristic speeds 
associated to the Jacobian $\partial\mathbf{F}^k/\partial\mathbf{U}$ are required,
for every direction $k$. In the ideal case we have \cite{luca_1}
\be
\lambda^k_\pm = \frac{(1-c^2_s)v^k\pm
\sqrt{c^2_s(1-v^2)[(1-v^2c_s^2)g^{kk}-(1-c^2_s){v^k}^2]}}{1-v^2c^2_s},
\label{lambdas}
\ee
where the sound speed is given by
\be
c^2_s=\frac{\partial\mathcal{P}}{\partial e} + \frac{n}{e+P}\frac{\partial\mathcal{P}}{\partial n}.
\label{sound}
\ee
The same numerical scheme has also been applied in the presence of viscosity.

\subsection{First and second order transport coefficients}

As seen in the previous section, the first-order transport 
coefficients $\eta$ and $\zeta$ (shear and bulk viscosity) appear as input parameters
 in the evolution equations of the viscous stress tensor. 
The ratio of these coefficients to the entropy density
 $s$ is known to be essential to understand to what extent dissipative processes 
are effective in slowing down the expansion of the fluid in response to pressure gradients. 
For vanishing baryon densities, we recall the fundamental relation
\be
sT = e + P,
\ee
providing the entropy density once the other thermodynamical quantities are known 
(see below for the choice of the equation of state).

Elliptic flow measurements at RHIC suggest a maximum 
value of $\eta/s\!\sim\!0.16$ \cite{phobos} for Glauber initial conditions~\cite{glauberinit}
 and of $\eta/s\sim 0.24$ for Color Glass Condensate (CGC) type initial conditions~\cite{cgc1,cgc2},
 which entail a larger spatial eccentricity with respect to optical-Glauber calculations.
In the present paper, where we employ a Glauber initialization, we let $\eta/s$ 
vary within the range $0.08\! \le \eta/s\! \le 0.16$. 
Concerning the bulk viscosity, we employ the relation 
\be
\frac{\zeta}{s}=2\frac{\eta}{s} \left(\frac{1}{3}-c_s^2\right),
\ee
as obtained from the study of strongly coupled gauge theories~\cite{adscft_bulk},
in which the sound speed is calculated as in Eq.~(\ref{sound}) assuming $P=\mathcal{P}(e)$.
The second order transport coefficients, {\it viz.} the shear and bulk
 relaxation times $\tau_\pi$ and $\tau_\Pi$, are known both from kinetic theory
~\cite{IS,bair1} and from AdS-CFT~\cite{roma_pi,adscft_tau}. 
Following~\cite{heinz,bozek}, in ECHO-QGP the choice is
\be
\tau_\pi=\tau_\Pi=\frac{3 \eta}{sT}.
\ee
As far as the coupling with the vorticity terms is concerned, as anticipated
we shall assume either $\lambda=0$ or $\lambda=1$.

Finally, note that for RHIC initial conditions, we initialize the ECHO-QGP 
code by choosing the LRF at $\tau_0$, where by definition $u^\mu\equiv (1, 0, 0, 0)$. 
However, in Bjorken coordinates, even when all spatial velocity components and their derivatives
vanish, $\theta$ and $\sigma^{\mu\nu}$ are not zero (see Appendix~B), thus $\Pi$ 
and the components of $\pi^{\mu\nu}$ must be initialized somehow.
We use the first-order expressions $\Pi=-\zeta \theta$ and $\pi^{\mu\nu}=-2 \eta\sigma^{\mu\nu}$, thus
\be
\label{rhicinit}
\Pi = - \zeta/\tau, \quad 2\pi^{xx} = 2\pi^{yy} = -\tau^2 \pi^{\eta\eta}= \tfrac{4}{3} \eta/\tau,
\ee
at $\tau=\tau_0$, with all other components set to zero.

\subsection{Equation of State}
\label{sec:EOS}

Solving hydrodynamic equations require the knowledge of the Equation of State (EOS) of the system,
and as anticipated, though the code is already designed to handle any form for $P=\mathcal{P}(e,n)$,
here we shall just consider the case $P=\mathcal{P}(e)$.
ECHO-QGP allows the use of any tabulated EOS of this kind, if provided in the format 
$(T, e/T^4, P/T^4, c_s^2)$ by the user, with $c_s^2\equiv dP/de$. 

However, some choices are already implemented in the code and are offered to the user. 
Test runs can be performed with the ultrarelativistic ideal gas EOS $P=e/3$. More precisely, 
we set in these cases
\be
P=e/3=\frac{g\pi^2}{90}T^4,\quad c_s^2=\frac{1}{3},
\ee
where $g=37$ for a non-interacting QGP with 3 light flavors.

More realistic QCD EOS's are included in the package, in the tabulated form mentioned above,
and can be selected by the user. The EOS in~\cite{laine}, arising
from a weak-coupling QCD calculation with realistic quark masses and 
employed in the code by Romatschke~\cite{rom}, will be often used in this paper.

\begin{figure}[th]
\begin{center}
\includegraphics[clip,width=.48\textwidth]{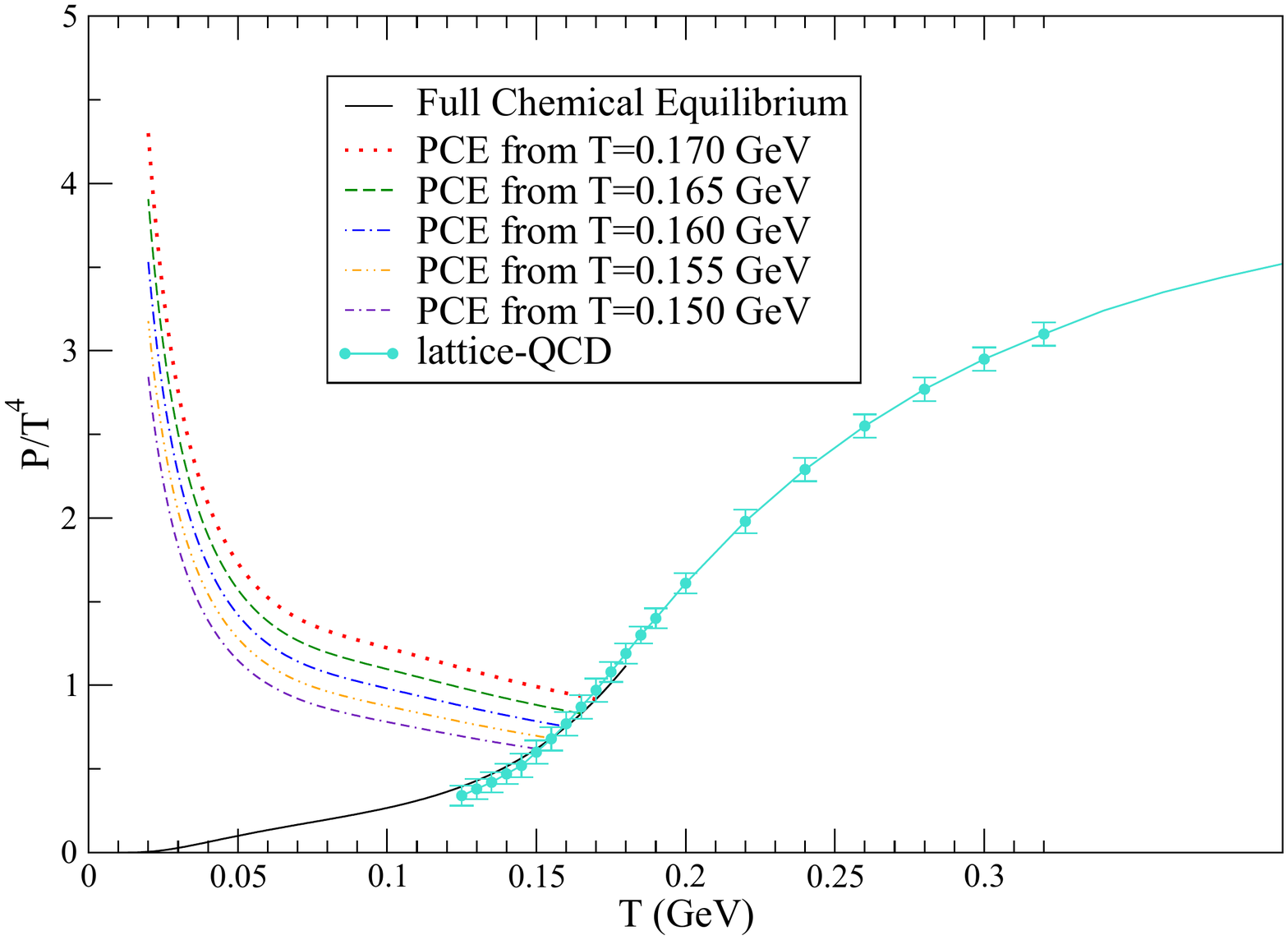}
\includegraphics[clip,width=.48\textwidth]{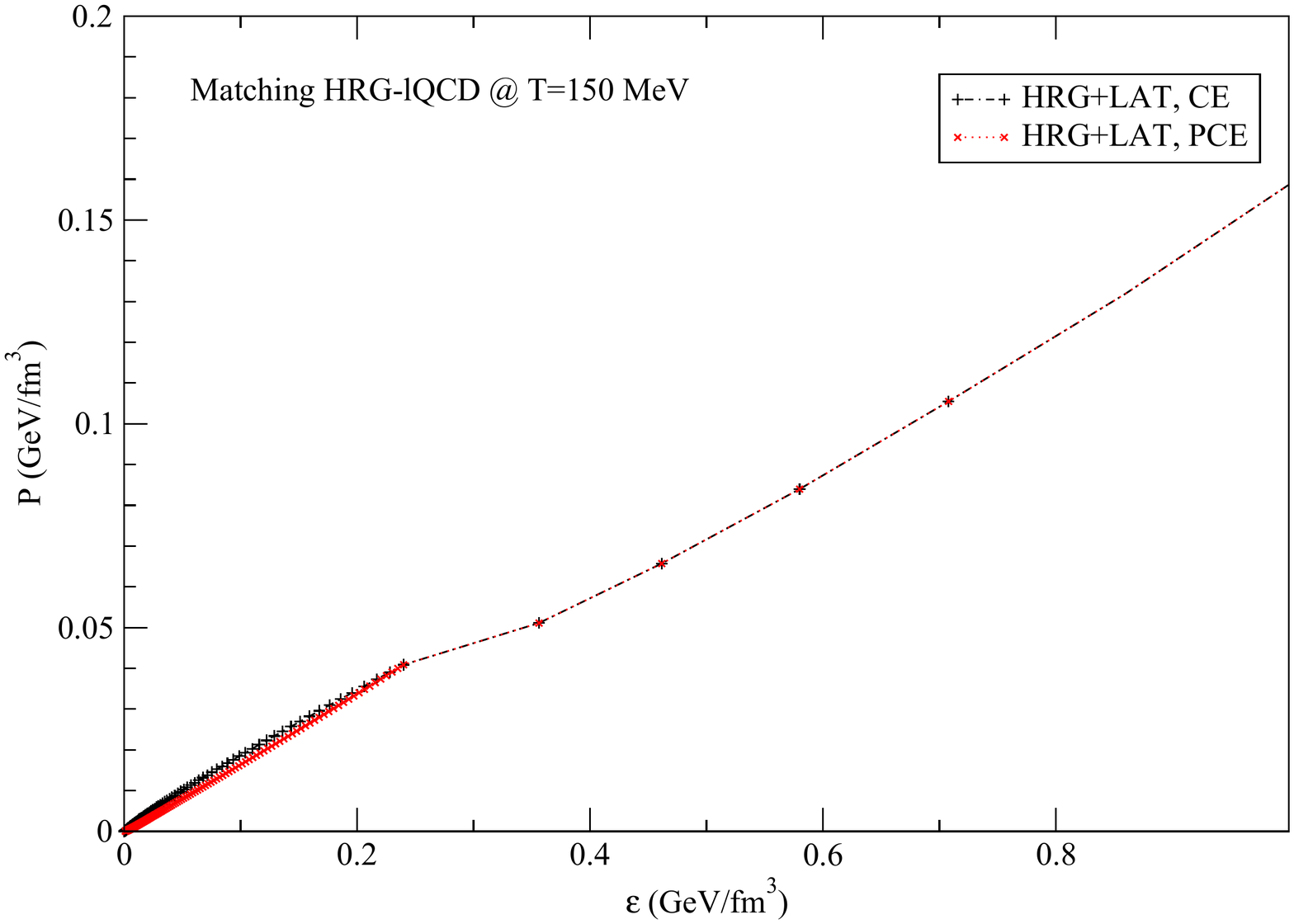}
\caption{Upper panel: HRG EOS, with chemical (CE, black continuous line) and partial 
chemical equilibrium (PCE, colored dotted/dashed lines),
 vs the lattice-QCD results in~\cite{wuppe} (turquoise points). 
 Lower panel: the EOS $P(e)$ resulting from the matching of HRG with lattice-QCD results, 
 in the CE (in black) and PCE (in red) cases. The matching has been performed 
 at the temperature $T\!=\!150$ MeV.}
\label{fig:EOS}
\end{center}
\end{figure}

ECHO-QGP includes also two tabulated EOS's obtained by matching a Hadron-Resonance-Gas 
EOS at low temperature with the continuum-extrapolated
 lattice-QCD results by the Budapest-Wuppertal collaboration~\cite{wuppe}.
The HRG EOS was obtained by summing the contributions of all hadrons and 
resonances in the PDG \cite{bluhm} up to a mass of 2 GeV: $P=\sum_r P_r$. 
In the classical limit $T\ll m_r$ (quantum corrections are included for pions, 
kaons and $\eta$'s) one has simply
\beq
P_r=g_r\,\frac{T^2m_r^2}{2\pi^2}\,e^{\mu_r/T}\,K_2\left(\frac{m_r}{T}\right),
\eeq
and the density of resonance $r$ in the cocktail is given by
\beq
n_r\equiv\left(\frac{\partial P}{\partial\mu_r}\right)_T=
g_r\,\frac{Tm_r^2}{2\pi^2}\,e^{\mu_r/T}\,K_2\left(\frac{m_r}{T}\right).
\label{eq:nr}
\eeq
In the Chemical Equilibrium case (CE) in the hadronic phase all
 the chemical potentials vanish ($\{\mu_r\!=\!0\}$) and the multiplicity of any resonance
 $r$ is simply set by the temperature through the ratio $m_r/T$.
On the other hand experimental data provide evidence that the chemical freeze-out -- in which 
particle ratios are fixed -- occurs earlier than the kinetic one, in which particle spectra gets frozen.
 A realistic EOS should in principle contain the correct chemical composition in the hadronic phase.
 This can be enforced in the following way. At the chemical freeze-out temperature $T_c$ 
the abundances $n_r$ of all the resonances are determined by Eq.~(\ref{eq:nr}) with $\mu_r\!=\!0$. 
Afterwards the fireball evolves maintaining Partial Chemical Equilibrium (PCE): elastic interactions mediated by 
resonances ($\pi\pi\to\rho\to\pi\pi$, $K\pi\to K^*\to K\pi$, $p\pi\to\Delta\to p\pi$...) are allowed,
 changing the abundance of the single resonances $r$, but conserving the ``effective multiplicity'' of stable 
hadrons $h$ ($\pi,K,\eta,N,\Lambda,\Sigma,\Xi$ and $\Omega$)
\beq
\overline{n}_h=n_h+\sum_{r\ne h}n_r\langle N_h^{r}\rangle,
\eeq   
where $\langle N_h^{r}\rangle$ represents the average number of hadrons $h$ 
coming from the decay of resonance $r$. Furthermore the multiplicity of resonance $r$ 
is fixed by the chemical potential $\mu_r\!\equiv\!\sum_h\langle N_h^{r}\rangle\mu_h$. 
Assuming an isoentropic expansion of the fireball, PCE is set fixing at each temperature 
the chemical potentials $\mu_h$ so to satisfy the relation
\beq
\frac{\overline{n}_h(T,\{\mu_{h'}\})}{s(T,\{\mu_{h'}\})}=\frac{\overline{n}_h(T_c,\{\mu_{h'}=0\})}{s(T_c,\{\mu_{h'}=0\})},
\eeq 
which amounts to the conservation of the entropy per (effective) particle throughout the medium evolution.
Both in the CE and in the PCE case the transition
from the lattice-QCD to the HRG description is performed at the temperature $T\!=\!150$ MeV
where the matching looks sufficiently smooth: results for the EOS are displayed in Fig.~\ref{fig:EOS}.
A tabulation of the HRG+lQCD EOS in the PCE case is also part of the ECHO-QGP code.

Finally, we set the acronyms for the different equations of state currently implemented in ECHO-QGP.
The ultrarelativistic ideal gas $P=e/3$ EOS will be labeled henceforth as EOS-I, and will be used 
mainly for testing purposes.
The EOS computed by Laine and Schr\"{o}der~\cite{laine} will be termed as EOS-LS.
The one with HRG+Lattice with CE (HRG+LAT+CE) will be termed as EOS-CE (though
it will be not used in this paper), while the analog one with partial chemical equilibrium
will be labeled as  EOS-PCE.

\subsection{Initial conditions}
\label{sec:init}

Various choices of initial conditions are implemented in the code and are
selectable by the user, including test problems used for the numerical
validation of the code.

For physical applications, ECHO-QGP will be mostly tested with smooth 
initial conditions based on the optical Glauber model. 
Initialization is done by setting either the energy-density or the
entropy-density distribution at the initial time $\tau_0$.
In the \2d case these quantities receive both a soft (proportional to 
$n_{\rm part}$) and a hard (proportional to $n_{\rm coll}$) contribution, with 
relative weight given by the coefficient $\alpha\!\in\![0,1]$ (see, e.~g.~\cite{Flo10}).
We set
\be
e(\tau_0,\x;b)=e_0\left[(1-\alpha)\frac{n_{\rm part}(\x;b)}{n_{\rm
      part}(\0;0)}+\alpha\frac{n_{\rm coll}(\x;b)}{n_{\rm coll}(\0;0)}\right], 
      \label{eq:alpha}
\ee
where $e(\tau_0,\x;b)$ stands for either the energy or the entropy-density and $e_0$ 
is the corresponding value at $\x\!=\!\0$ and $b\!=\!0$, $\x$ and $b$ being the coordinates 
in the transverse plane and the impact parameter, respectively.
In the optical Glauber model the density of participants 
$n_{\rm part}(\x;b)\equiv n_{\rm part}^A(\x;b)+n_{\rm part}^B(\x;b)$ and of
binary collisions in the transverse plane are given by
\beqa
n_{\rm part}^A(\x;b)& \! = \! A\,\widehat{T}_{A}(\x+\b/2)
\left\{1 \! - \! [1\! -\! \widehat{T}_{B}(\x-\b/2)\sigma^{\rm in}_{NN}]^B\right\} ,
\nonumber\\ 
n_{\rm part}^B(\x;b)& \! = \!B\,\widehat{T}_{B}(\x-\b/2) 
\left\{1\! - \! [1\!-\! \widehat{T}_{A}(\x+\b/2)\sigma^{\rm in}_{NN}]^A\right\},
\nonumber\\ 
\eeqa
and
\beq
n_{\rm coll}(\x;b)=AB\,\sigma^{\rm in}_{NN}\,\widehat{T}_{A}(\x+\b/2)\widehat{T}_{B}(\x-\b/2),
\eeq
$\sigma^{\rm in}_{NN}$ being the inelastic nucleon-nucleon cross-section and
the nuclear thickness function (normalized to 1) being defined by 
\beq
\widehat{T}_{A/B}(\x)\equiv\int_{-\infty}^{\infty}\!\!\!dz\, \rho_{A/B}(\x,z).
\label{eq:nucl_thick}
\eeq
In Eq.~(\ref{eq:nucl_thick}) $\rho$ is a Fermi parameterization of the nuclear 
density distribution \cite{DeV87}.
Tunable parameters are the maximum initial energy density in central collisions
$e_0$ and the hardness fraction $\alpha$.

In the 3D case the initialization is performed using 
the model for the density distribution as in~\cite{Hir06,Adi05}
\begin{multline}
e(\tau_0,\x,\eta_s;b)=\tilde{e}_0\,\theta(Y_b\!-\!|\eta_s|)\,
  f^{\rm pp}(\eta_s) \left[\alpha n_{\rm coll}(\x;b)\right.\\
\left. +(1-\alpha)\left(\frac{Y_b-\eta_s}{Y_b}n_{\rm part}^A(\x;b) 
  +\frac{Y_b+\eta_s}{Y_b}n_{\rm part}^B(\x;b)\right)\right]
\label{ene3D}
\end{multline}
(note that here $\tilde{e}_0$ does not represent the energy or entropy-density
at $\x\!=\!\0$ and $b\!=\!0$). 
The initial entropy density vanishes at space-time rapidity $\eta_s$ larger
than the beam-rapidity $Y_b\!\approx\!\ln(\sqrt{s_{\rm NN}}/m_p)$;
particles produced by the participants of nucleus A/B tend to follow the
rapidity of their respective source, the effect being parametrized by the
factors $(Y_b\pm\eta_s)/Y_b$. $\tilde{e}_0$ is an overall normalization factor, 
whereas the function $f^{\rm pp}(\eta_s)$ describes the rapidity profile in p-p
collisions
\be
f^{\rm pp}(\eta_s)=\exp\left[-\theta(|\eta_s|-\Delta_\eta/2)
  \frac{(|\eta_s|-\Delta_\eta/2)^2}{2\sigma_\eta^2}\right]. 
  \label{init3d}
\ee
This is a flat profile for $|\eta_s|\!\le\!\Delta_\eta/2$ and displays a gaussian damping at
forward/backward rapidities. The extension of the rapidity plateau $\Delta_\eta$
and the width $\sigma_\eta$ of the gaussian falloff are the two further
parameters describing the rapidity dependence in the 3D case. Any other
functional form can be implemented by the user.

ECHO-QGP includes also the possibility of performing event-by-event hydro
calculations with fluctuating initial conditions. A simple Glauber Monte
Carlo routine is provided with the code: 
\begin{itemize}
\item A sample of $N_{\rm conf}$ nuclear configurations is generated,
  extracting randomly the positions of the nucleons of the A and B nuclei from
  a Woods-Saxon distribution. The transverse positions of the nucleons in each
  nucleus is then reshuffled into the respective center-of-mass frame. 
\item For a given configuration a random impact parameter $b\!\in\![0,b_{\rm
    max}]$ is extracted from the distribution {$dP\!=\!2\pi bdb$}. Nucleons $i$
  (from nucleus A) and $j$ (from nucleus B) collide if
  {$(x_i\!-\!x_j)^2\!+\!(y_i\!-\!y_j)^2<\sigma_{\rm NN}/\pi$}. If at least a
  binary nucleon-nucleon collision occurred the event is kept and the
  information ($\x_{\rm part}^A$, $\x_{\rm part}^B$ and $\x_{\rm coll}$) is
  stored, otherwise not. The procedure is repeated $N_{\rm trials}$ times for
  each configuration of the incoming nuclei. 
\item Each participant nucleon and collision, with a gaussian smearing of
  variance $\sigma$, is a source of energy density (with the parameter $\alpha$
  setting the hardness fraction): 
\begin{multline}
e(\tau_0,\x)=\frac{K}{2\pi\sigma}\left\{(1-\alpha)\sum_{i=1}^{N_{\rm part}} 
  \exp\left[-\frac{(\x-\x_i^{\rm part})^2}{2\sigma^2} \right] \right.\\ 
\left.+\alpha\sum_{i=1}^{N_{\rm coll}}\exp\left[-
  \frac{(\x-\x_i^{\rm coll})^2}{2\sigma^2}\right]\right\}.
\end{multline}
The model has been employed in~\cite{esk} and tuned, with a pure
dependence on participants ($\alpha\!=\!0$), to Au-Au data at RHIC. 
The rapidity dependence in the 3D case can be inserted \emph{a posteriori} as
in the  optical-Glauber initialization of Eq.~(\ref{ene3D}). 
Storing information both on $\x_{\rm part}^A$ and on $\x_{\rm part}^B$ it is
even possible to account for the different rapidity dependence of the
contributions of the participants from the two nuclei (leading to a direct flow
$v_1$ far from mid-rapidity).
\end{itemize}

Initial conditions for the flow are chosen in both \2d and \3d cases in order
to have, at $\tau=\tau_0$, zero transverse flow velocities and a longitudinal
flow given by the Bjorken's solution ($Y=\eta_s$, $Y$ being the fluid rapidity).
Other choices can be easily implemented.

\begin{figure}[th]
\begin{center}
\includegraphics[clip,width=.52\textwidth]{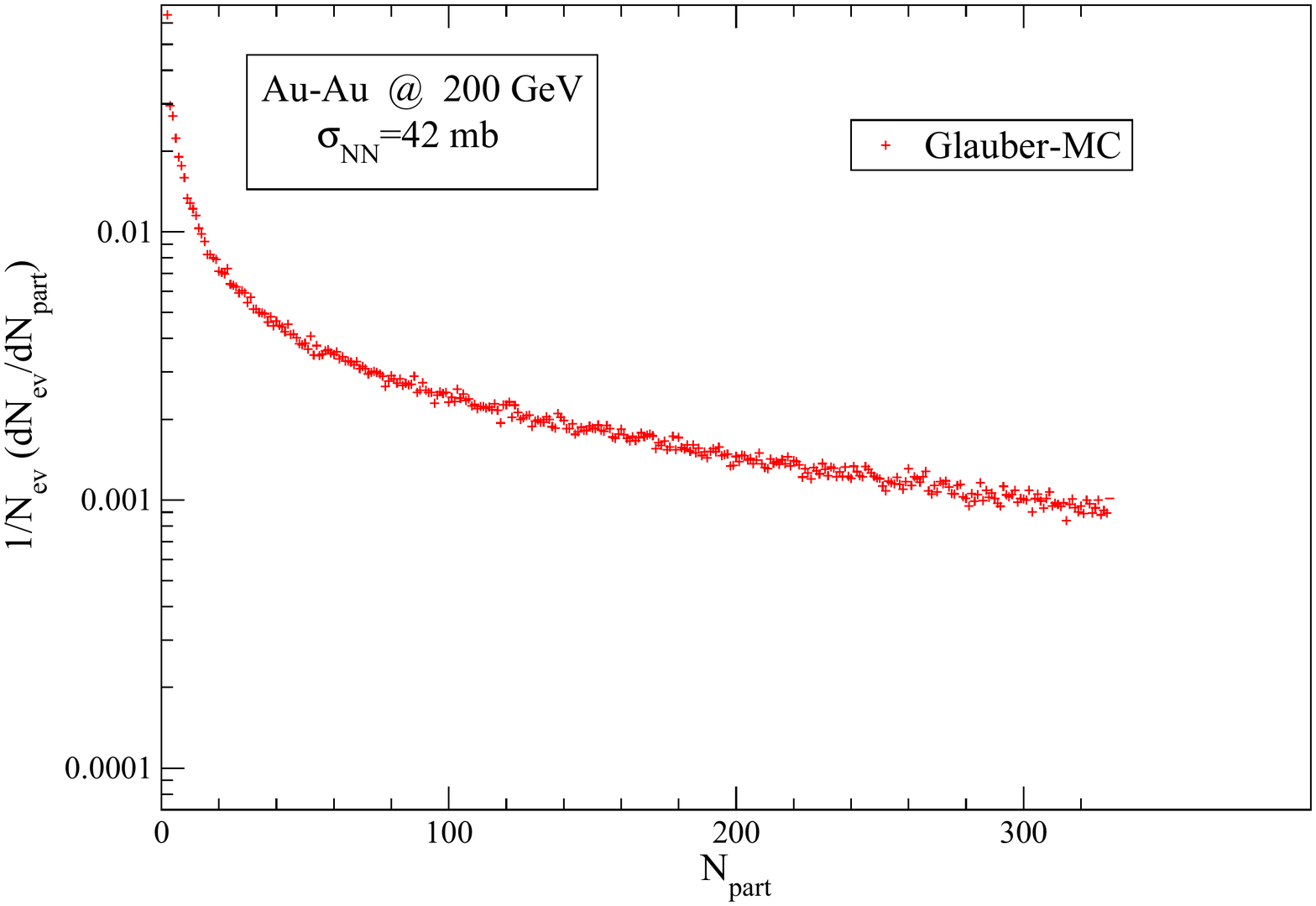}
\includegraphics[clip,width=.52\textwidth]{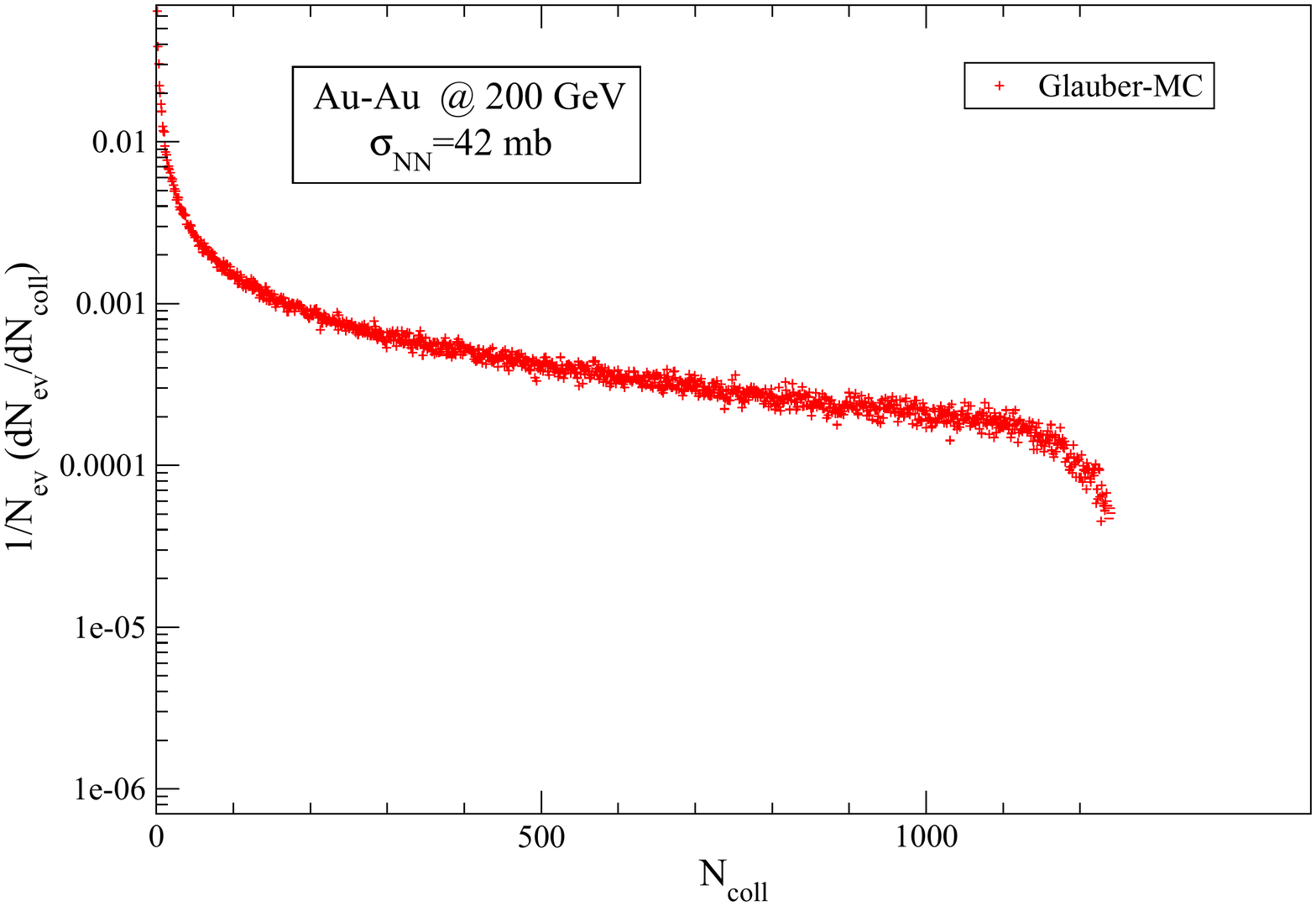}
\caption{The participant and binary-collision distribution from the Glauber-MC
  simulations of Au-Au events at $\sqrt{s_{NN}}\!=\!200$ GeV.}
\label{fig:dNdcentr}
\end{center}
\end{figure}

\section{Numerical tests}

\subsection{Numerical scheme and algorithms}

The ECHO-QGP code has been built upon the original ECHO scheme for
relativistic hydrodynamics and MHD (in any GR metric, even time dependent
like for Bjorken coordinates).  Therefore, it shares with it the finite-difference discretization, 
the conservative approach, and the shock-capturing techniques. 
The reader is referred to~\cite{luca_1} for details, here we just summarize 
the main procedures (see also~\cite{luca02}).
\begin{itemize}
\item The spatial grid is discretized along all the directions of interest
as a $N_x\times N_y\times N_z$ set of cells ($N_\eta$ in Bjorken coordinates). 
Lower dimensionality runs are always admitted, for example, 2-D
tests with boost invariance in Bjorken coordinates are performed by choosing $N_\eta =1$.
\item Physical \emph{primitive variables} are initialized for $t=0$ 
(or typically $\tau=1$ for Bjorken coordinates) as point values at cell centers.
Here, as anticipated, we choose the following set of 12 variables:
\be
\mathbf{P}=(n,v^i,P,\Pi,\pi^{ij}).
\ee
\item For each direction, primitive variables are reconstructed obtaining \emph{left} 
($\mathbf{P}_L$) and \emph{right} ($\mathbf{P}_R$) states at cell interfaces, where
the corresponding conservative variables $\mathbf{U}$ and fluxes $\mathbf{F}$ 
are also calculated according to Eq.~(\ref{echoeq1}).
\item For each component and at each intercell, upwind fluxes $\hat{F}^k$
(along direction $k$) are worked out using the so-called HLL two-state formula as
\be
\hat{F}^k = \frac{a^k_+ F^k(\mathbf{P}_L) + a^k_- F^k(\mathbf{P}_R) - a^k_+a^k_-
[U(\mathbf{P}_R) - U(\mathbf{P}_L) ]}{a^k_+ + a^k_-},
\ee
where
\be
a^k_\pm = \mathrm{max}\{ 0, \pm\lambda^k_\pm ( \mathbf{P}_L),  
\pm\lambda^k_\pm ( \mathbf{P}_R) \},
\ee
and the local fastest characteristic speeds are worked out according
to Eq.~(\ref{lambdas}), providing an approximated solution 
to the local Riemann problem.
\item High order derivatives of fluxes are calculated for each direction and source terms
are added to provide the right hand side of the evolution equations. Time derivatives
contained in some of the source terms (like in the expansion scalar) are simply 
calculated by using their values at the previous timestep.
\item The evolution equations are updated in time via a second or third order
Runge-Kutta time-stepping routine.
\item At each temporal sub-step, from the updated set of conservative variables
we must derive the set of primitive variables. We use the method described
earlier (an external cycle on $v^i$ components with a nested
Newton-Raphson root-finding method for the pressure $P$), but other choices
are possible.
\item Output of primitive variables and other diagnostic quantities are provided
for selected times.
\end{itemize}

When not otherwise specified, in the following tests we well use a second order
Runge-Kutta method for time integration and a fifth order 
routine for spatial reconstruction with \emph{monotonicity preserving} filter
(MP5, see~\cite{luca_1} and references therein). 
Notice that in viscous runs, the timestep must be also limited
by the viscous relaxation timescales~\cite{takamoto}. When these are much smaller
than the corresponding hyperbolic evolution times, the system may become
\emph{stiff} and implicit time integration may be needed. Future improvements
will adopt the same techniques used for resistive MHD described in~\cite{bucciantini2}.
Finally, the ECHO code is parallelized in order to be able to run on any supercomputing platform.


\subsection{Mildly relativistic shear flow in \1d}

The diffusion of a one-dimensional shear flow profile may be followed in time and checked
against an analytical solution, provided the flow is only mildly relativistic.
A similar case has already been studied for testing numerical algorithms for
relativistic viscosity \cite{takamoto} and also resistive magnetohydrodynamics \cite{bucciantini2}.

\begin{figure}
\begin{center}
\includegraphics[scale=0.52]{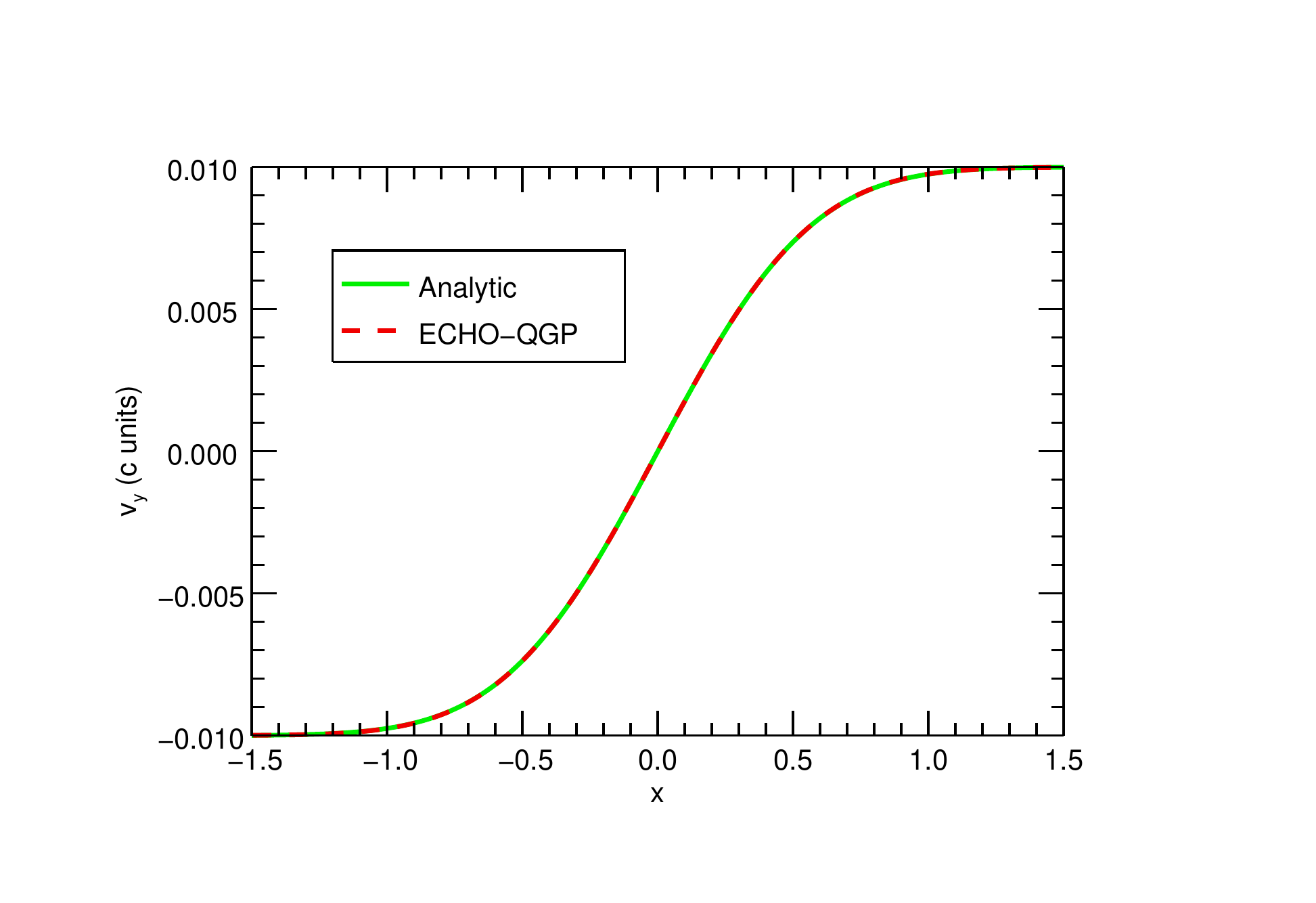}
\caption{\label{fig:takamoto} Spatial dependence of the velocity shown along with the 
analytic result at $t=10$~fm/c. The grid is made by 301 cells, ranging from $x=-1.5$ to 1.5~fm.}
\end{center}
\end{figure}

\begin{figure*}[th]
\begin{center}
 \includegraphics[clip,width=.52\textwidth]{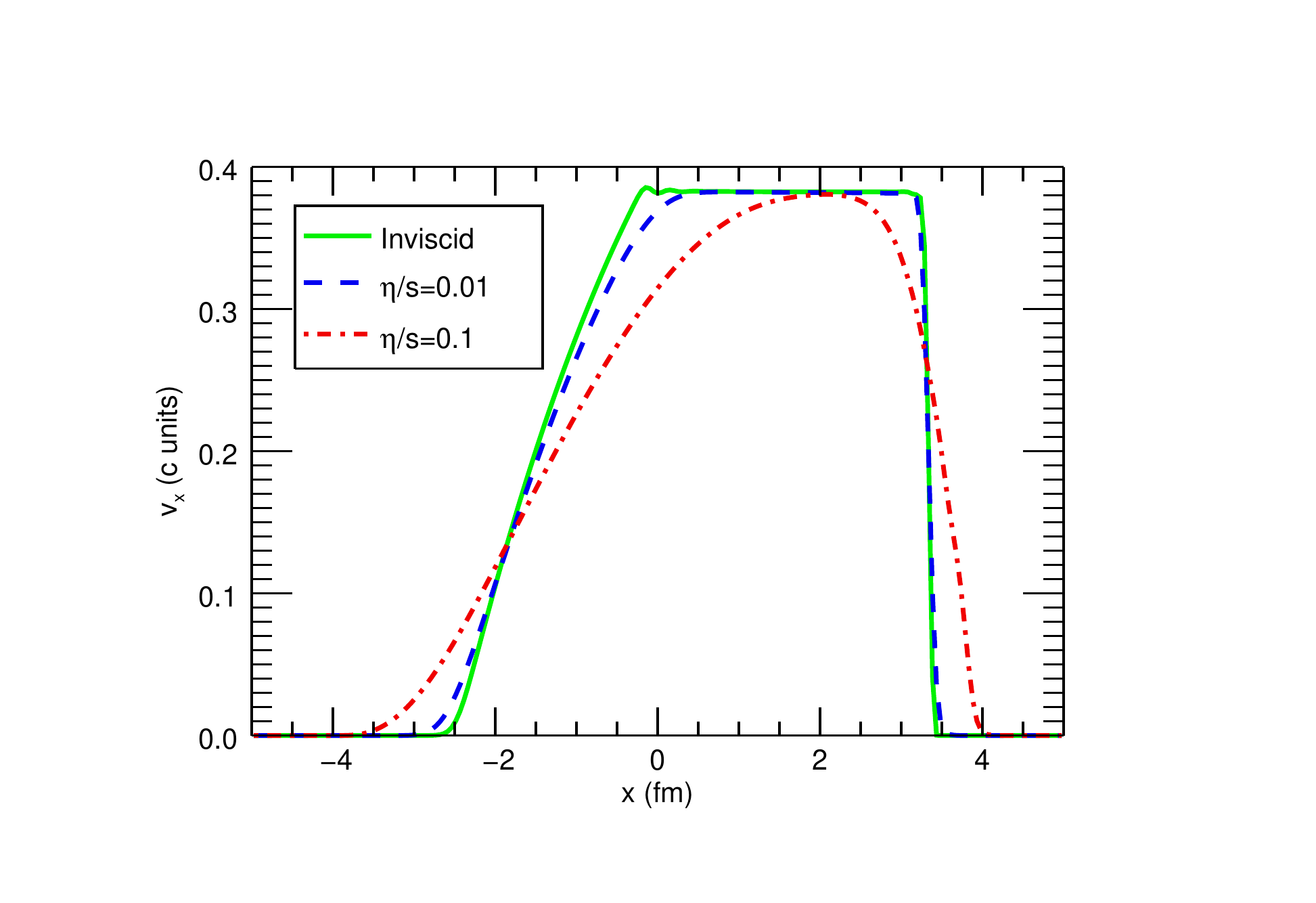}\hspace{-12mm}
\includegraphics[clip,width=.52\textwidth]{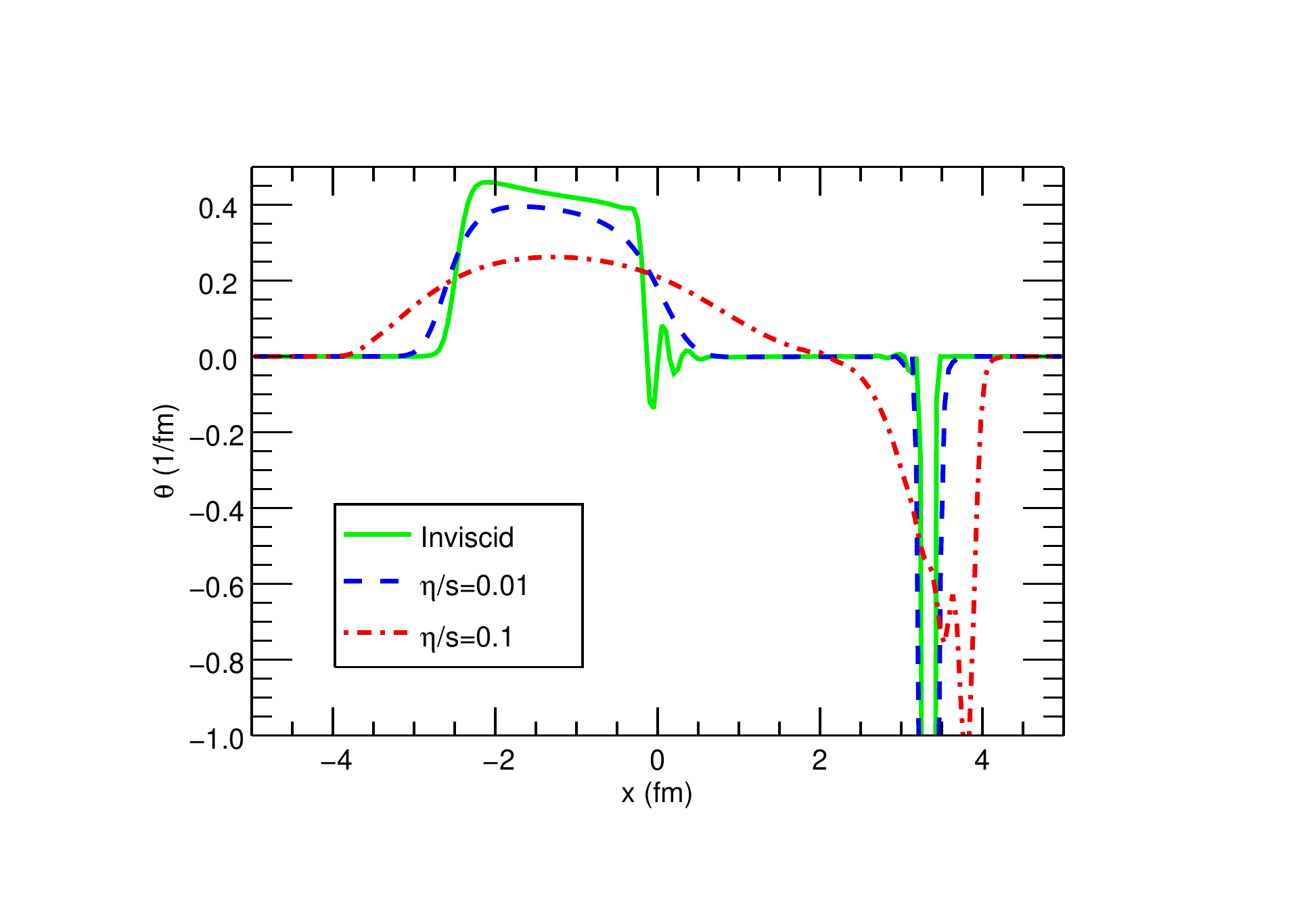}\vspace{-5mm}
\includegraphics[clip,width=.52\textwidth]{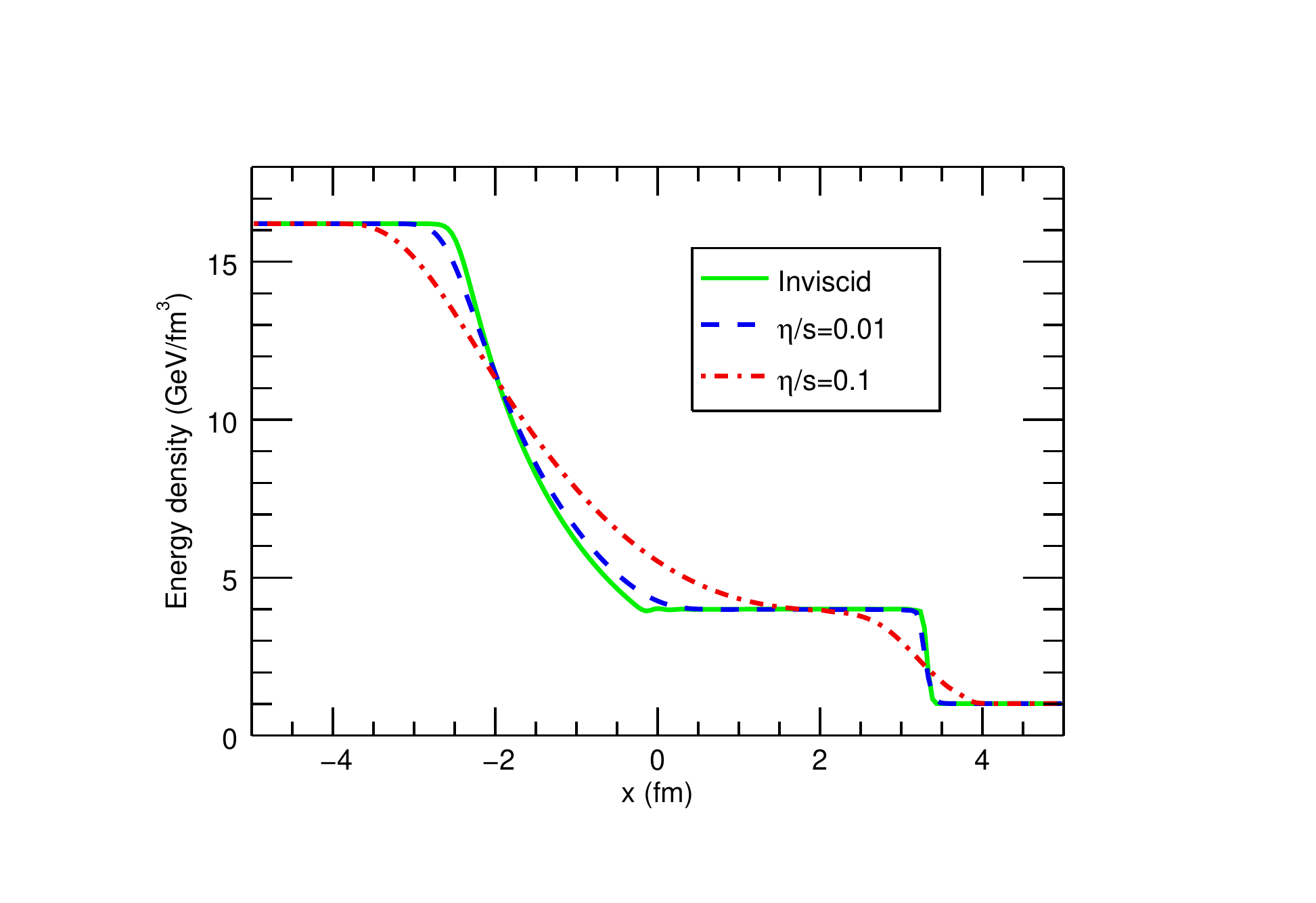}\hspace{-12mm}
\includegraphics[clip,width=.52\textwidth]{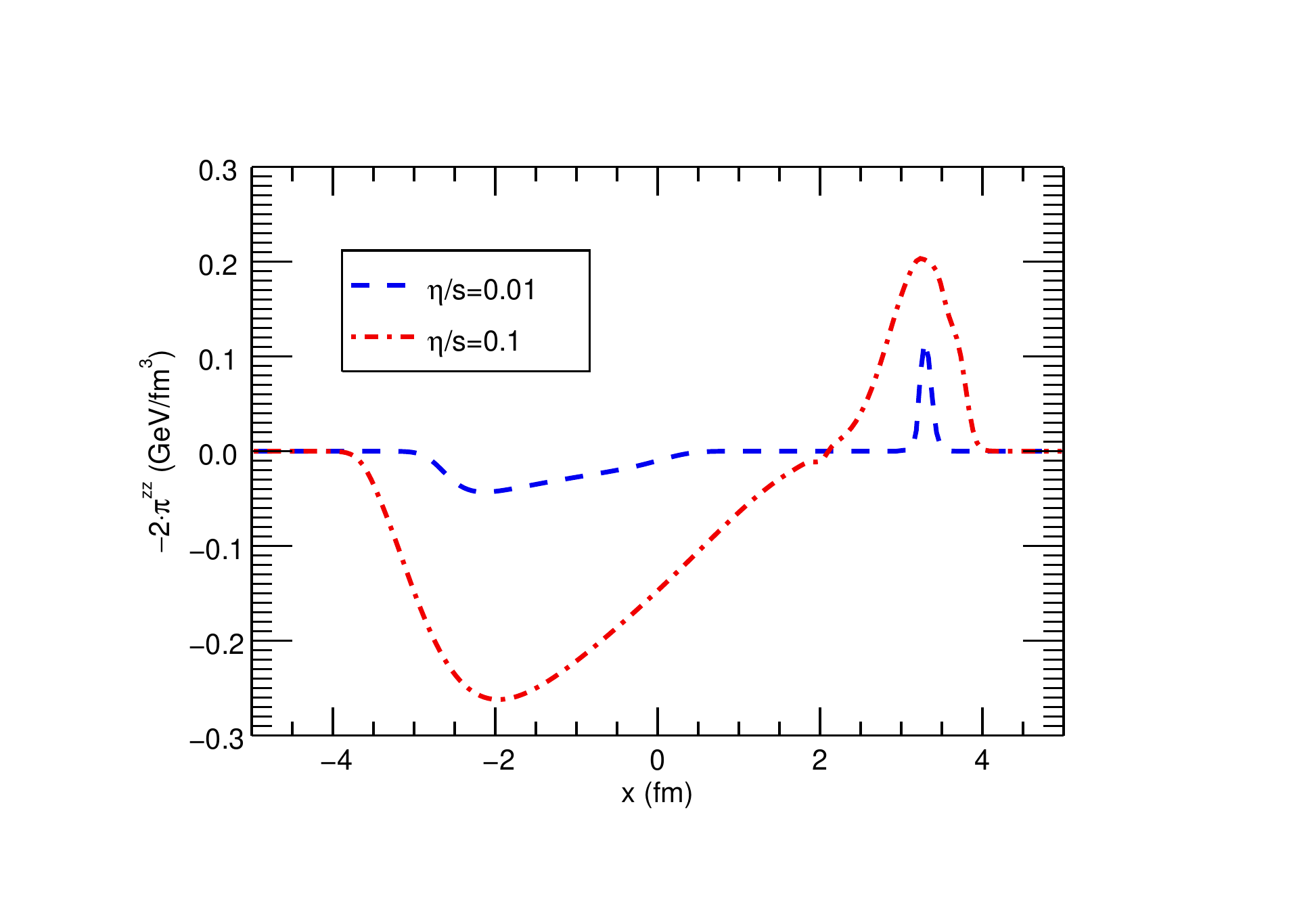}
\caption{\label{figvxth} The velocity 
component $v_x$, the expansion rate $\theta$, the energy density $e$, and $-2\pi^{zz}$ 
as a function of $x$ for $\eta/s=0,\,0.01,\, 0.1$ at $t=4$~fm/c.
The grid is made by $201\times 201$ regularly spaced cells,
with $x$ and $y$ coordinates ranging from $-5$ to $5$~fm using Minkowski coordinates.
}
\end{center}
\end{figure*}

We perform this \1d test in Minkowskian Cartesian coordinates,  choosing a velocity profile 
$v^y=v^y(x)$. For sub-relativistic speeds and a constant background
state in terms of energy density and pressure (here we use EOS-I for simplicity),
at any time $t$ of the evolution only $v^y$ will change due to shear viscosity
(the bulk viscosity does not play a role since $\theta\equiv 0$), always
preserving $\gamma \approx 1$ and $e+P\approx\mathrm{const}$. 
In such Navier-Stokes limit, the momentum 
equation along $y$ reads
 \be
 (e+P)\partial_t v^y + \partial_x \pi^{xy}=0, \quad \pi^{xy} = -2\eta\sigma^{xy} =-\eta \partial_x v^y,
 \ee
 which leads, for a constant $\eta$ coefficient, to the classical 1D diffusion equation
 \be
 \partial_t v^y = D_\eta \partial_x^2 v^y,\quad D_\eta=\eta/(e+P),
 \ee
with (constant) diffusion coefficient $D_\eta$.
If we now assume that $v^y(x)$ has a step function profile for $t=0$,
with constant values $-v_0$ for $x<0$ and $v_0$ for $x>0$, the 
exact solution at any time $t$ is known to be
\be
v^y=v_0 \erf \bigg[\frac{1}{2}\sqrt{\frac{x^2}{D_\eta t}}\bigg].
\ee
We run the ECHO-QGP code with these conditions and, instead of starting from the discontinuous
solution, we assume the above profile at the initial time $t=1$~fm/c, comparing
the analytical result with the output at a later time $t=10$~fm/c.
The spatial dependence of the fluid velocity is shown in Fig.~\ref{fig:takamoto}, where
we have assumed $v_0=0.01$, $e+P=4P=1$~GeV/fm$^3$, 
$\eta = 0.01\,\mathrm{GeV/fm}^2$, hence $D_\eta=0.01$~fm.
Our result matches the analytic solution throughout the evolution. The numerical grid
is $x=[-1.5,1.5]$~fm, and $N_x=301$ numerical cells have been employed.

\subsection{Shock-tube problem in \2d}

Shock-capturing numerical schemes, as in the classical hydrodynamical case,
are designed to handle and evolve discontinuous quantities invariably arising
due to the nonlinear nature of the fluid equations. In order to validate these codes,
typical tests are the so-called shock-tube 1D problems. 
Two constant states are taken on the left and on the right with respect of an imaginary 
diaphragm, which is supposed to be initially present and then removed.
Typical patterns seen in the subsequent evolution are 
shocks and rarefaction waves. Here we present a relativistic blast wave explosion problem,
characterized by an initial static state with temperature and pressure much higher 
in the region on the left, namely $T^L=0.4$~GeV ($P^L=5.40$~GeV/fm$^3$)
and $T^R=0.2$~GeV ($P=0.34$~GeV/fm$^3$), as in \cite{molnar}, though the
EOS used is not coincident (thus also results are quantitatively different).

For a more stringent test of our code, we employ this shock-tube test by placing the initial
diaphragm along the diagonal of a square box
(201 points and size $10~\mathrm{fm}$ along $x$ and $y$)
adopting Minkowskian Cartesian coordinates, and we let the system evolve from 
$t=1$ up to $t=4$~fm/c with EOS-I and different values of the shear viscosity $\eta/s$.

In Fig.~\ref{figvxth} we show $v_x$, the expansion scalar $\theta$, $e$, and $-2\pi^{zz}$, as in \cite{molnar},
at the final time as a function of $x$ and along $y=0$. 
Notice the high accuracy of the results and the absence of numerical spurious oscillations
near the shock front in the ideal case. Gibbs-like effects are visible
just in the expansion scalar, because spatial derivatives in $\theta$ and
$\sigma^{ij}$ are calculated with central schemes (known to fail in the presence of discontinuities),
but these quantities are not used in the ideal case.
When increasing $\eta/s$, quantities are clearly damped with respect to the ideal case, as expected.

\subsection{Boost-invariant expansion along $z$-axis}

As a first validation of ECHO-QGP in Bjorken coordinates we consider
a test with no dependence on the transverse coordinates $(x, y)$
(then the vorticity vanishes), and we assume boost invariance along $z$-direction,
thus quantities do not depend on $\eta_s$ either, and we are actually dealing
with a  (0+1)-D test case. Evolution of uniform quantities will be then just due
to the $\tau$ dependence of the $g_{\eta\eta}$ term in the metric tensor,
in the absence of velocities.
The energy-momentum tensor simplifies to
\be
T^{\mu\nu}\equiv \mathrm{diag}\{e, \ P+\Pi+\pi^{xx}, \ P+\Pi+\pi^{yy},\ (P+\Pi)/\tau^2+\pi^{\eta\eta}\}
\ee
with $\pi^{xx}$, $\pi^{yy}$, and $\pi^{\eta\eta}$ the only non-vanishing components of $\pi^{\mu\nu}$.
Owing to the tracelessness of $\pi^{\mu\nu}$ and to the assumed symmetries, 
one can write $2\pi^{xx}\!=\!2\pi^{yy}\!=\!-\tau^2 \pi^{\eta\eta}\!\equiv\! \phi$.
Therefore, we need only one independent component to specify the shear viscous tensor $\pi^{\mu\nu}$,
though we remind here that in ECHO-QGP all 6 spatial components are evolved
and the trace-free condition is never imposed.
A constant initial energy density profile is chosen at $\tau=1$~fm/c and
the system is then left free to evolve.

\begin{figure}[thb]
\begin{center}
\includegraphics[scale=.52]{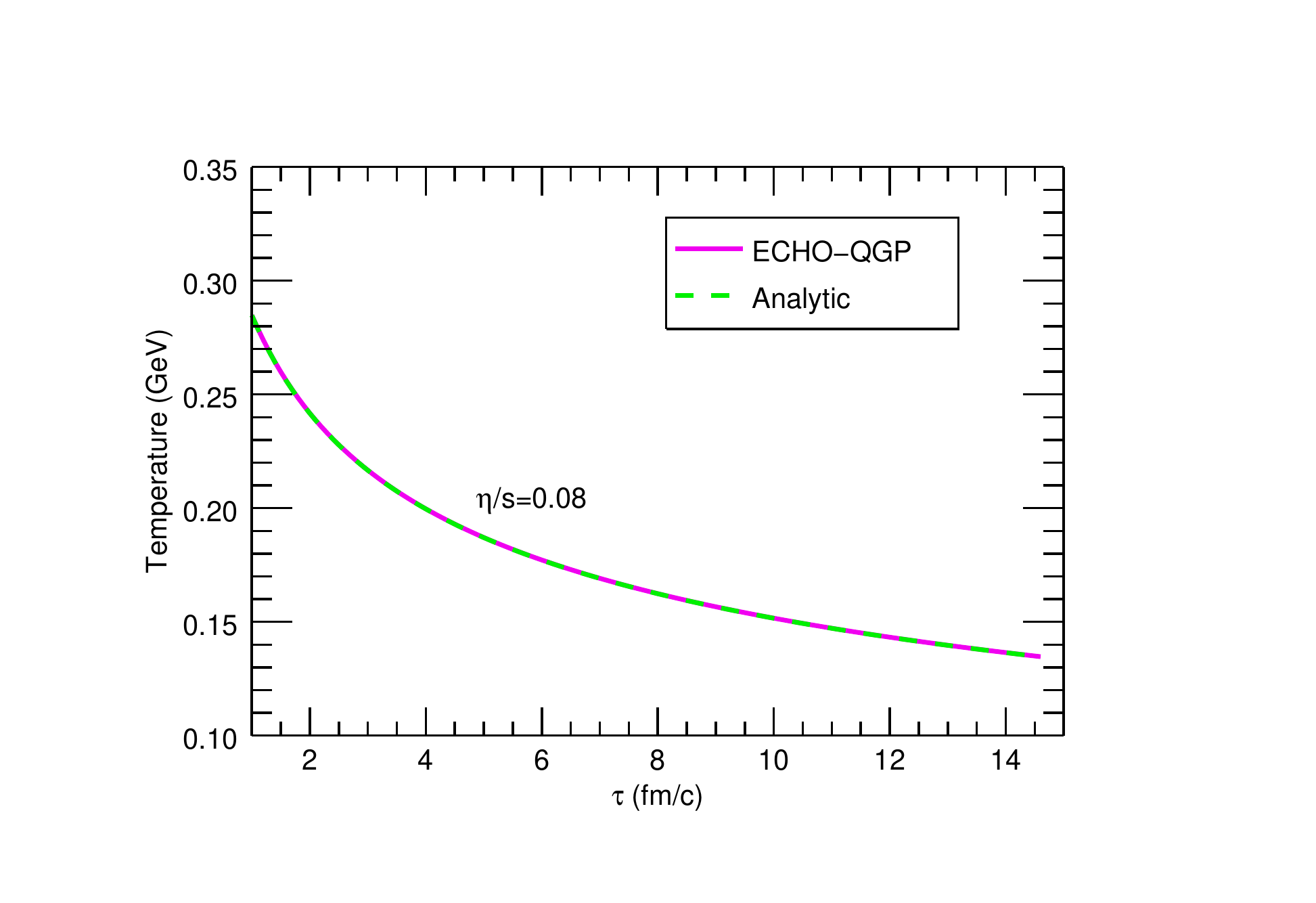}
\caption{Comparison between the evolution of $T(\tau)$ computed by ECHO-QGP
applying the first order Navier-Stokes expressions for viscous fluxes and the analytic solution. 
The initial time is set to $\tau_0\!=\!1$~fm/c and the viscosity-to-entropy ratio is $\eta/s\!=\!0.08$.}
\label{fig:tempns}
\end{center}
\end{figure}

\begin{figure}[thb]
\begin{center}
\vspace{-5mm}
\includegraphics[scale=0.52]{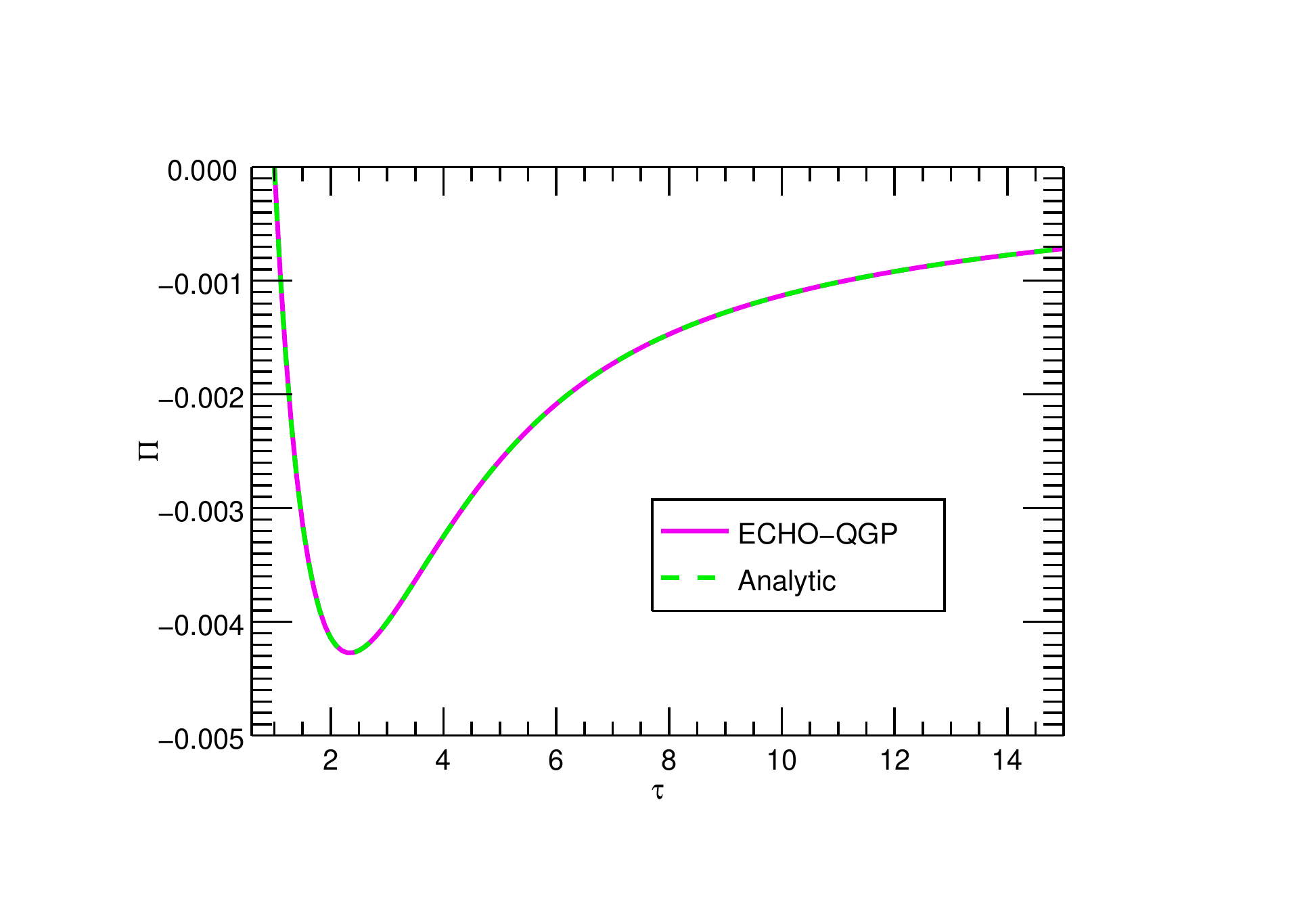}
\vspace{-10mm}
\caption{Comparison between the evolution of $\Pi(\tau)$ computed by ECHO-QGP 
and the semi-analytic solution describe in the text. The parameters are 
$\tau_0=1$~fm/c, $\zeta=0.01\,\mathrm{GeV/fm}^2$, and $\tau_{\pi}=1$~fm/c.}
\label{fig:bulktau}
\end{center}
\end{figure}

The code outputs in time can be actually checked against an analytical solution,
provided first order theory applies.
The energy equation is then enough to describe the overall evolution
\be
\label{eqpi}
\frac{\partial e}{\partial \tau}=-\frac{e+P +\Pi-\phi}{\tau},
\ee
where in the first-order theory $\Pi$ and $\phi$ are obtained from their Navier-Stokes (NS) values
\be
\Pi=-\frac{\zeta}{\tau},\qquad \phi=\frac{4\eta}{3 \tau}.
\ee
In this case, employing EOS-I
and assuming constant values for $\eta/s$ and for $\zeta/s$, Eq.~(\ref{eqpi}) 
admits the following analytic solution~\cite{mura,roma1,chandra_quarkonia} for the
temperature as a function of the proper time:
\be
T(\tau)=T_0 \left(\frac{\tau_0}{\tau}\right)^{\frac{1}{3}} 
\left[1+\frac{4\eta/3s+\zeta/s}{2\tau_0 T_0}
\left(1-\left(\frac{\tau_0}{\tau}\right)^{\frac{2}{3}}\right)\right],
\ee
where $T_0$ is the temperature at the initial time $\tau_0$.
ECHO-QGP reproduces this analytic solution in the NS limit, as displayed in Fig.~\ref{fig:tempns}
(we have chosen $\zeta/s=0$ here).

On the other hand, within the second-order theory, these equations do not admit any analytic solution.
However, the evolution equations of $\Pi$ and $\phi$, in the case in which their evolution is 
simply governed by the relaxation part of the source terms 
(for simplicity we set as usual $\tau_\pi\!=\!\tau_\Pi$), are
\be
\label{anapi}
\frac{\partial \Pi}{\partial \tau}=-\frac{1}{\tau_\pi}\left(\Pi+\frac{\zeta}{\tau}\right), \quad
\frac{\partial \phi}{\partial \tau}=-\frac{1}{\tau_\pi}\left(\phi-\frac{4\eta}{3 \tau}\right).
\ee
Assuming $\eta$, $\zeta$ and $\tau_\pi$ to be independent of the temperature,
 Eqs.~(\ref{anapi}) admits the semi-analytic solution for $\Pi$ and $\phi$
\ba
\Pi(\tau)&=&\Pi(\tau_0) \exp[-(\tau-\tau_0)/\tau_\pi]
\nn&&+\frac{\zeta}{\tau_\pi}\exp(-\tau/\tau_\pi) [{\rm Ei}(\tau_0/\tau_\pi)-{\rm Ei}(\tau/\tau_\pi) ],\nonumber\\
\phi(\tau)&=&\phi(\tau_0) \exp[-(\tau-\tau_0)/\tau_\pi]
\nn&&-\frac{4 \eta}{3\tau_\pi}\exp(-\tau/\tau_\pi) [{\rm Ei}(\tau_0/\tau_\pi)-{\rm Ei}(\tau/\tau_\pi) ],
\ea
where, ${\rm Ei}(x)$ denotes the exponential integral function.
The solution for $\Pi$ obtained from ECHO-QGP under the same assumptions 
is plotted in Fig.~\ref{fig:bulktau}, and it perfectly agrees with the analytic result. 

\subsection{\2d tests with azimuthal symmetry}

Let us now consider a couple of inviscid tests in Bjorken coordinates, again
assuming EOS-I, $P=e/3$, boost invariance, thus $\partial_\eta\equiv 0$, 
but here also evolution in the transverse plane. However, when the
initial state at $\tau_0$ is azimuthally invariant, the \2d evolution
with ECHO-QGP can be compared with analytic results in \1d.

\begin{figure*}[th]
\begin{center}
\includegraphics[clip,width=.52\textwidth]{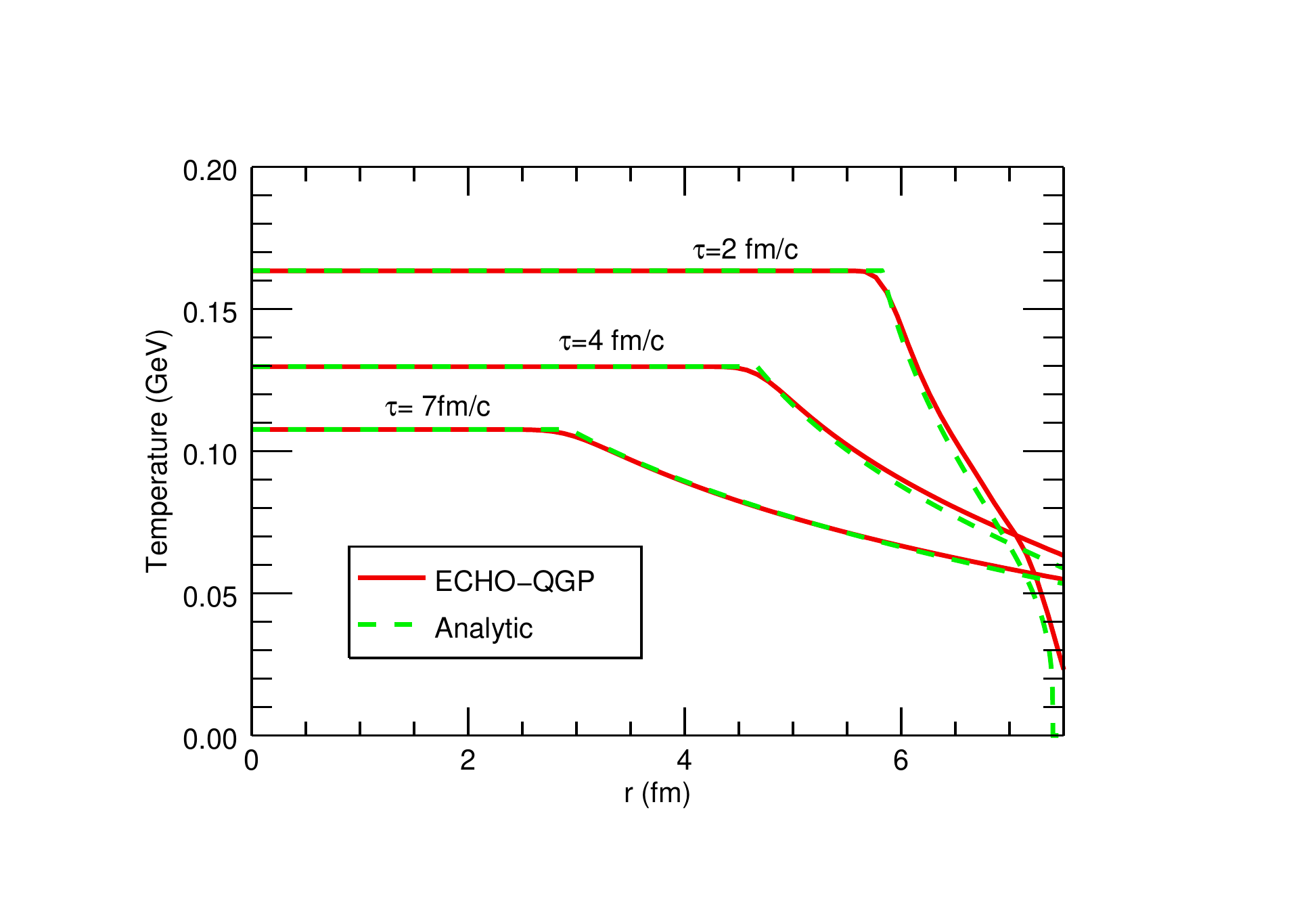}\hspace{-10mm}
\includegraphics[clip,width=.52\textwidth]{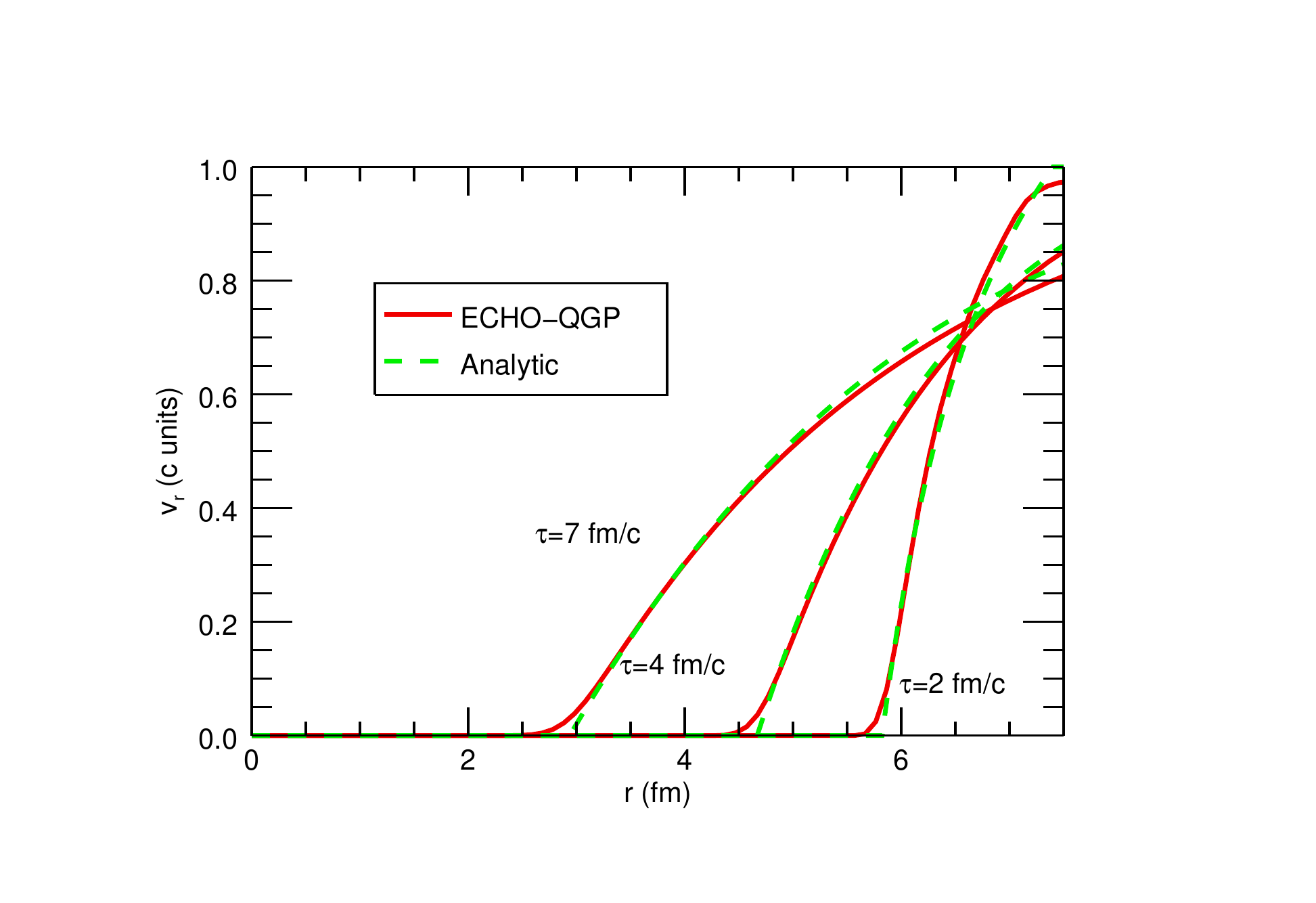}
\caption{\label{figbym}
Spatial dependence of the temperature and of the radial velocity at different times 
along with the analytic solution in the case of a Woods-Saxon initial
condition with cylindrical symmetry. Results obtained with ECHO-QGP in \2d
agree very well with the analytical solution by Baym {\it et al.}~\cite{baym}.
}
\end{center}
\end{figure*}

\begin{figure*}[th]
\begin{center}
\includegraphics[clip,width=.52\textwidth]{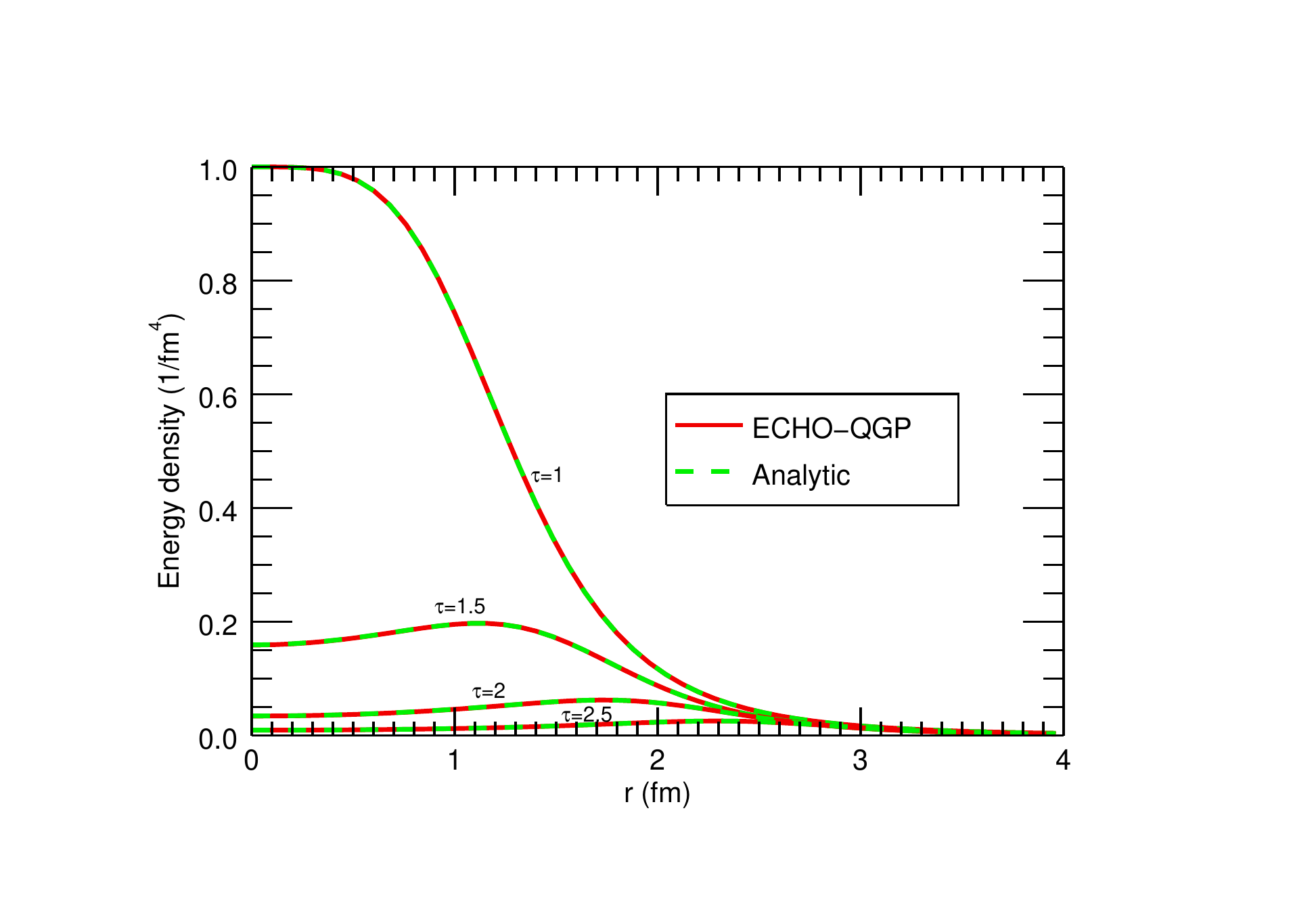}\hspace{-10mm}
\includegraphics[clip,width=.52\textwidth]{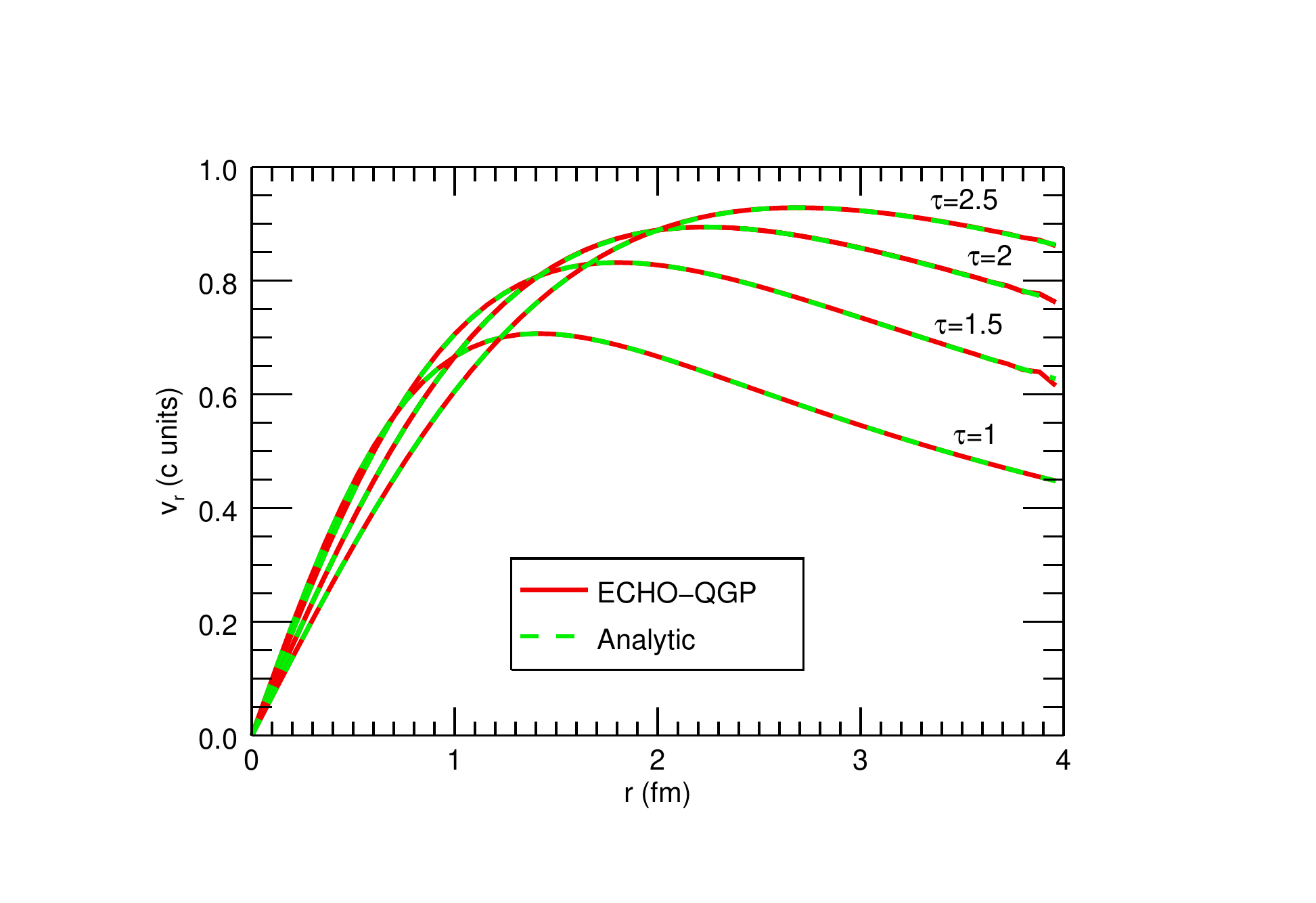}
\caption{\label{fig:gubser} Comparison of the radial dependence of
the energy density $e$ (left panel) and of the radial velocity $v_r$ (right panel)
with the Gubser flow~\cite{gubser} (inviscid case) at $\tau=1.0,\ 1.5,\ 2.0, \ 2.5$ fm/c.
ECHO-QGP outcomes show a perfect matching with the analytical results. 
The EOS is chosen to be EOS-I.}
\end{center}
\end{figure*}

In the first case, a Woods-Saxon profile for the initial energy density,
as appropriate for central nucleus-nucleus collisions, is assumed
 \be
e(r,\tau_0)=\frac{e_0}{1+\exp{[(r-R)/\sigma]}},
\label{woodsaxon}
\ee
where $\tau_0$ is the initial time, $r=(x^2+y^2)^{1/2}$ is the radius in the transverse plane, 
and $R$ can be thought of as the radius of the nuclei. The analytical solution for 
the subsequent evolution, as a function of $\tau$ and $r$, was found in~\cite{baym} 
and it will be compared with our numerical results. The run parameters are
$R\!=\!6.4$~fm, $\sigma\!=\!0.02$~fm and an 
initial temperature $T_0\!=\!0.2$~GeV at $r=0$.
As shown in Fig.~\ref{figbym}, there is a perfect agreement between the ECHO-QGP 
results at any time $\tau$ and the analytic solution.

Recently, Gubser~\cite{gubser} has derived another analytic solution for 
a \1d conformal fluid (thus with $P=e/3$) with azimuthal symmetry in the transverse plane.  
Then also this test can be used for a further numerical validation of ECHO-QGP in \2d.
The analytic solution reads
\ba
e &=& \frac{\hat{e}_0}{\tau^{4/3}} (2q)^{8/3}\bigg[1+2q^2(\tau^2+r^2)+ q^4(\tau^2-r^2)^2\bigg]^{-4/3},\nn
u^\tau &=&\cosh[k(\tau,r)] , \quad u^\eta  = 0, \nn
u^x &=&\frac{x}{r}\sinh[k(\tau,r)], \quad u^y= \frac{y}{r}\sinh[k(\tau,r)],
\ea
where
\be
k(\tau,r)\!=\! {\rm arctanh\,}\left(\frac{2 q^2 \tau r}{1+q^2\tau^2+q^2 r^2}\right),
\ee
and the parameter $q$ is a parameter with the dimension of an inverse length (we set it to 1~fm$^{-1}$).
To perform the test we choose $\hat{e}_0=1$ and we set
the initial profiles from the above solutions at $\tau_0=1$~fm/c; then 
we record outputs for the energy density and the radial velocity 
$v_r\!=\! (u_x^2+u_y^2)^{1/2}/u^\tau$ for increasing values of $\tau$. 
The ECHO-QGP results along with the analytic solution are 
shown in Fig.~\ref{fig:gubser} and they show perfect agreement.

\subsection{\3d test with spherical symmetry}

\begin{figure*}[th]
\begin{center}
\includegraphics[clip,width=.52\textwidth]{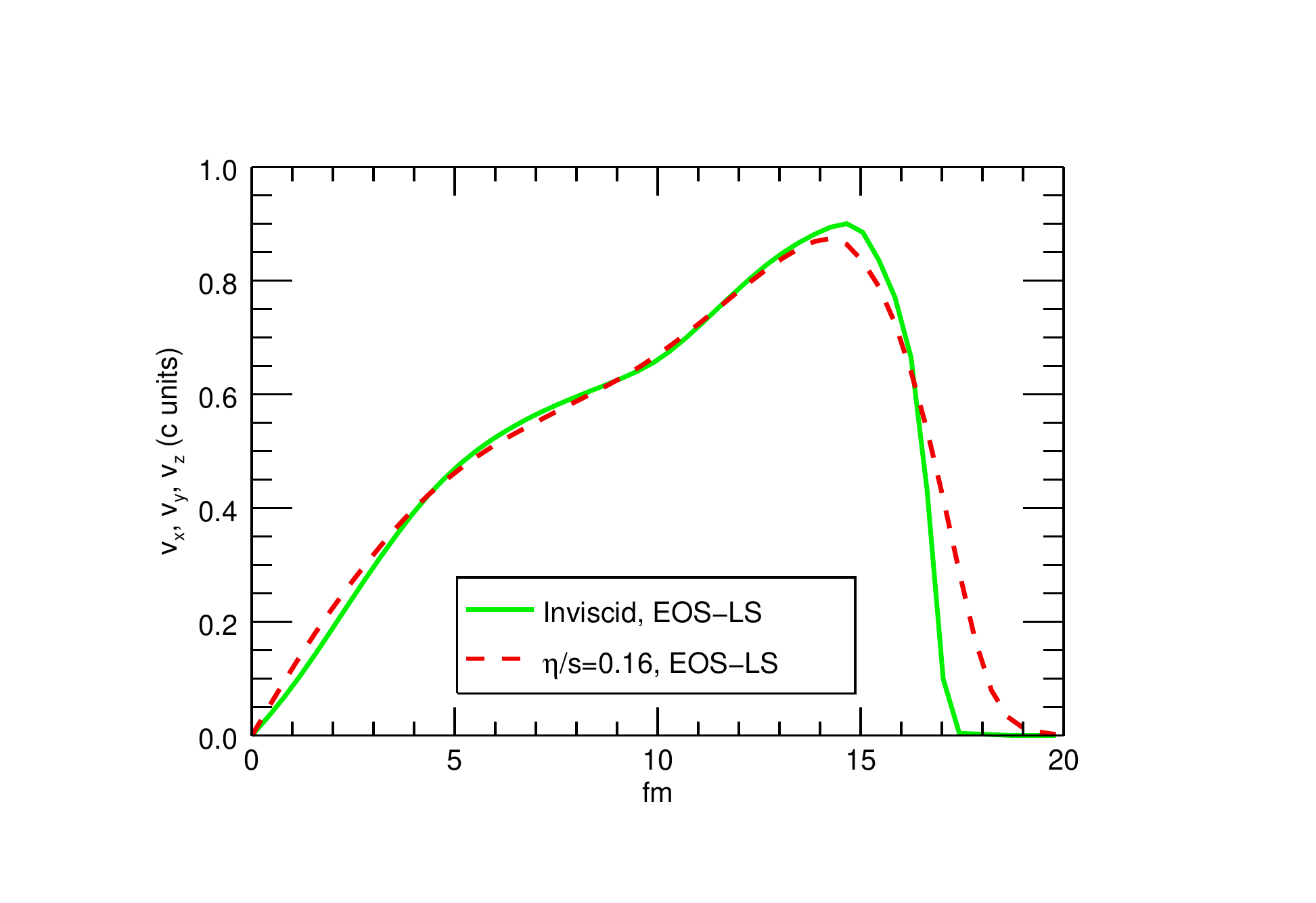}
\includegraphics[clip,width=.52\textwidth]{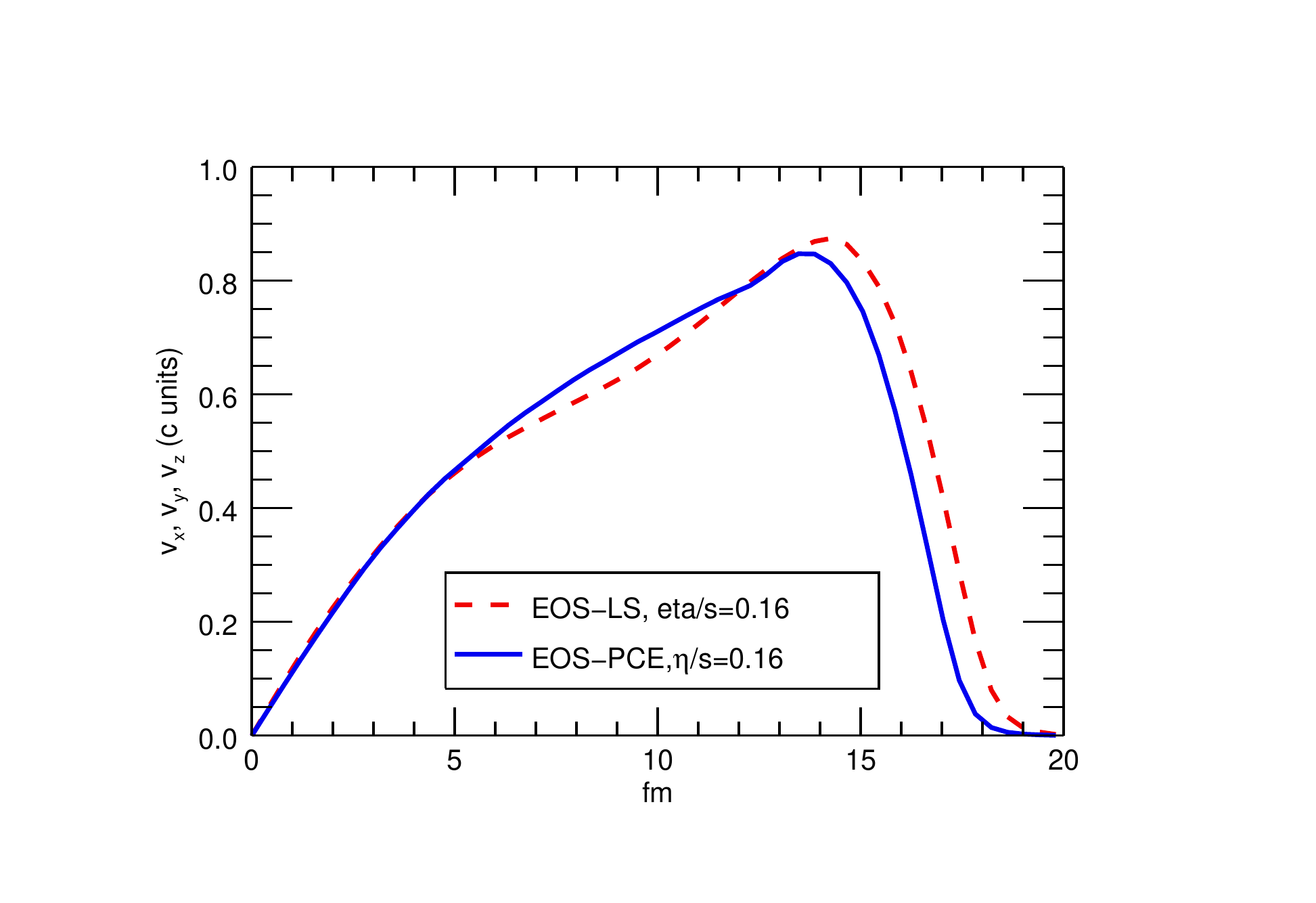}
\caption{\label{fig:3dsph} Comparison at $t=10$~fm/c of  the spatial components of fluid 
velocity in a 3D run in Minkowski coordinates. 
In the left panel we compare inviscid and viscous runs, while EOS-LS and EOS-PCE 
are compared in the right panel, for a viscous fluid with $\eta/s=0.16$.}
\end{center}
\end{figure*}

We test the ECHO-QGP code in the \3d case in the presence of 
a spherically symmetric initial pressure or energy-density profile. 
This test is essential to check the correctness of the viscous implementation 
by checking the symmetries which are preserved by the velocity components 
during the whole fireball evolution. Here, we
investigate the spatial components of the velocity, $v_x$, $v_y$, and $v_z$.
Since this system possesses spherical symmetry we expect a pure radial dependence 
of the fluid velocity $\vec{v}\!=\!v(r,t)\vec{r}/r$ throughout all the medium evolution, for both inviscid and viscous fluids.

To perform this test, the initial pressure profile is
chosen to be of Woods-Saxon type as in Eq.~(\ref{woodsaxon}), with $P$ and $P_0$
replacing $e$ and $e_0$, now in flat Cartesian
coordinates with $r=(x^2+y^2+z^2)^{1/2}$. The other parameters are chosen as
$\sigma=0.5$~fm, $R=6.4$~fm, $P_0=4$~GeV/fm$^3$, such that the initial temperature is $0.307$ GeV,
and tests are performed with either EOS-LS and EOS-PCE, precisely to investigate
the behavior of different EOS's in a realistic \3d case.
The grid extends from $-20$ to $20$~fm, with 101 cells, along all three directions.
The fluid 4-velocity at the initial time is $u^\mu=(1,0,0,0)$, 
and in the viscous case we initialize $\pi^{\mu\nu}\equiv 0$ (here $\zeta/s=\Pi=0$),
since we do not have boosting effects in Minkowski.
Viscous runs are performed with $\eta/s=0.16$  for both EOS-LS and EOS-PCE.
Simulations are performed until 
$t=10$~fm/c, and we have recorded $v_x$, $v_y$, $v_z$ along their respective axes 
and plotted them in Fig.~\ref{fig:3dsph}. All of them perfecly
lie on top of each other, both for the inviscid and the viscous cases. 
Shear viscous effects play the usual role of smoothening the velocity profiles, as expected.

\section{The algorithm for particle spectra}

Before illustrating ECHO-QGP results of physical interest for
heavy-ion collisions, it is mandatory to implement a
decoupling routine accounting for the transition from the fluid 
description to the final hadronic observables 
to compare with the data and other authors' results.

The process of decoupling of hadrons from the fireball and their
subsequent propagation in space-time is very complex and there are
different recipes to model it. The most used scheme is based on the
notion of freeze out. Since the particle mean free paths 
strongly depend on the temperature of the medium
one can assume that below a certain temperature $T_{freeze}$
particles stop interacting within the fireball and they propagate as
free streaming particles. 
This is the so called kinetic freeze-out and
corresponds to the end of the hydrodynamical evolution of the system.
In this scheme the hadron spectra are calculated using the
Cooper-Frye prescription \cite{Cooper:1974mv}: 
from the temperature profiles obtained within 
the hydrodynamic simulation 
one first determines the hypersurface $\Sigma$
of constant temperature $T=T_{freeze}$ and the total emission of primary
particles is then calculated as a sum of the thermal emission of
cells lying on the freeze-out hypersurface. Corrections to the
particle spectra related to the decay of unstable particles have been shown 
to be significant and they must be included to reproduce the 
experimental data \cite{Sollfrank:1990qz,Sollfrank:1991xm}.

In the last years, hybrid approaches have been proposed in which the
decoupling is treated as a switch, at a certain temperature
$T_{switch}$, from a hydrodynamical description of the fireball to a
particle transport description~\cite{Bass:2000ib,Teaney:2001av,Hirano:2005xf,Nonaka:2006yn,Petersen:2008dd,Werner:2010aa,Song:2010mg,Karpenko:2012yf,Hirano:2012kj,Huovinen:2012is}. 

For the sake of simplicity and for performing our first tests of ECHO-QGP,
we adopt here the freeze-out scheme and retain the hybrid approach
as an important outlook of our work.
Let us now briefly review the formalism used for calculating 
the particle spectra within the Cooper-Frye scheme.
The momentum spectrum of hadrons of species $i$ is written as
\be
 E\dfrac{d^3N_i}{dp^3}=\dfrac{d^3N_i}{d	\mathrm{y} p_Tdp_Td\phi}=
\dfrac{g_i}{{(2\pi)}^3}\int_\Sigma{ \dfrac{-p^\mu d^3\Sigma_\mu}
{{\exp\left[{-\frac{u^\mu p_\mu + \mu_i}{T_{\mathrm{freeze}}}}\right]}\pm 1}},
\label{cooperfrye}
\ee
where the index $i$ refers to the hadrons such as pions, kaons, protons etc,
$g_i$ and $\mu_i$ are the corresponding degeneracy and chemical potential and 
finally $p_{\mu}$ is the four momentum of the particle.

An improvement with respect to the pure kinetic freeze-out distinguishes between
the temperature at which elastic interactions between particles cease,
$T_{\mathrm{freeze}}$, and the chemical freeze out temperature
$T_{\mathrm{c}}$ at which just the inelastic interactions cease.
As explained before, below $T_{\mathrm{c}}$ a PCE equation of state
is computed which allows to determine the chemical potentials
$\mu_i$ of each ``frozen particle'' at decoupling.
In this work we will use the PCE EOS shown in Fig.~\ref{fig:EOS}
and also the one presented in~\cite{Kolb:2002ve}.

The use of Eq.~(\ref{cooperfrye}) requires to evaluate the hypersurface
$\Sigma$ of constant temperature. In \3d, determining such a surface
is computationally quite demanding because of the many different possibilities in which
the 3D hypersurface can intersect the 4D hypercubes of the
hydrodynamical simulation grid \cite{Huovinen:2012is}. Here we follow a simpler method also
used in \cite{Hirano:2012kj}: we can imagine the hypersurface to
be the collection of the hypercubes' faces of those neighbours cells
which are respectively above and below the threshold
$T_{\mathrm{freeze}}$. In this case the $d^3\Sigma^\mu$ is composed by 
the sum (in Bjorken coordinates)
\begin{align} d^3\Sigma^\mu=\left(
\begin{array}{r}
  dV^{\perp \tau} \\
  dV^{\perp x} \\
  dV^{\perp y} \\
  dV^{\perp \eta}
\end{array}\right)
=  \left(
\begin{array}{r}
 \tau \: \Delta x\Delta y\Delta \eta_s \: s^\tau \\
 \tau \: \Delta y\Delta \eta_s \Delta \tau  \: s^x \\
 \tau \: \Delta \eta_s \Delta \tau \Delta x \: s^y \\
  \frac{1}{\tau}\:\Delta \tau \Delta x \Delta y \: s^\eta
\end{array}\right),
\label{eq:dsigma}
\end{align}
where each volume element of the hypersurface is oriented by the vector
\begin{align}
s^\mu= - \mathrm{sign} \left( \dfrac{\partial T}{\partial x^\mu} \right).
\end{align}
In this way, we associate
to each of these cells a normal unitary vector oriented toward
the direction of negative temperature gradient.

In most cases only one of the components of $d^3\Sigma^\mu$ is
different from zero, since the $dV^{\perp\mu}$ is added only if the
freeze out condition is fulfilled. Let us label with $T_A$ the
temperature in an arbitrary cell, and $T_B$ the temperature of its
neighbour in the positive $\mu$ (with $\mu$ running over the four
dimensions). As a first approximation,
if $(T_A-T_{\mathrm{freeze}})(T_{\mathrm{freeze}}-T_B)>0$ then the
hypersurface contains the element $dV^{\perp \mu}$ relative to those
cells and direction $\mu$. A more refined procedure that we here adopt is to
construct a cell with values of temperature and four velocity
interpolated between the cells A and B. This construction allows
to compute the scalar product in the numerator of (\ref{cooperfrye})
at each hypersurface cell and could give a positive or negative
contribution to the total spectrum depending on the orientation of the cell
and the orientation of the four momentum of the particle.

Once the hypersurface is determined, one can calculate
the spectra as functions of the four momentum $p^{\mu}$ which,
in the Bjorken coordinates, reads
\begin{align}
p^\mu=\left(m_\mathrm{T}\cosh(\mathrm{y} -\eta_s ), p_\mathrm{T} \cos\phi, 
p_\mathrm{T} \sin\phi, \frac{1}{\tau} m_\mathrm{T}\sinh(\mathrm{y} -\eta_s )\right),
\end{align}
where $\mathrm{y}$ is the rapidity, $p_\mathrm{T}$ the transverse momentum
and \mbox{ $m_{\mathrm{T}}={\left({p_\mathrm{T}^2+m^2}\right)}^{1/2}$} the transverse mass.

The observables we will consider in this paper are the transverse spectrum 
at midrapidity ($\mathrm{y}=0$) averaged over the \mbox{angle $\phi$}
\begin{align}
\frac{1}{2\pi}\int_0^{2\pi}\dfrac{d^3N_i}{p_{\mathrm{T}}
dp_{\mathrm{T}}d\mathrm{y} d\phi}(\mathrm{y}=0,p_{\mathrm{T}},\phi)d\phi,
\end{align}
the elliptic flow coefficient $v_2$
\begin{align}
v_2=\frac{\int_0^{2\pi}\dfrac{d^3N_i}{p_{\mathrm{T}}dp_{\mathrm{T}}d\mathrm{y} d\phi}(\mathrm{y}=0,p_{\mathrm{T}},\phi)\cos(2\phi)d\phi}
{\int_0^{2\pi}\dfrac{d^3N_i}{p_{\mathrm{T}}dp_{\mathrm{T}}d\mathrm{y} d\phi}(\mathrm{y}=0,p_{\mathrm{T}},\phi)d\phi},
\end{align}
and the rapidity spectrum
\begin{align}
\frac{dN_i}{d\mathrm{y}}=\int_0^{2\pi}\int_0^{+\infty}\dfrac{d^3N_i}
{p_{\mathrm{T}}dp_{\mathrm{T}}d\mathrm{y} d\phi}(\mathrm{y},p_{\mathrm{T}},\phi)d\phi p_{\mathrm{T}}dp_{\mathrm{T}}.
\label{rapidityspectrum}
\end{align}

\subsection{ECHO-QGP and AZHYDRO freeze-out routines: comparison in ideal \2d cases}

Our procedure to determine the freeze-out surface is somehow simpler than
that employed within other \3d codes (MUSIC and 
CORNELIUS~\cite{Schenke:2010nt,Huovinen:2012is}). It is therefore essential
to compare our results for the particles spectra with the results
obtained by using other codes. For the \2d case there are several
available codes such as AZHYDRO~\cite{Kolb:2000sd,Kolb:2002ve} (ideal
hydrodynamics) and UVH2+1~\cite{Romatschke:2007jx} (viscous
hydrodynamics). We present here comparisons with results obtained 
by using AZHYDRO in which a triangular mesh is determined to approximate
the hypersurface.
In particular, we have simulated the hydrodynamical
stage of heavy ions collisions with AZHYDRO and for calculating the
spectra of primary particles at decoupling we have used the routines
for the freeze-out included in AZHYDRO and freeze out routine of ECHO-QGP 
described above. Notice that within AZHYDRO the particle distribution function 
is assumed to be a Maxwell distribution and the boost invariance allows to compute
analytically the integral on the $\eta$ variable in the Cooper-Frye formula.
Following this procedure, the integral on the $\eta$ variable 
of Eq.~(\ref{cooperfrye}) leads to modified Bessel functions. 
The parameter set used for AZHYDRO can be found in Tab.~\ref{tab:Heinz};
the equation of state is EOS-Q of~\cite{Kolb:2000sd,Kolb:2002ve}. 

\begin{table}[!ht]
\centering
  \begin{tabular}{ccccccc}
  \hline
 $\sigma_{NN}$  & $\tau_0$ &  $e_0$  & $\alpha$  & b 	&$\mu_{\pi}$ & $T_{freeze}$ 	\\
 mb	& fm/c & Gev fm$^{-3}$  & & fm 	& GeV & GeV		\\	
 \hline                                                           
 40	 & 0.6 &  24.5 & 1  & 0, 3, 6, 9, 12 & 0.0622 & 0.120  	\\
 \hline 
\end{tabular}
\caption{Parameter set used within AZHYDRO for testing the ECHO-QGP freeze out routine. 
The pion chemical potential 
is taken from \cite{Kolb:2002ve}). 
The grid spacing here used is: $\Delta x=\Delta y=0.4$ fm \mbox{$\Delta \tau=0.16$ fm.}}\label{tab:Heinz}
\end{table}

\begin{figure*}[!th]
\vskip 0.5cm
\begin{centering}
\includegraphics[width=0.42\textwidth]{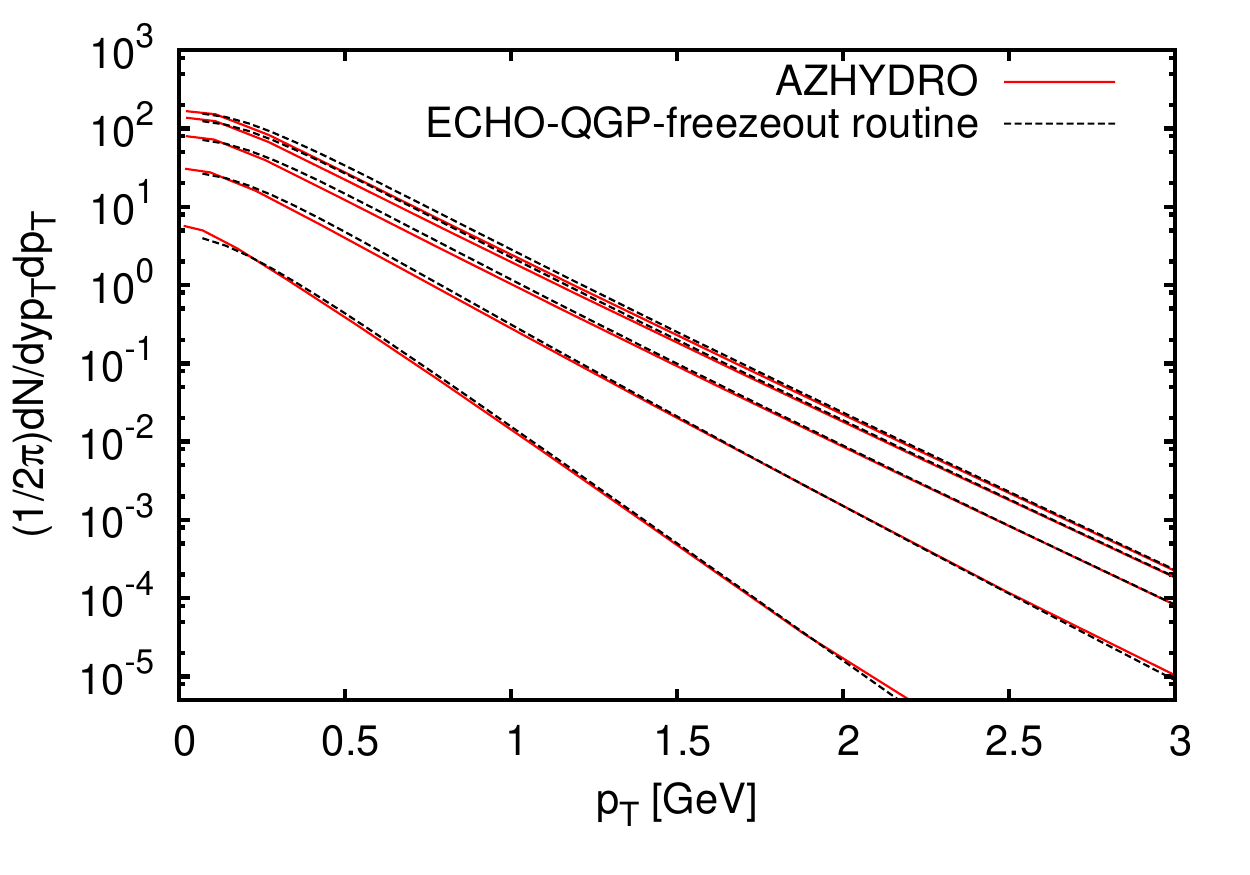}\hspace{10mm}
\includegraphics[width=0.42\textwidth]{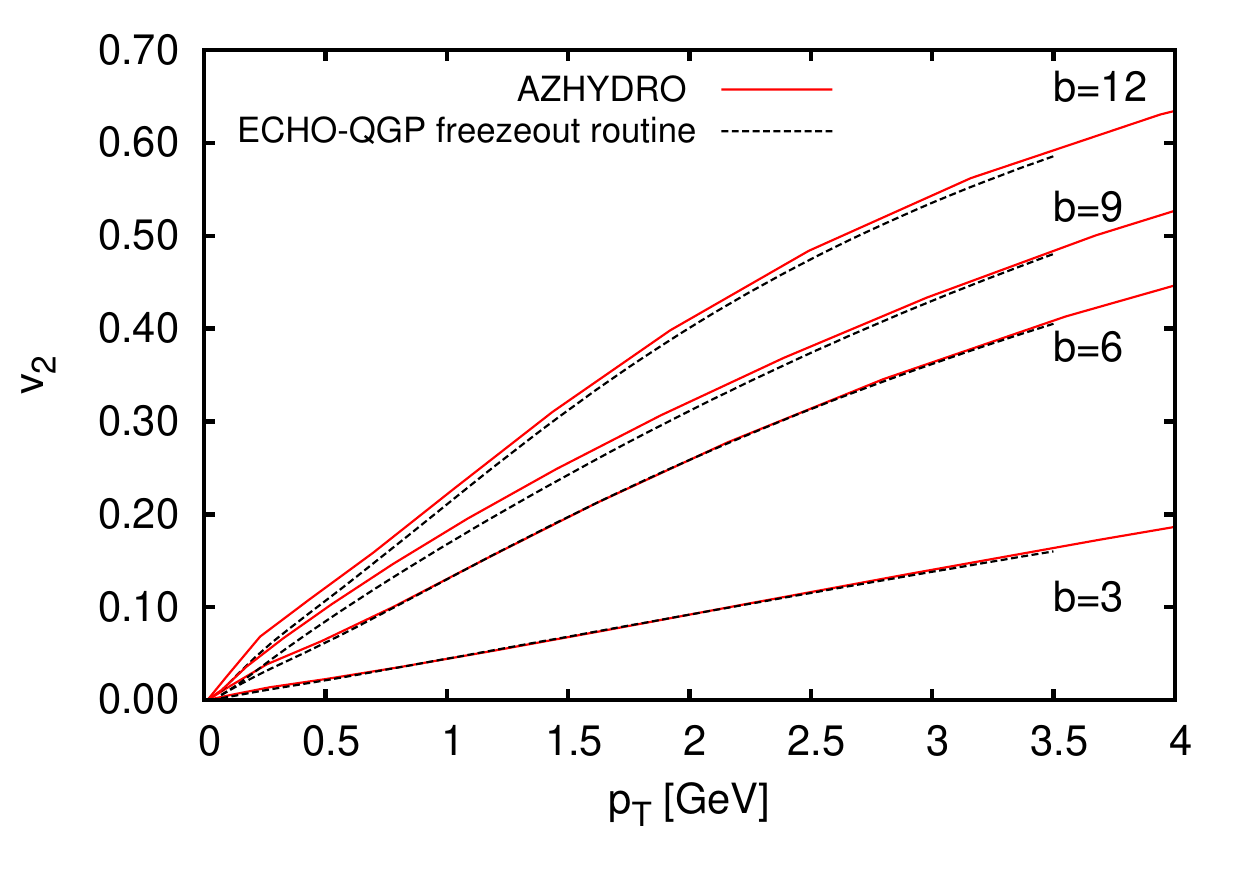}
\caption{Comparison between the AZHYDRO and ECHO-QGP freeze-out routines. 
The hydrodynamical evolution has been calculated with AZHYDRO.
Left panel: transverse momentum pion spectra at different values of impact parameters (from the top: $b=0,3,6,9,12$~fm). 
Right panel: $p_{\mathrm{T}}$ and $b$ dependence of the pion elliptic flow coefficient $v_2$.
}
\label{fig1freeze}
\end{centering}
\end{figure*}

\begin{figure}[!h]
\vskip 0.5cm
\begin{centering}
\includegraphics[width=0.42\textwidth]{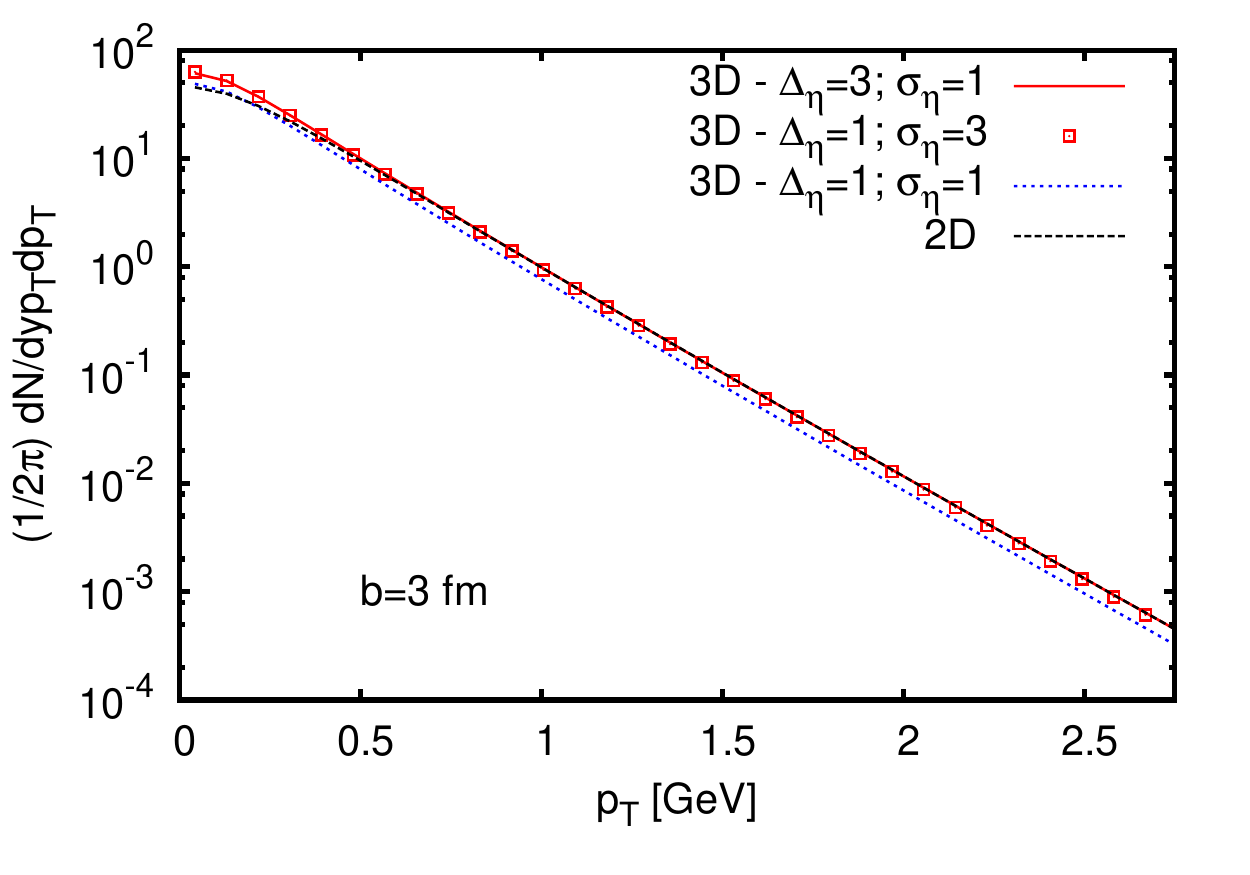}
\includegraphics[width=0.42\textwidth]{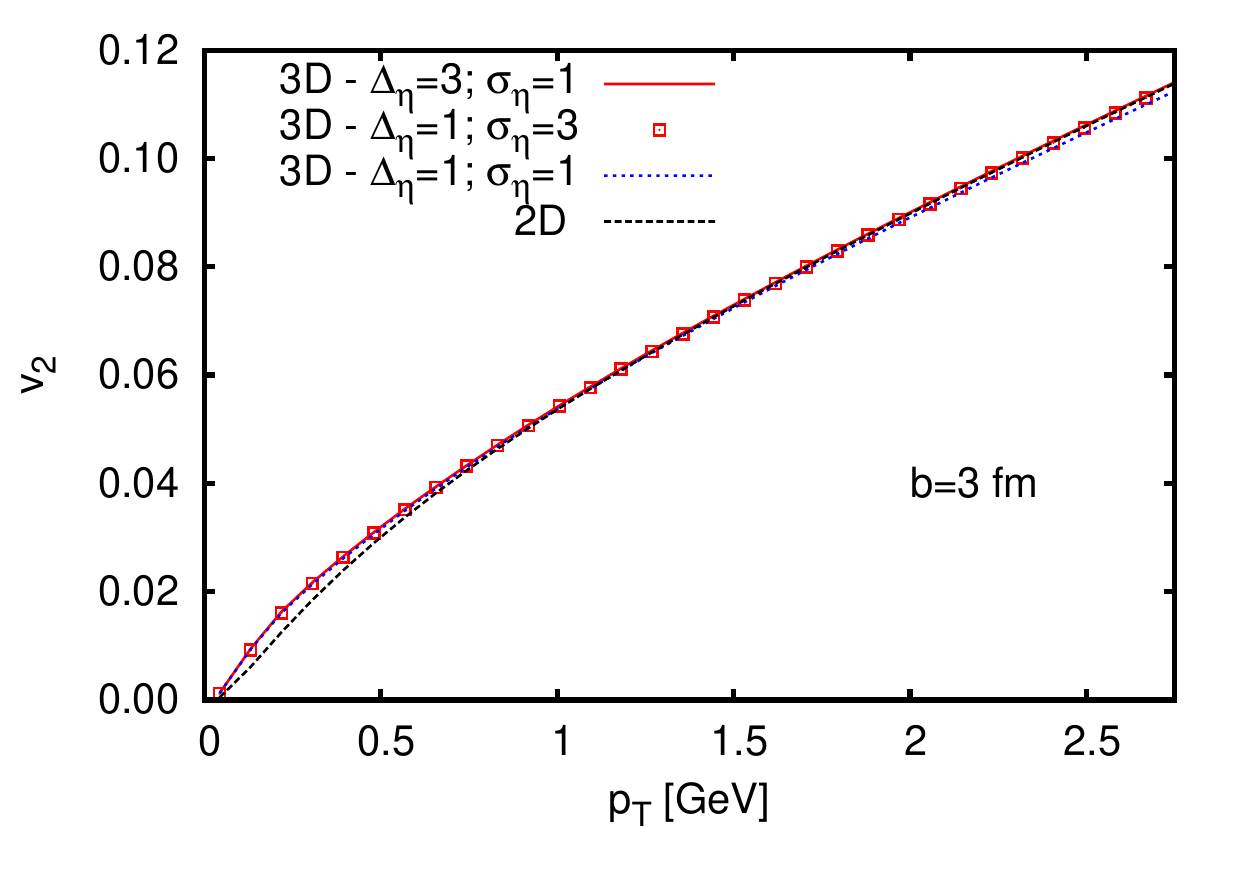}
\caption{Comparison of the pion transverse momentum spectra (top panel)
and of the elliptic flow (lower panel) obtained by ECHO-QGP in \2d and \3d for different
parametrizations of the initial energy density profile in the $\eta_s$ direction.}
\label{fig3freeze}
\end{centering}
\end{figure}

\begin{figure}[!h]
\vskip 0.5cm
\begin{centering}
\includegraphics[width=0.4\textwidth]{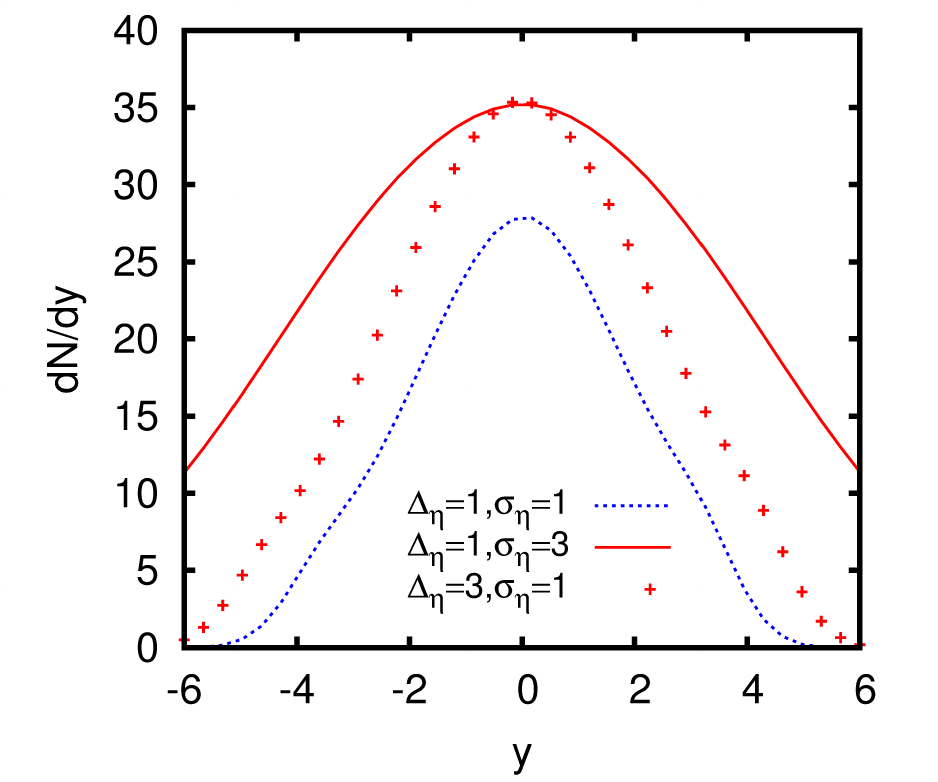}
\caption{Pion rapidity spectra with  different 
parametrizations of the initial energy density profile in the $\eta_s$ direction.
The setup here used is given in tab. \ref{tab:echo1}}
\label{fig4freeze}
\end{centering}
\end{figure}

In Fig.~\ref{fig1freeze} we compare results for primary
pions transverse momentum spectra and $v_2$ at several impact
parameters. The agreement between our results and the ones obtained
within AZHYDRO is quite satisfying for values of $b$ which are
relevant from the experimental point of view ($b\lesssim 6-7$ fm).  For larger
values, deviations of the order of $20\%$ are present in the $v_2$
spectra at low transverse momenta, $p_{\mathrm{T}} \sim 0.1$ GeV.
It has also been remarked~\cite{Schenke:2010nt,Hirano:2012kj}
that in spite of its simplicity, this method to pinpoint the freeze out hypersurface
is sufficiently accurate for computing particle spectra and $v_2$.
While we have here shown only the comparisons
of pion spectra, the spectra of other species such as 
kaons and protons can be computed just by modifying 
the specific particle properties i.e. the mass, the spin and the chemical potential
in the thermal distribution function of the Cooper-Frye formula.
The corresponding results show the same good agreement between the
AZHYDRO and ECHO-QGP freeze-out routines.

\subsection{Consistency between \2d and \3d}

The comparison between the ECHO-QGP and the AZHYDRO results on
particle spectra in \2d presented in the previous section is a
crucial test before extending our calculations to \3d. Having a good
agreement with AZHYDRO, we can now use the ECHO-QGP \2d results as a
benchmark for the \3d calculations.

In order to perform this test
we have used the initial conditions specified in Sec. \ref{sec:init} for which
the energy profile along the $\eta_s$ direction 
is flat up to $\Delta_{\eta}$ and
has then a smooth gaussian drop for larger values of $\eta_s$.
In the transverse direction, the energy profile is 
the same for the \2d and the \3d simulations.
The lack of boost invariance in \3d implies that the hydrodynamical quantities in
Eq.~(\ref{cooperfrye}), temperature and four velocity, depend on
$\eta_s$ and thus the integral on this variable must be calculated
numerically. Also, the hypersurface depends now on $\eta_s$.

\begin{table}[!ht]
\centering
 \begin{tabular}{ccccccc}
  \hline
 $\sigma_{NN}$  & $\tau_0$ & $e_0$ & $\alpha$ & b &$\mu_{\pi}$ & $T_{freeze}$ 	\\
 mb	 & fm/c & Gev/fm$^3$  & & fm & GeV & GeV		\\	
 \hline                                                           
 40	& 0.6	 &  24.5 & 1  & 3.0 & 0.0622 & 0.120  	\\
 \hline 
\end{tabular}
\caption{Parameter set used for comparing particle spectra obtained from the  \2d and \3d 
ECHO-QGP ideal hydrodynamics outputs. Different values 
of $\Delta_{\eta}$ and $\sigma_{\eta}$ are used 
(see Fig.~\ref{fig3freeze}). 
}\label{tab:echo1}
\end{table}

Tab.~\ref{tab:echo1} reports the parameter set used for 
the above mentioned test of consistency between \3d and \2d simulations
(fully performed with ECHO-QGP). Being a test of hydrodynamics, 
we used a simple pion gas equation of state.

As one can notice in Fig.~\ref{fig3freeze} (top panel), for
$\Delta_{\eta}=1$ and $\sigma_{\eta}=3$ and $\Delta_{\eta}=3$ and
$\sigma_{\eta}=1$ the \3d results lie on top of the \2d results
apart from the region at low $p_\mathrm{T}$ where in the latter case
the thermal distributions are approximated by Maxwell distributions in
order to analytically perform the integral over $\eta_s$. The
\2d spectrum is thus underestimated. For $\Delta_{\eta}=1$ and
$\sigma_{\eta}=1$ on the other hand the \3d curve is lower than the
\3d curve due to the lower extension of the hypersurface.
In the bottom panel we  display the elliptic flow coefficient $v_2$,
computed at $b=3$ fm in \2d and \3d. Also in this case the results are compatible with each other, 
with a very slight discrepancy at low \pt$\sim 0.5$ due again to the use of the Maxwell distribution 
in the \2d runs.

In \3d, another interesting observable is the rapidity spectrum that
we show in Fig.~\ref{fig4freeze}. Although we do not present here a
comparison with experimental data, the dependence on $\mathrm{y}$ is
qualitatively very similar to the one obtained, for instance, in
\cite{Huovinen:2012is}, see their Fig.~21.  At $\mathrm{y}=0$ these spectra
represent just the integral of $p_{\mathrm{T}}$ of the transverse
momentum spectra in Fig.~\ref{fig4freeze}, and we have obtained,
consistently, that the $\Delta_{\eta}=1$,  $\sigma_{\eta}=3$ and
$\Delta_{\eta}=3$, $\sigma_{\eta}=1$ cases both provide the same result,
while for $\Delta_{\eta}=1$, $\sigma_{\eta}=1$ a lower value of the spectrum 
is obtained. As $\mathrm{y}$ is
shifted, one probes the tails of the freeze-out hypersurface along the
$\eta_s$ direction. Thus the larger the value of $\sigma_{\eta}$ the
harder is the spectrum (see the curves corresponding to
$\Delta_{\eta}=3$, $\sigma_{\eta}=1$ and $\Delta_{\eta}=1$,
$\sigma_{\eta}=3$). 

We would like to state here that in the present section the freeze-out
procedure employed does not include the appropriate viscous 
corrections to the particle distributions in the Cooper-Frye algorithm.
We are aware that such corrections are important for 
the final $v_n$ coefficients, and such an improvement 
will be certainly implemented and tested before applying
the model in any realistic situations and comparing to data. 

\section{Physics results}

In this section we present some selected physics results referring to initial conditions representative of Au-Au 
collisions at RHIC (with $\sqrt{s_\mathrm{NN}}=\! 200 \!$~GeV, $\sigma_{\rm NN}\!=\!42$ mb, 
$R\!=\! 6.38$~fm, $e_0=30$~GeV/fm$^3$, $\alpha=0.15$ as in Eq.~(\ref{eq:alpha})) 
obtained with ECHO-QGP, along with comparison 
with other existing RHD viscous codes (concerning mainly the freeze-out routine and the final particle spectra). 
Results are presented for various cases, in \2d and \3d, with or without viscosity and for 
various impact parameters. They refer to the temporal evolution of the temperature, of the spatial 
and momentum anisotropy ($e_x$ and $e_p$ respectively) and to the particle spectra and elliptic-flow. 

Finally, some specific results with fluctuating Glauber-MC initial conditions are also highlighted, 
in order to demonstrate the capability of ECHO-QGP of treating all kinds of complex initializations,
leaving the detailed analysis of the higher order flow harmonics for the future.

\subsection{Temperature and eccentricity evolution}
\label{sec:rhic}

We start considering the time evolution of the central temperature $T(\tau)$ (obtained from the 
local energy density through the EOS) both for central ($b=0$) and non-central ($b\neq 0$) 
Au-Au collisions with RHIC-type initial conditions. We assume \2d evolution, to be followed with ECHO-QGP. 
Simulations are performed in Bjorken coordinates with a grid size in the transverse ($x-y$) plane of 
$201\times 201$ cells and physical dimensions ranging from $-20$~fm to $20$~fm. For \3d
runs we use $101\times 101$ cells ranging from $-20$~fm to $20$~fm in the transverse
plane, and 151 point along $\eta_s$, going from $-11$ to $11$~fm.

\begin{figure*}[!th]
\begin{center}
\includegraphics[clip,width=.50\textwidth]{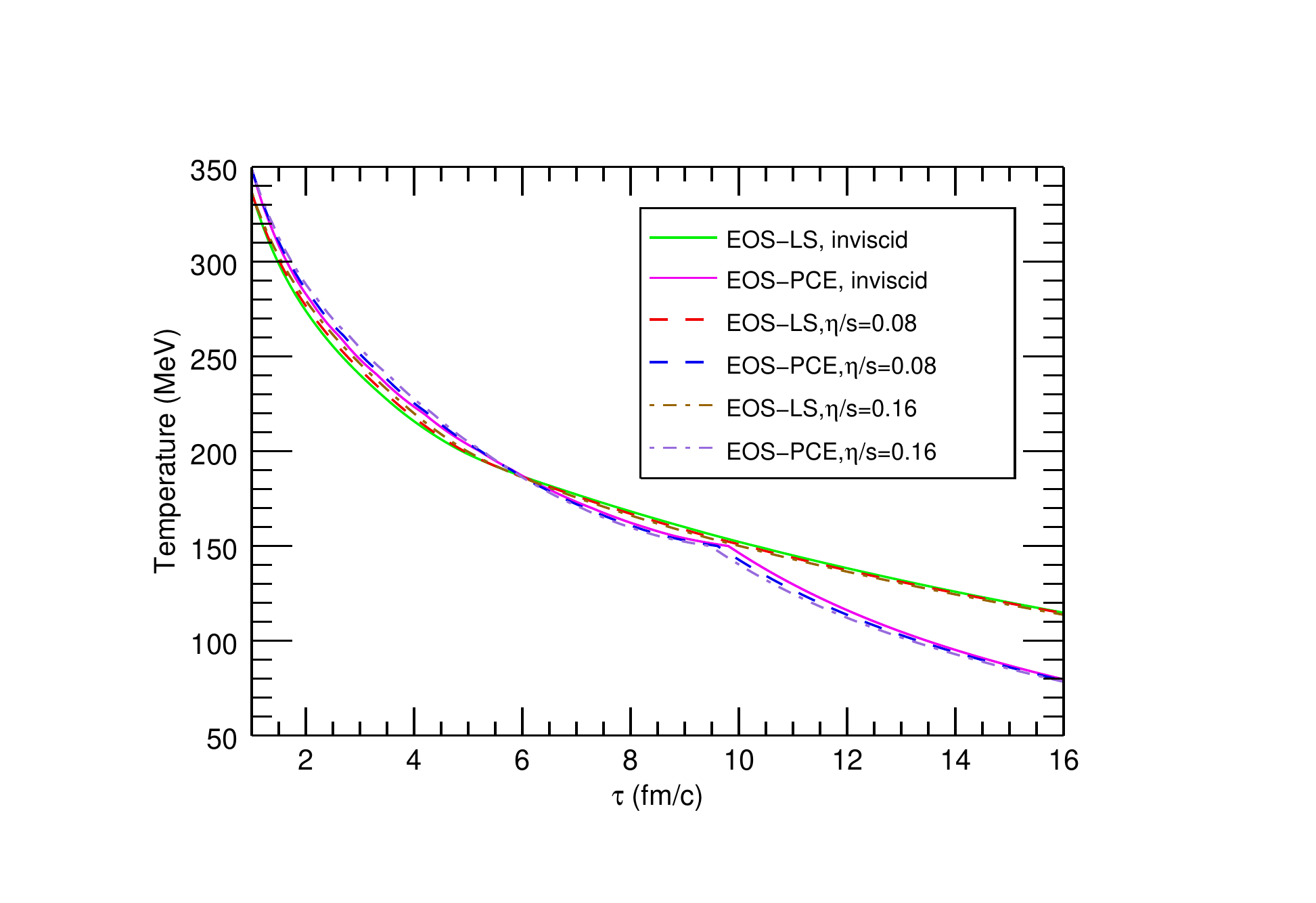}\hspace{-10mm}
\includegraphics[clip,width=.50\textwidth]{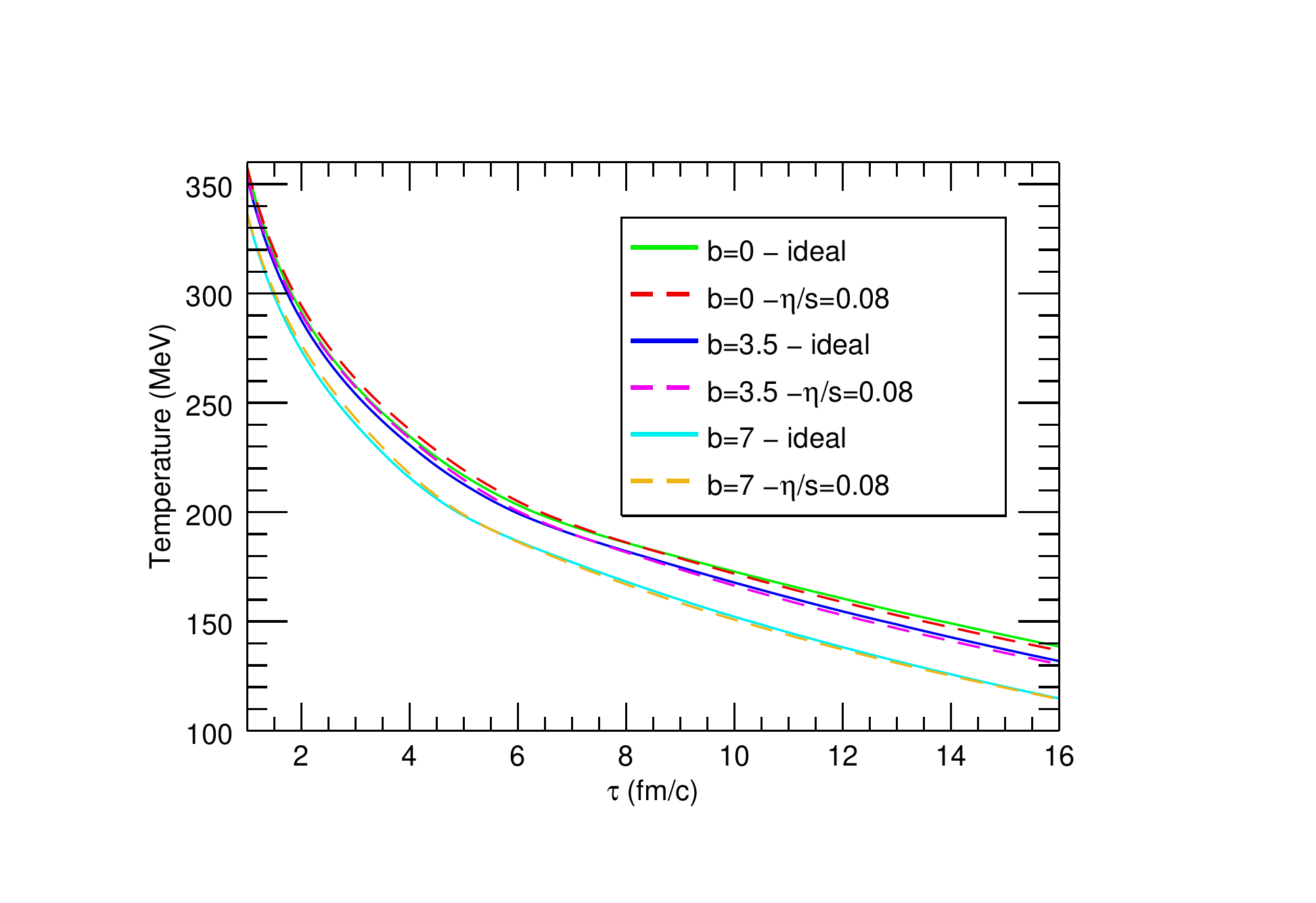}\hspace{-10mm}
\caption{Temperature as a function of $\tau$ at the center of the fireball 
with RHIC-type initial conditions. In the left panel we compare EOS-LS 
against EOS-PCE, and ideal RHD against viscous runs with $\eta/s=0.08$, 
or 0.16 for  $b=7$~fm. In the right panel we study the influence of shear viscosity ($\eta/s$=0.08) against
ideal RHD for $b=0$, $b=3.5$~fm, and $b=7$~fm.} 
\label{fig:temp2d}
\end{center}
\end{figure*}

\begin{figure}[b]
\begin{center}
\vspace{-5mm}
\includegraphics[clip,width=.50\textwidth]{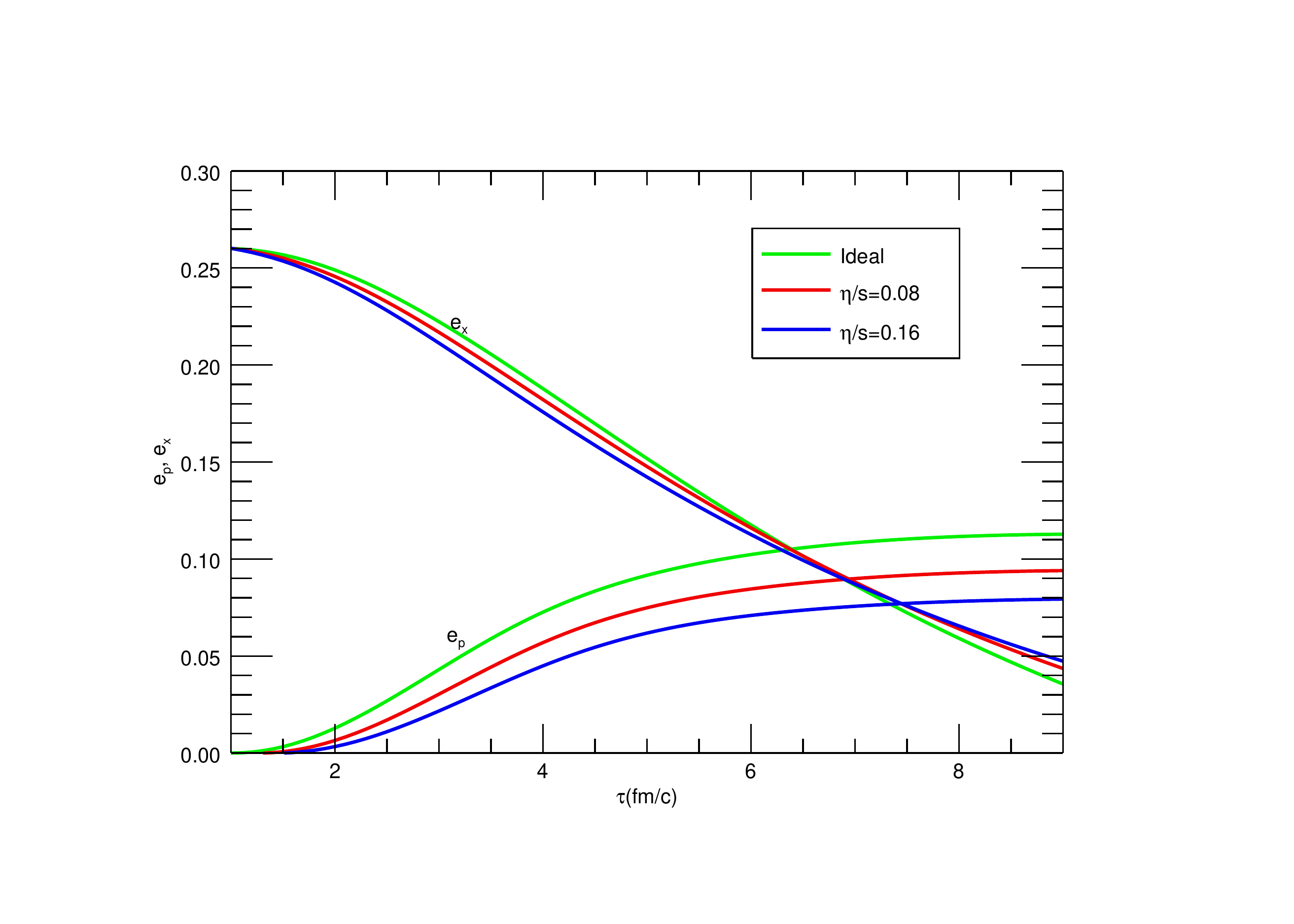}\hspace{-10mm}\vspace{-10mm}
\caption{Spatial anisotropy $e_x$ and momentum anisotropy $e_p$ as a function of $\tau$
in RHIC-type \2d simulations with ECHO-QGP, using EOS-LS. 
We compare runs with $b=7$~fm and for $\eta/s=0,\,0.08,\,0.16$. 
}
\label{fig:exep7}
\end{center}
\end{figure}

The dependence of $T(\tau)$ on the EOS, on the impact parameter and on the shear viscosity is 
displayed in Fig.~\ref{fig:temp2d}. 
The temperature is sensitive to the equations of the state chosen throughout 
the evolution. As expected, the differences are 
more pronounced in the later stages, when the temperature drops below $T=150$~MeV 
and the effects of the partial chemical equilibration plays crucial role. 
Concerning the dependence on the shear viscosity, we notice that its effect is very 
limited in the central region, where the fluid velocity is small.
More important is the dependence on the impact parameter.
It is clear that  the larger the value of $b$, the earlier the occurrence of freeze-out. 
This is mainly guided by the impact parameter dependence of the initial energy density profile.

We then move to consider the evolution of the eccentricity in non-central collisions.
{Hydrodynamics translates the initial spatial eccentricity of the system -- arising essentially 
from the non-vanishing impact parameter of the A-A collision and giving rise to asymmetric 
pressure gradients -- into a final anisotropy in the momentum spectra of the produced hadrons. 
The spatial anisotropy in the transverse plane is usually quantified, in the case of smooth initial conditions, in terms of the 
coefficient~\cite{glauberinit}}
\be
e_x= \frac{\langle y^2-x^2\rangle_e}{\langle y^2+x^2\rangle_e},
\ee
where $\langle..\rangle_e$ denotes a spatial average over the transverse plane, 
with the local energy-density $e$ (or entropy density $s$, depending on the choice done in the initialization stage) as a weight.
The momentum anisotropy is estimated, following~\cite{kolb_ep},
in terms of the components of $T^{\mu\nu}$, as
\be
e_p=\frac{\langle T^{xx}-T^{yy}\rangle}{\langle T^{xx}+T^{yy}\rangle},
\ee
where $\langle.. \rangle$ denotes a spatial averaging (over the transverse plane) with weight factor unity.

\begin{figure*}[th]
\begin{center}
\includegraphics[clip,width=.50\textwidth]{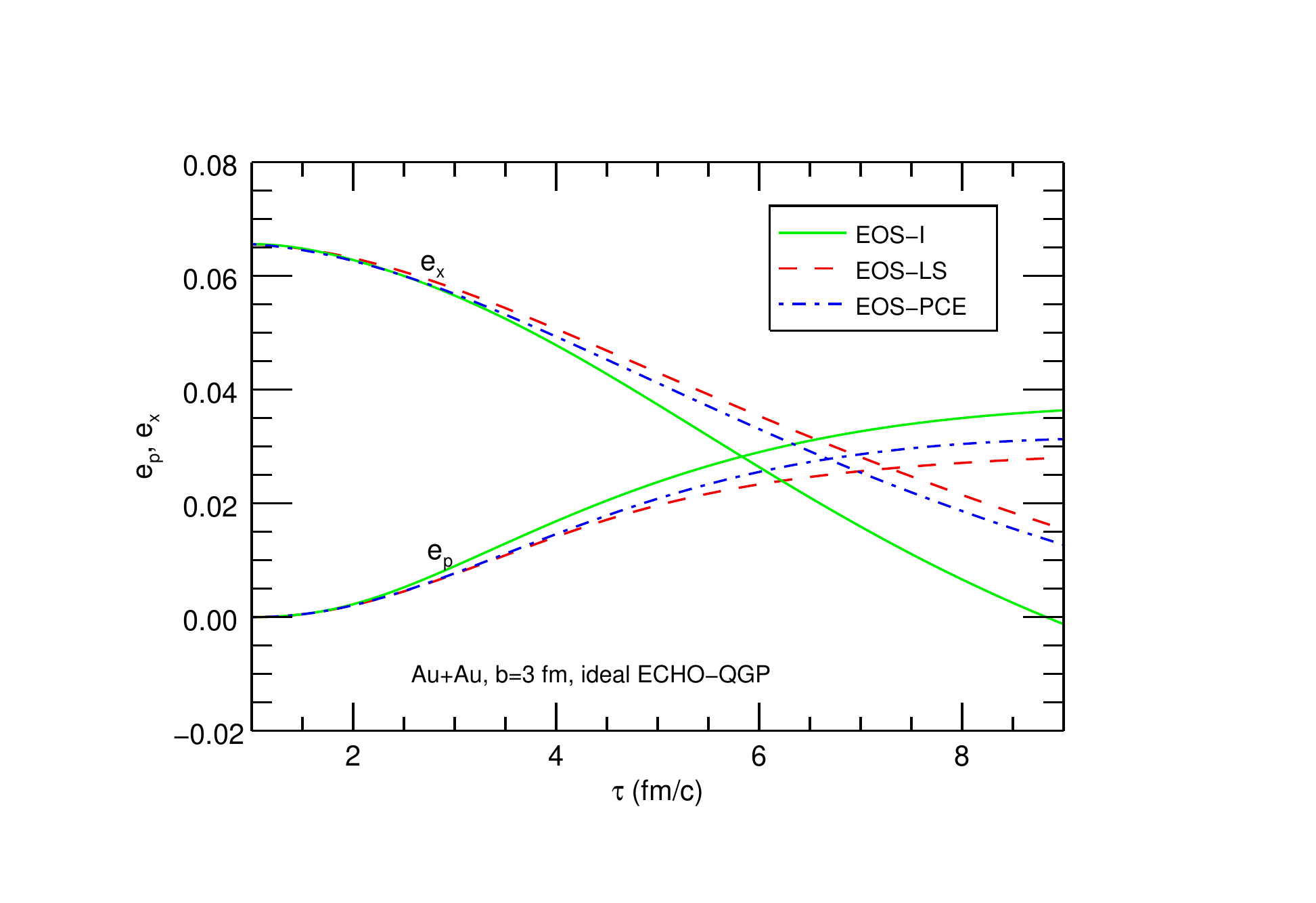}\hspace{-10mm}
\includegraphics[clip,width=.50\textwidth]{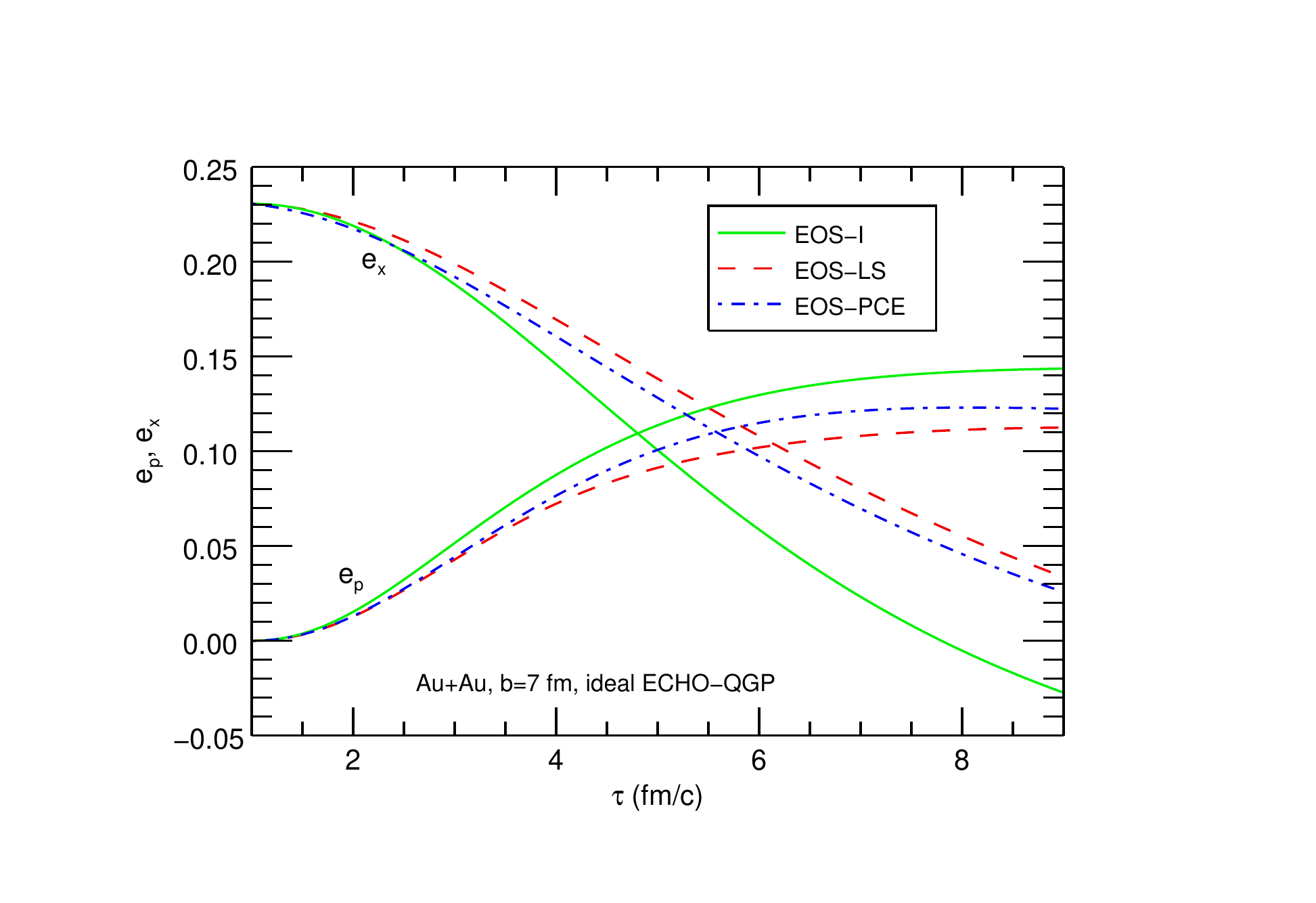}\vspace{-10mm}
\includegraphics[clip,width=.50\textwidth]{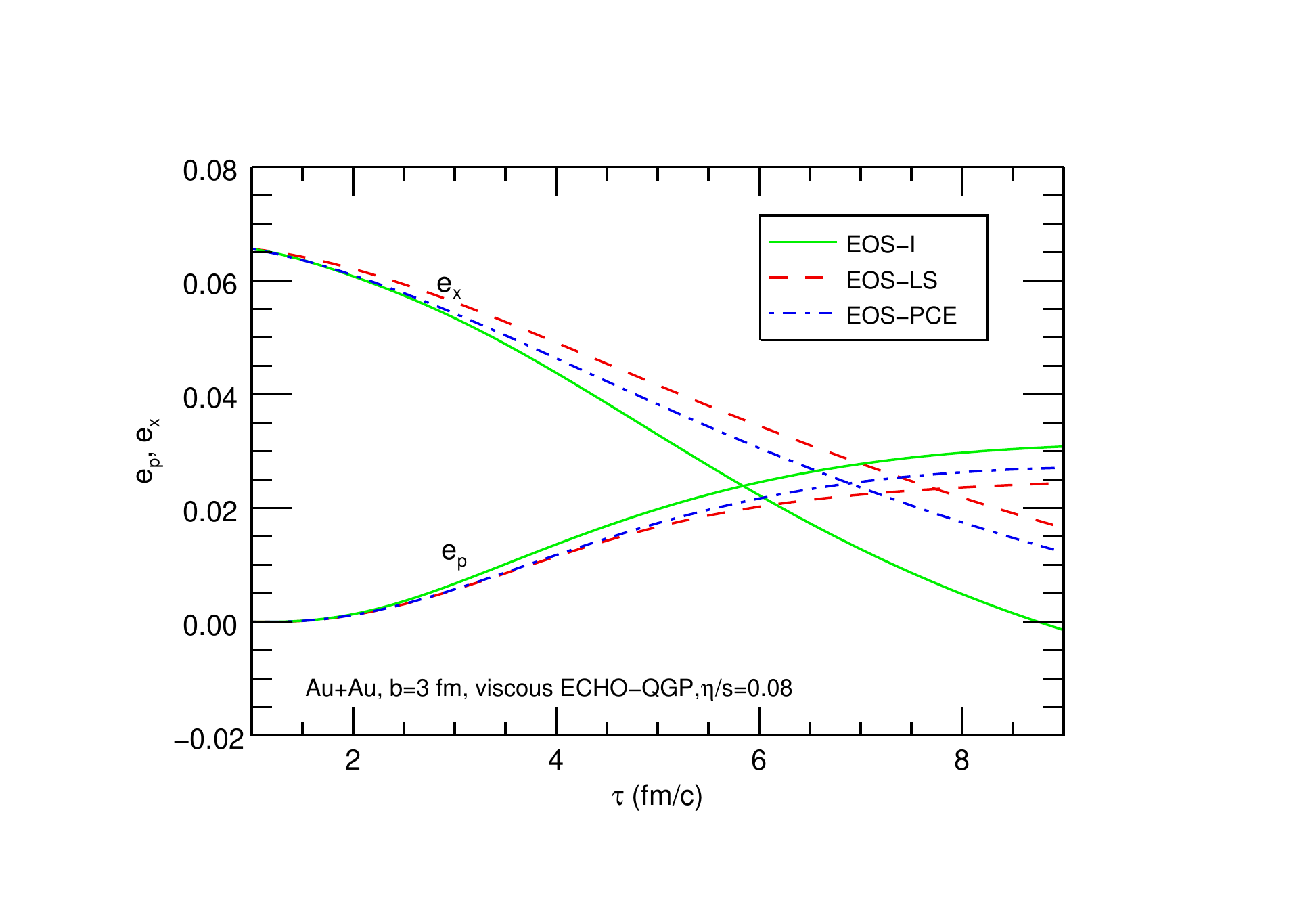}\hspace{-10mm}
\includegraphics[clip,width=.50\textwidth]{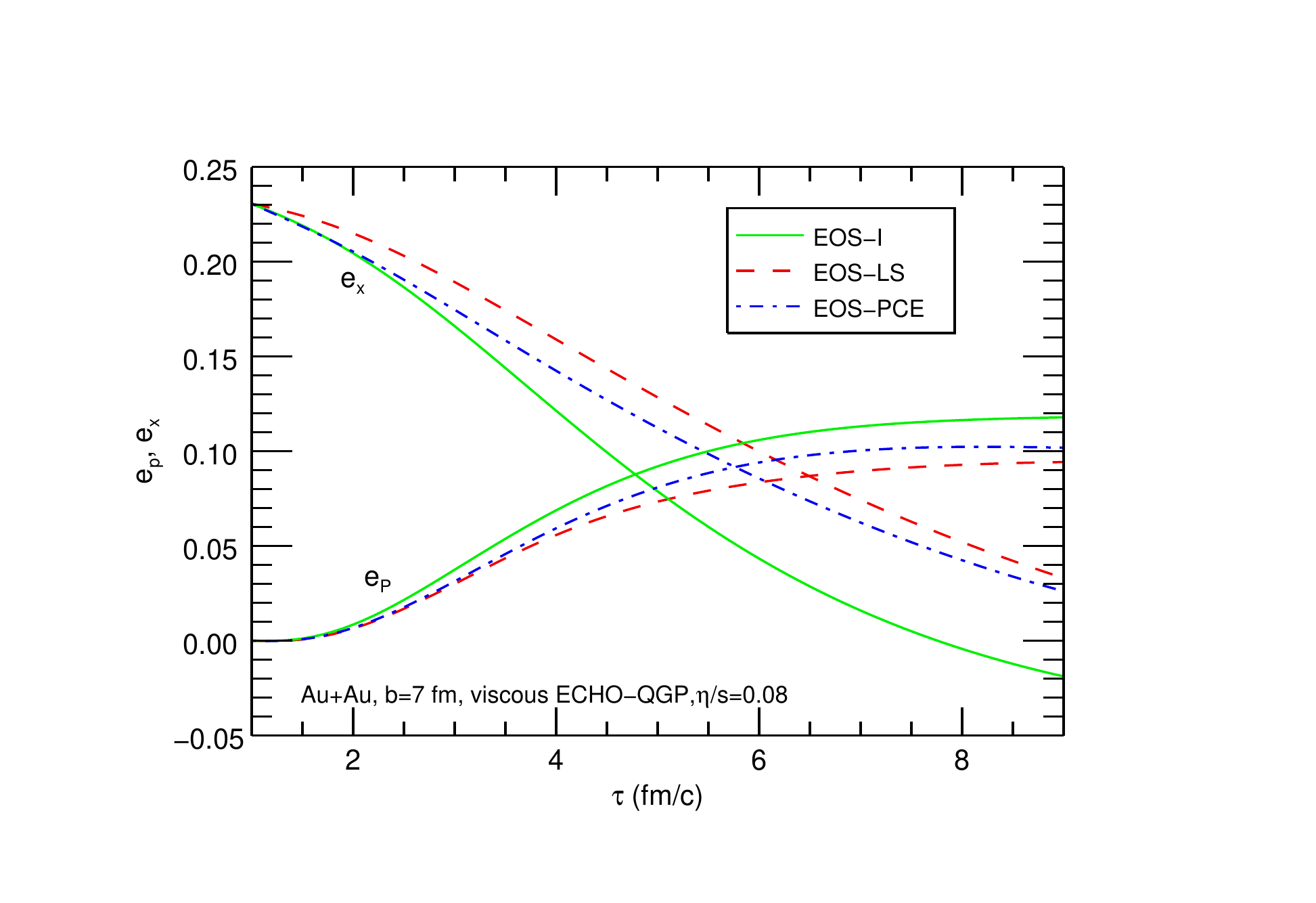}
\caption{Spatial and momentum anisotropies for different EOSs and impact parameters employed
in \2d simulations. The ideal case (the viscous case with $\eta/s=0.08$) is shown in top (bottom) panels.}
\label{fig:exepi}
\end{center}
\end{figure*}

\begin{figure*}[th]
\begin{center}
\hspace{-8mm}
\includegraphics[clip,width=.45\textwidth]{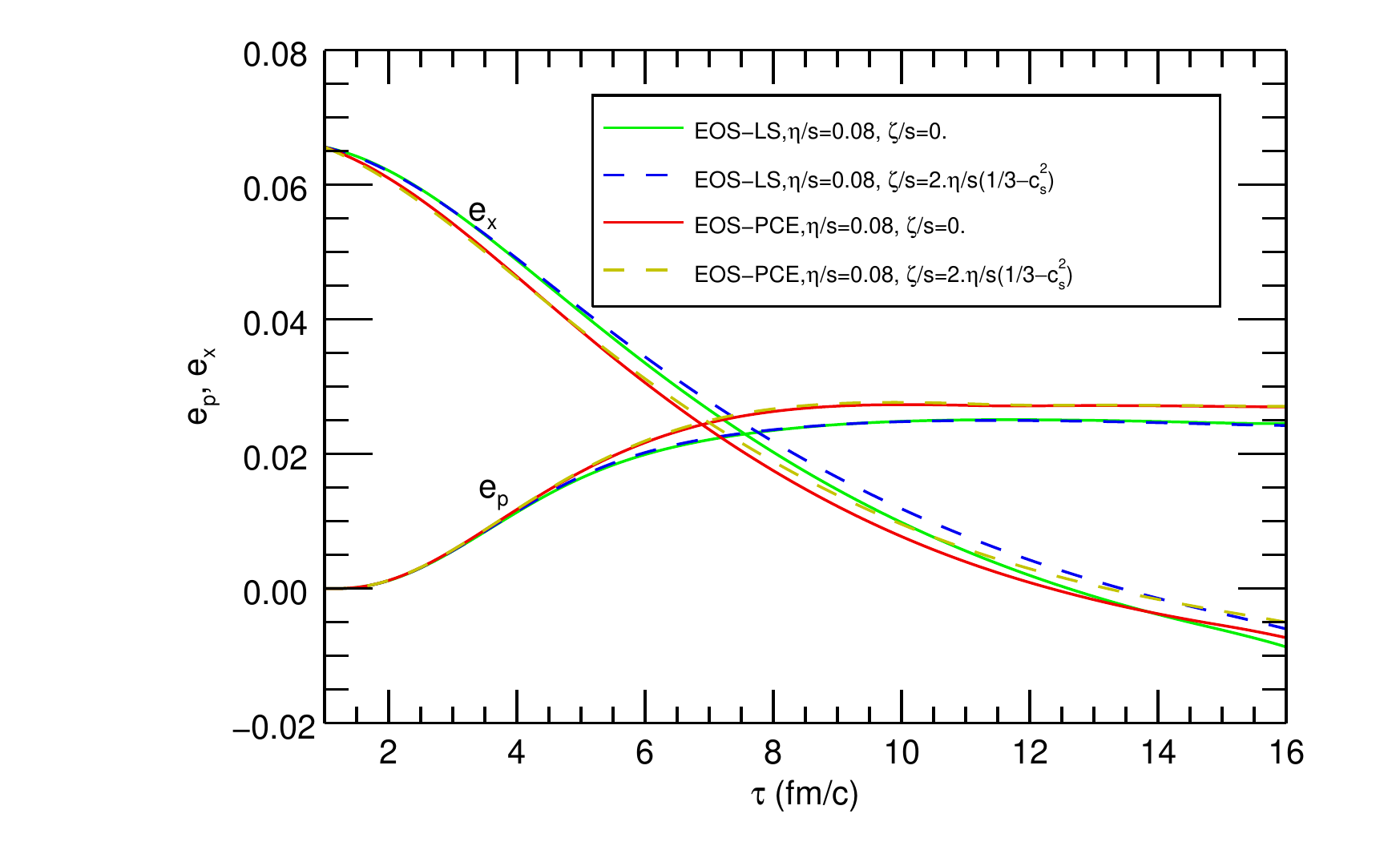}
\includegraphics[clip,width=.45\textwidth]{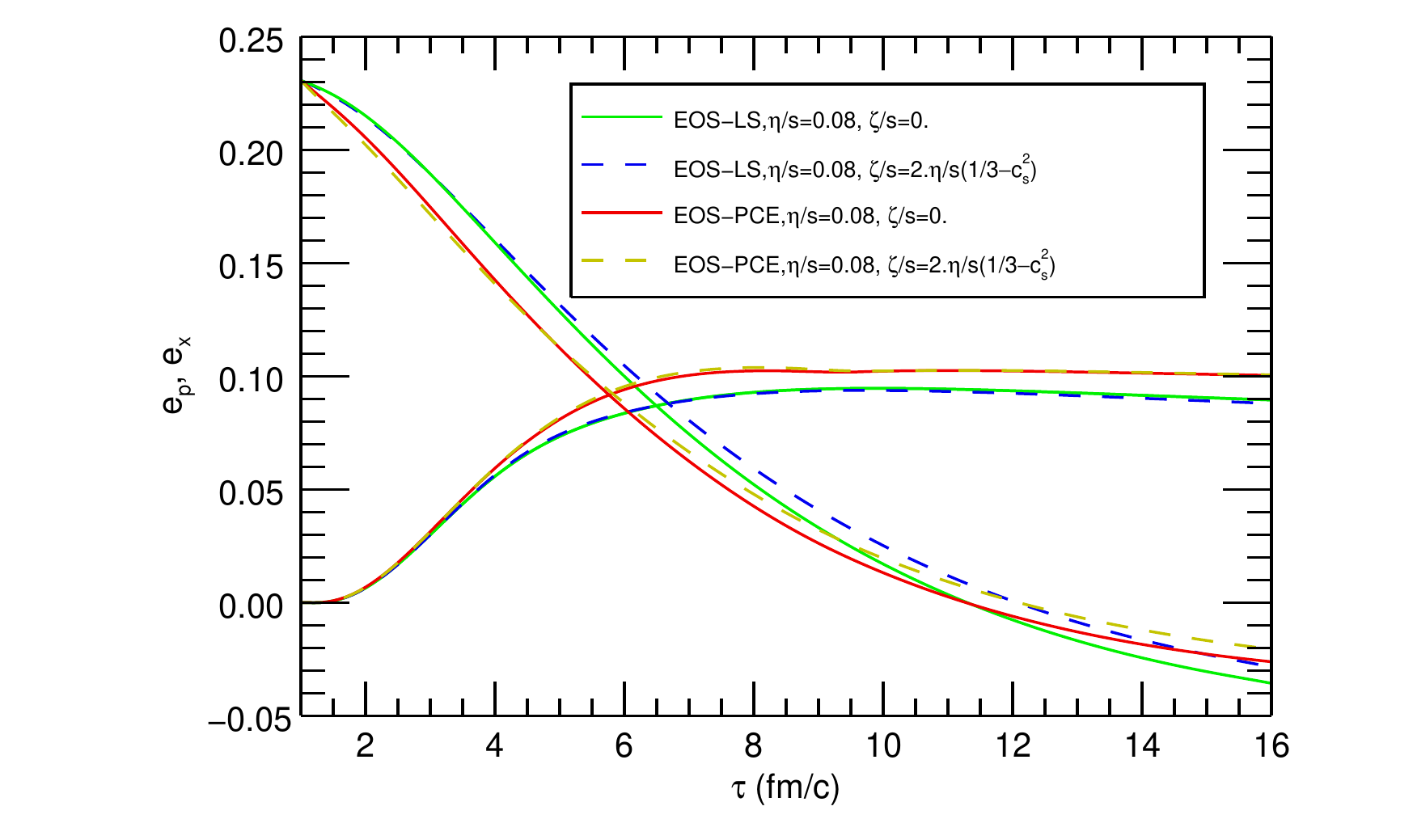}
\caption{Spatial and momentum anisotropies for different values
of the  $\eta/s$ and $\zeta/s$ parameters, for $b=3$ and $b=7$~fm, and
with EOS-LS or EOS-PCE.}
\label{fig:etztepex}
\end{center}
\end{figure*}

Because of the larger pressure gradients along the reaction-plane, during the hydrodynamic 
evolution of the system the momentum anisotropy $e_p$ is expected to increase at the 
expense of the spatial eccentricity $e_x$.
The temporal evolution of $e_x$ and $e_p$ at RHIC, along with their sensitivity to the 
EOS and the magnitude of the viscous effects, are shown in Fig.~\ref{fig:exep7} for $b=7$~fm in \2d. 
We can observe that, with higher values of $\eta/s$, the growth of the 
momentum anisotropy is lower throughout the time evolution, reflecting the role 
of dissipative effects in taming the collective response of the system to the pressure gradients.

\begin{figure}[th]
\begin{center}
\includegraphics[clip,width=.50\textwidth]{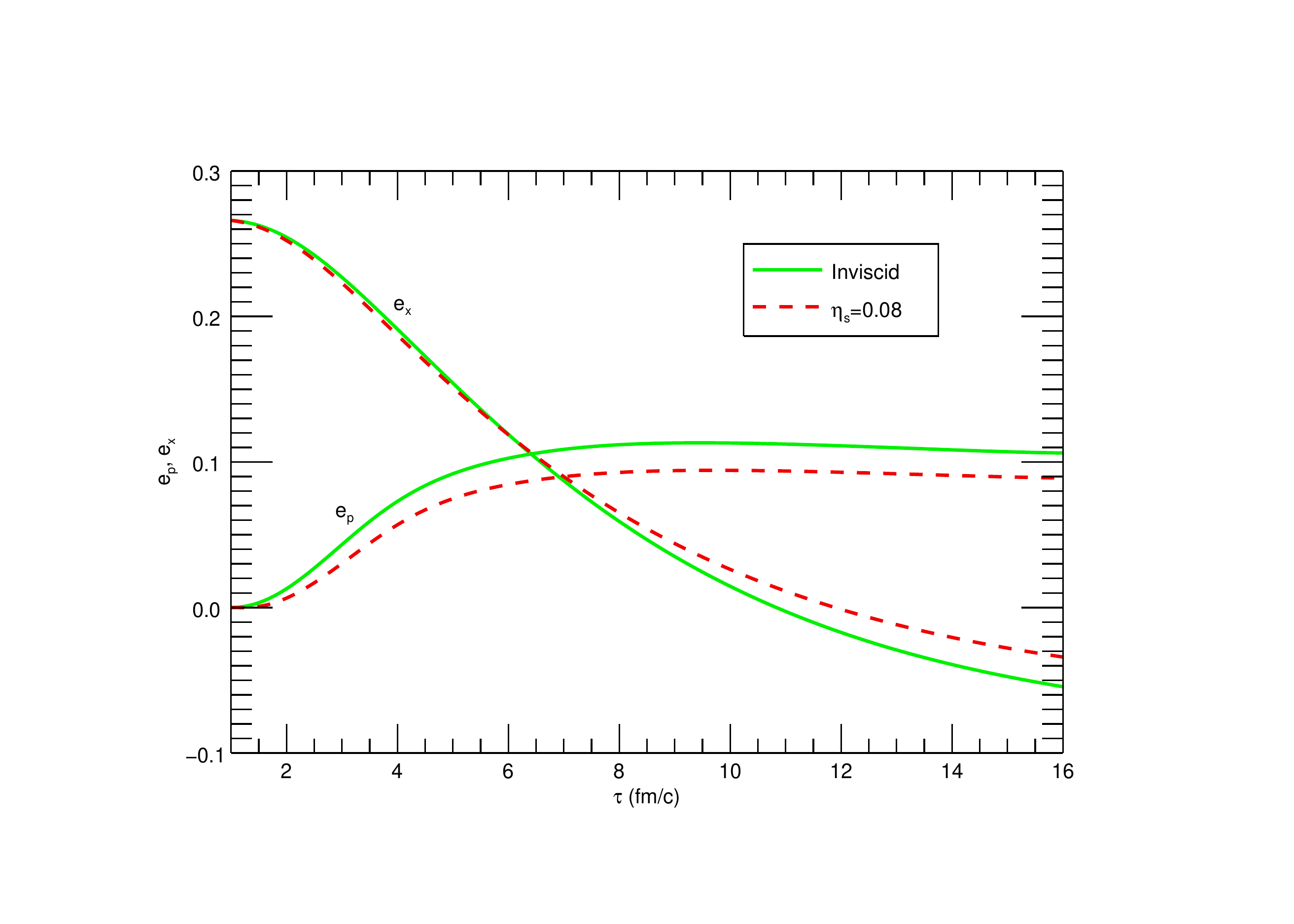}\hspace{-10mm}
\caption{Spatial anisotropy $e_x$ and momentum anisotropy $e_p$ as a function of $\tau$
in RHIC-type \3d simulations, using EOS-LS. 
We compare runs with $b=7$~fm and for $\eta/s=0$ and $0.08$ in the plane
with $\eta_s=0$ space-time rapidity. 
}
\label{fig:3dbj}
\end{center}
\end{figure}

Next, we consider the sensitivity of $e_x$ and $e_p$ to the EOSs employed and the impact parameter.
The time evolution of the spatial and momentum anisotropies {is shown, for $b\!=\!3$~fm and $b\!=\!7$}~fm, 
with EOS-I, EOS-LS and EOS-PCE, with and without shear viscosity (here we have switched 
off the bulk viscous effects), in Fig.~\ref{fig:exepi}.
Both the spatial and momentum anisotropies are quite sensitive to the EOS 
employed in the simulations. The differences among the different choices can be 
observed at the later stages of the collisions both at $b\!=\!3$~fm and at $b\!=\!7$~fm.
Differences are more pronounced for the more peripheral collisions.
All these observations still hold in the presence of viscosity.

We now investigate the role of bulk viscosity 
as far as the time evolutions of $e_x$ and $e_p$ is concerned. 
We have plotted $e_x$ and $e_p$ at $b=3$~fm and $b=7$~fm
with and without $\zeta/s$  for two of the tabulated equations of state,
EOS-LS and EOS-PCE in Fig.~\ref{fig:etztepex}. In both cases, we set 
$\eta/s=0.08$. The value of $\zeta/s$ is set to $2\eta/s(1/3-c_s^2)$.
We observe that the non-vanishing $\zeta/s$ has a negligible impact at the 
initial times as compared to role played by $\eta/s$. There are some mild effects seen at 
lower temperatures (later stages of the evolution). This is not surprising, since the temperature 
behavior $\zeta/s$ is governed by the factor  
$1/3-c_s^2$. All the above observations are valid for both EOS-LS and EOS-PCE.

Finally, also \3d simulations, with the same set up (RHIC-type initialization
with $b=7$~fm, EOS-LS, Bjorken coordinates), have been performed.
Expansion now occurs also in the $\eta_s$ direction, as expected,
and in Fig.~\ref{fig:3dbj} we show the time evolution of the $e_x$ and $e_p$ 
quantities calculated at $\eta_s=0$, for both the ideal case and the viscous one, with $\eta/s=0.08$.
The behavior is similar to the corresponding \2d case, and different
cuts in the space-time rapidity $\eta_s$ also produce similar results.
We have found that in some cases the expanding front along $\eta_s$
shows instabilities in \3d viscous runs (only for Bjorken coordinates).
To cure this problem, we adopt a similar strategy as in~\cite{trick}, where
viscous tensor components (and the bulk viscous pressure) are decreased
proportionally to $P/P_\mathrm{cut}$ when $P<P_\mathrm{cut}$, where
the assumed threshold corresponds to a temperature $T\simeq 45$~MeV
(for the chosen EOS-LS), well below the freeze-out limit.

\subsection{Particle spectra in ideal \3d and viscous \2d cases}

We will show now results for particle spectra and elliptic flow obtained using the
PCE EOS: the parameters are reported in Tab.~\ref{tab:3D_qcd}.

\begin{table}[!h]
\centering
 \begin{tabular}{ccccccc}
  \hline
 $\sigma_{NN}$  & $\tau_0$  & $e_0$ & $\alpha$ & $b$  &  $\mu_{\pi}$ & $T_{freeze}$ \\
 mb	 & fm/c & Gev/fm$^3$ & & fm & GeV & GeV \\	
 \hline                                                           
 40	 & 0.6 &  25.0  & 0.25   & 3, 5, 7	 & 0.03217 & 0.130 \\
 \hline 
\end{tabular}
\caption{\label{tab:3D_qcd}
Parameter set used for the \3d ideal hydrodynamics results of ECHO-QGP. 
In addition $\Delta_\eta=5.0$, $\sigma_\eta = 0.8$ in Eq.~(\ref{init3d}).
}
\end{table}

\begin{figure}
\begin{center}
\includegraphics[width=0.45\textwidth]{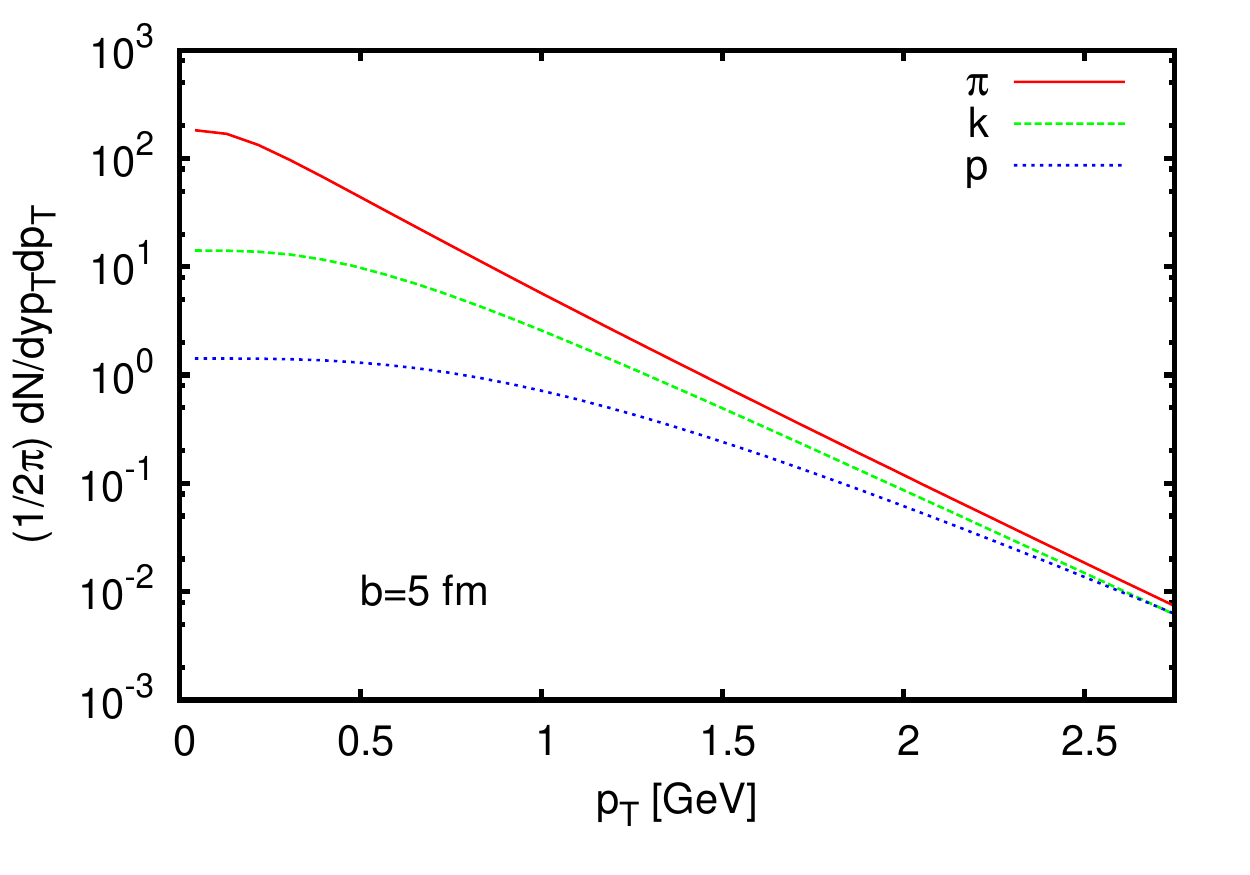}
\caption{\pt \: spectra of pions, kaons and protons as obtained in \3d 
ideal hydrodynamics. Parameters are specified in Tab.~\ref{tab:3D_qcd}. 
The grid steps are 
$\Delta x=\Delta y=0.2$~fm $\Delta\eta_s=0.2$, $\Delta \tau=0.1$~fm/c.}
\label{fig:spe_3D_qcd_b5}
\end{center}
\end{figure}

We remark again that for a comparison with the experimental data one needs to include 
the contribution of unstable particles to the final particle spectra. In this work such contributions 
are not implemented. In Fig.~\ref{fig:spe_3D_qcd_b5} we display the transverse momentum spectra of (direct)
pions, kaons and protons. For large values of \pt \: our results are compatible with results obtained in the \3d
code developed in~\cite{Schenke:2010nt} (see their Fig.~1) where also a fit of the experimental data is presented.
A value of the pion spectrum of about $0.1$~GeV$^{-2}$ at \pt $\sim 2$~GeV is obtained.
The agreement with \cite{Schenke:2010nt} is lost at low \pt due to lack of resonance feed-down in our scheme.
Indeed in~\cite{Sollfrank:1991xm} an enhancement of a factor of 4 is obtained for the pion spectra
at \pt~$=0$. We are thus confident that including the resonance decay will allow to correctly 
reproduce the experimental data. 

\begin{figure}
\begin{center}
\includegraphics[width=0.45\textwidth]{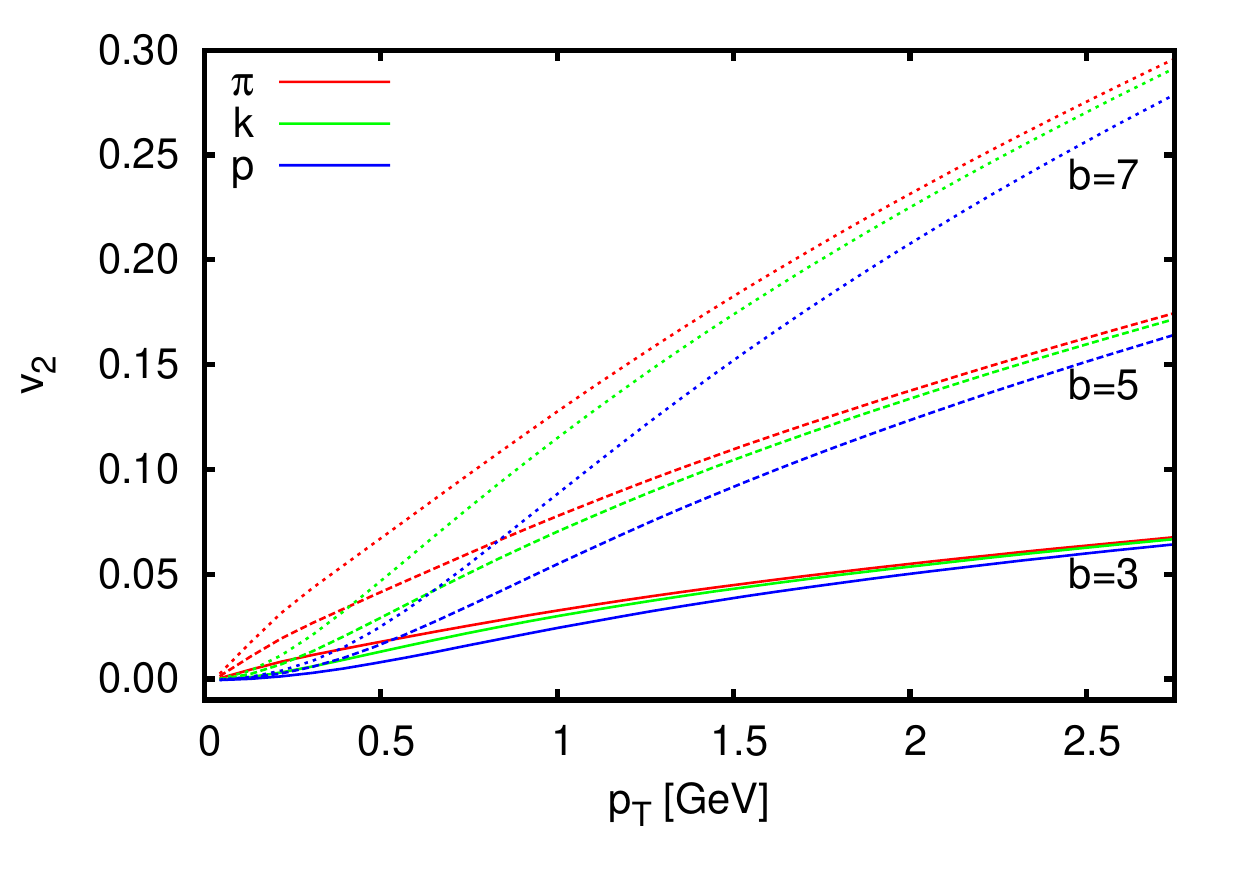}
\caption{\pt\: and $b$ dependence of the elliptic flow $v_2$ as obtained in \3d ideal hydrodynamics. 
Parameters are chosen the same as in the previous figure.}
\label{fig:v2_3D}
\end{center}
\end{figure}

\begin{figure}
\begin{center}
\includegraphics[width=0.45\textwidth]{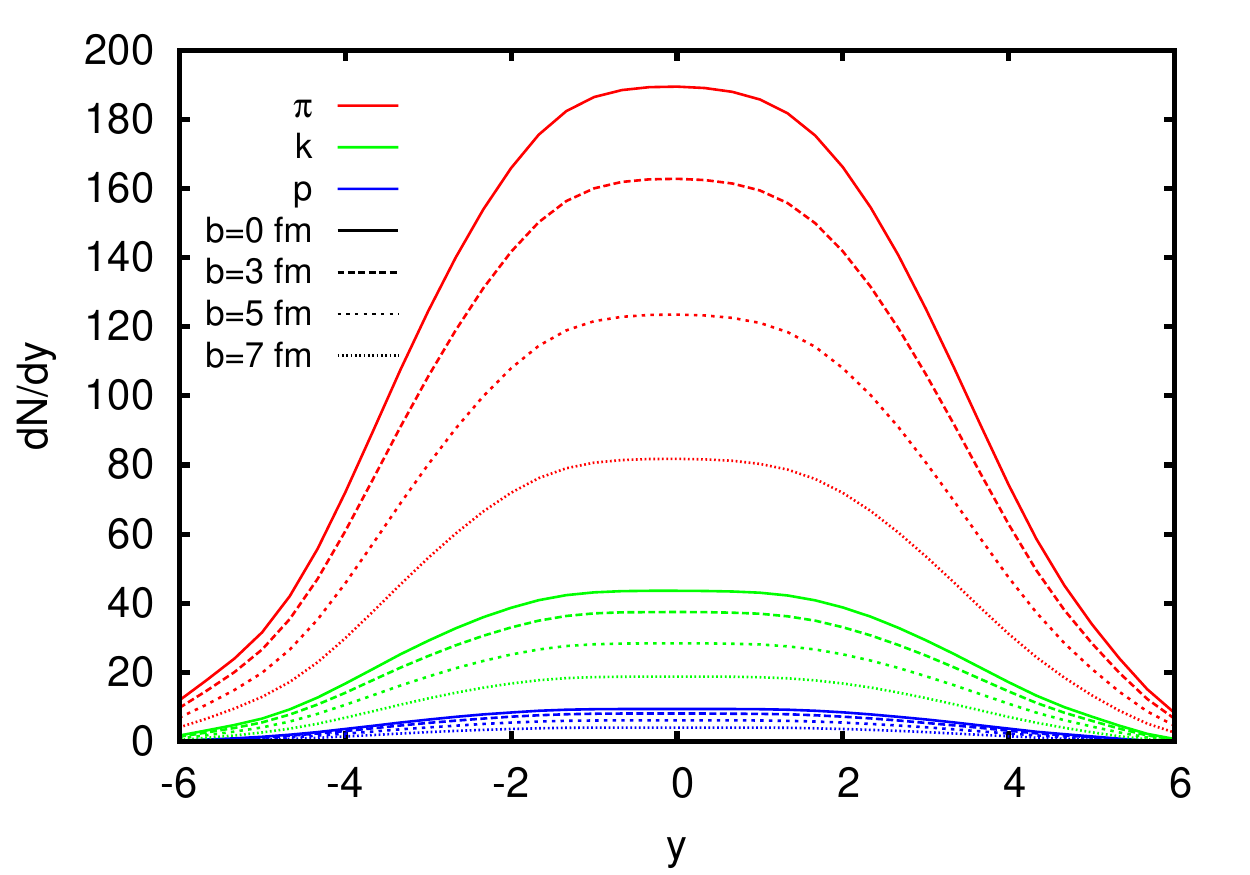}
\caption{Rapidity spectra of pions, kaons and protons as obtained in \3d ideal hydrodynamics. }
\label{fig:rapidity_3D}
\end{center}
\end{figure}

\begin{figure}[th]
\begin{center}
\includegraphics[width=0.45\textwidth]{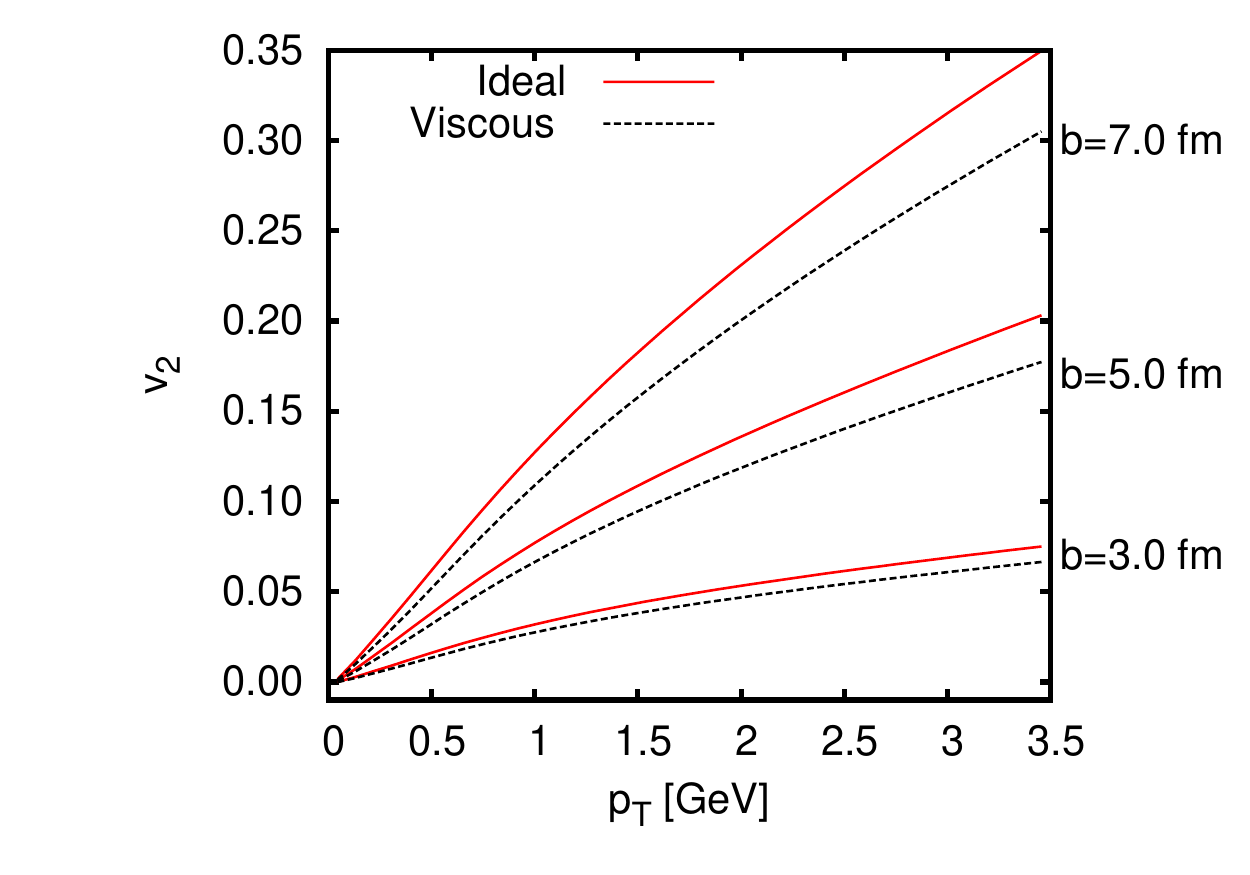}
\caption{$p_{\mathrm{T}}$ and $b$ dependence of the pion elliptic flow coefficient $v_2$  
within \2d ideal and viscous hydrodynamics. Parameters are reported in  Tab.~\ref{tab:2+1vis}}
\label{fig:v2_pionvis}
\end{center}
\end{figure}

\begin{figure}[t]
\begin{center}
\includegraphics[width=0.45\textwidth]{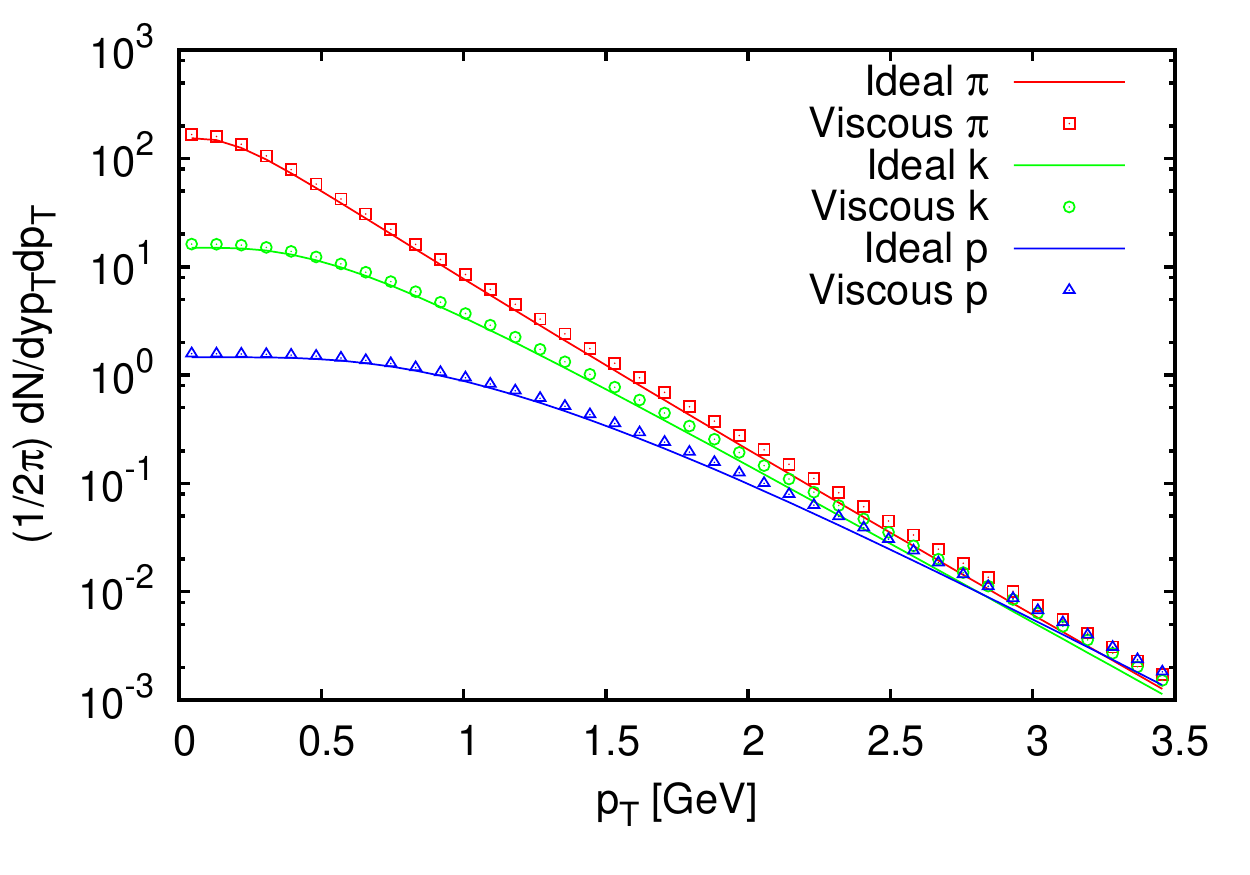}
\caption{ $p_{\mathrm{T}}$ spectra for pions, kaons and protons. Comparison 
between results obtained in \2d ideal (sold lines) and viscous (dotted lines) hydrodynamics. 
Here $b=5$~fm, all the other simulation parameters are reported in~\ref{tab:2+1vis}.}
\label{fig:pkp_7}
\end{center}
\end{figure}

On the other hand, the elliptic flow results are not so much affected by the resonance feed-down
because it is a ratio of spectra. Considering for instance the pion $v_2$, we obtain a value of 
$\sim 0.16\%$ at \pt$=1.5$~GeV, from Fig.~\ref{fig:v2_3D}, which is
quite close to the value obtained in~\cite{Schenke:2010nt} (see Fig.~5 therein). 
Finally in Fig.~\ref{fig:rapidity_3D}, we show the rapidity spectra of pions, kaons and protons.
This observable will be important for future developments of ECHO-QGP when comparing with the experimental 
data. In particular, it will allow one to better constrain the initial conditions of the hydrodynamical evolution.

Viscosity provides important corrections to the particle spectra
particularly evident in the $p_{\mathrm{T}}$ and $b$ dependence of the elliptic flow coefficient $v_2$.
For this section we limit our discussion to \2d simulations with the parameters of Table~\ref{tab:2+1vis},
neglecting viscous corrections to the particle distributions in the Cooper-Frye algorithm.
\begin{table}[!h]
\centering
\begin{tabular}{ccccccc}
\hline
$\sigma_{NN}$  & $\tau_0$ & $e_0$ & $\alpha$  & $b$  & $\mu_{\pi}$ & $T_{freeze}$ \\
mb & fm/c	 & Gev/fm$^3$ 	 &  & fm &  GeV & GeV		\\	
  \hline                                                           
  42 & 1.0	 &  30.0 & 0.15  & 3, 5, 7 &  0.03217 & 0.130  	\\
  \hline 
 \end{tabular}
 \caption{Parameter set used in the comparison between \2d ideal and viscous ($\eta/s=0.08$) simulation 
(see Fig.\ref{fig:v2_pionvis}-\ref{fig:pkp_7}).} 
 \label{tab:2+1vis}
\end{table}

The chosen equation of state is the one in Sec.~\ref{sec:EOS}. As shown in Fig.~\ref{fig:v2_pionvis},
we obtain the standard result of a suppression of the $v_2$ when including viscosity.
At $p_\mathrm{T}=1.5$~GeV, $\eta/s=0.08$ and $b=7.0$~fm, the suppression is of the $v_2$ 
of the order of $10\%$, which is in agreement with \cite{roma09} (see Fig.~10).
Finally, in Fig.~\ref{fig:pkp_7} we display results for the transverse momentum spectra
of pions kaons and protons with $b=3$ fm. The effect of the viscosity 
is qualitatively consistent with previous results \cite{bair2}: up to \pt$\sim 1$~GeV
spectra are slightly suppressed with respect to the ideal case and at larger \pt\: are instead 
enhanced (almost doubled at \pt$\sim 2$~GeV).
This enhancement with large \pt\:  is due to the growth of the transverse
expansion in presence of viscosity compared to the ideal case, as discussed in~\ref{sec:rhic}.

\subsection{MC-Glauber initial conditions: a test case}

\begin{figure*}[t]
\begin{center}
\includegraphics[clip,width=.30\textwidth]{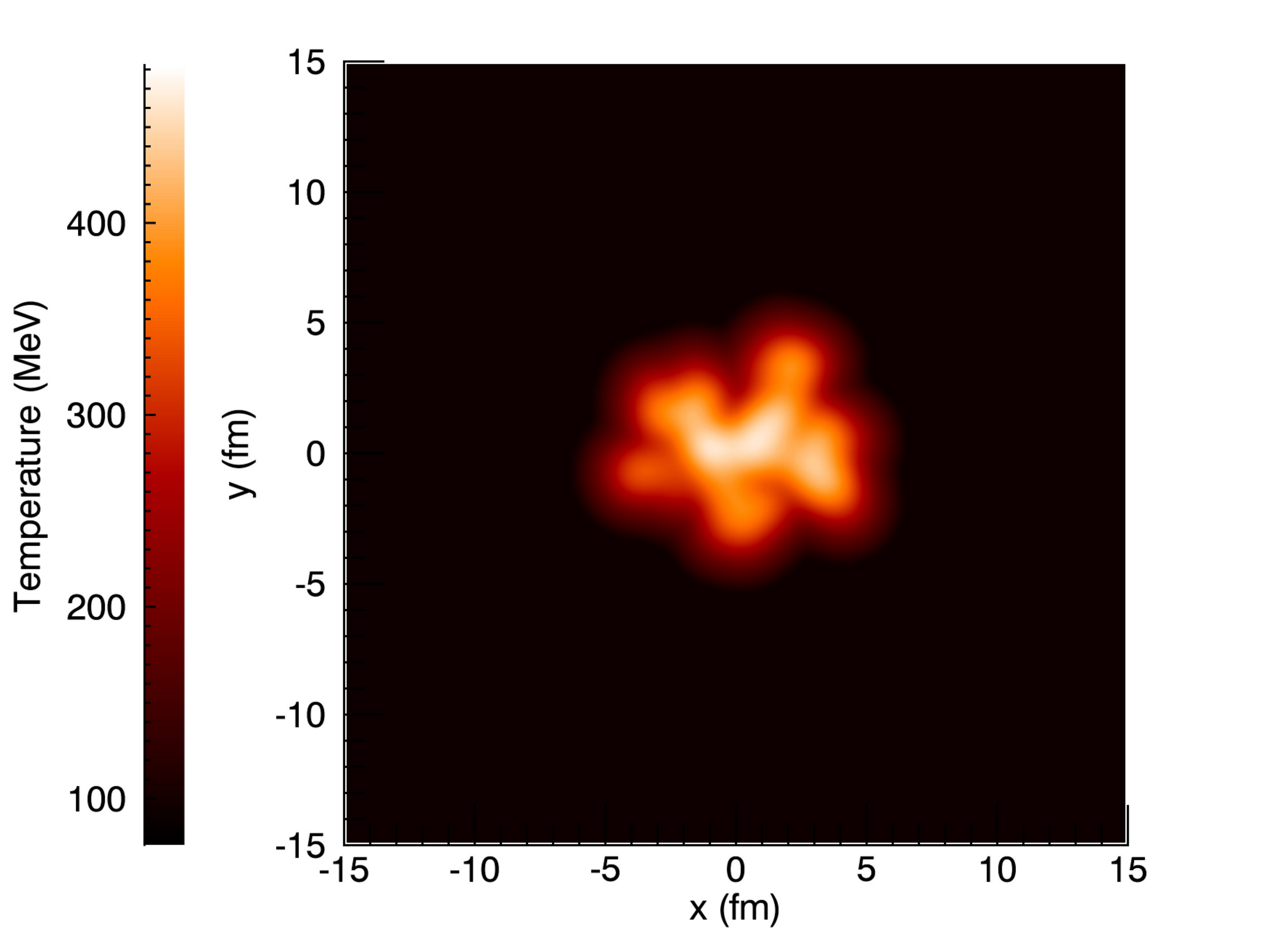}
\includegraphics[clip,width=.30\textwidth]{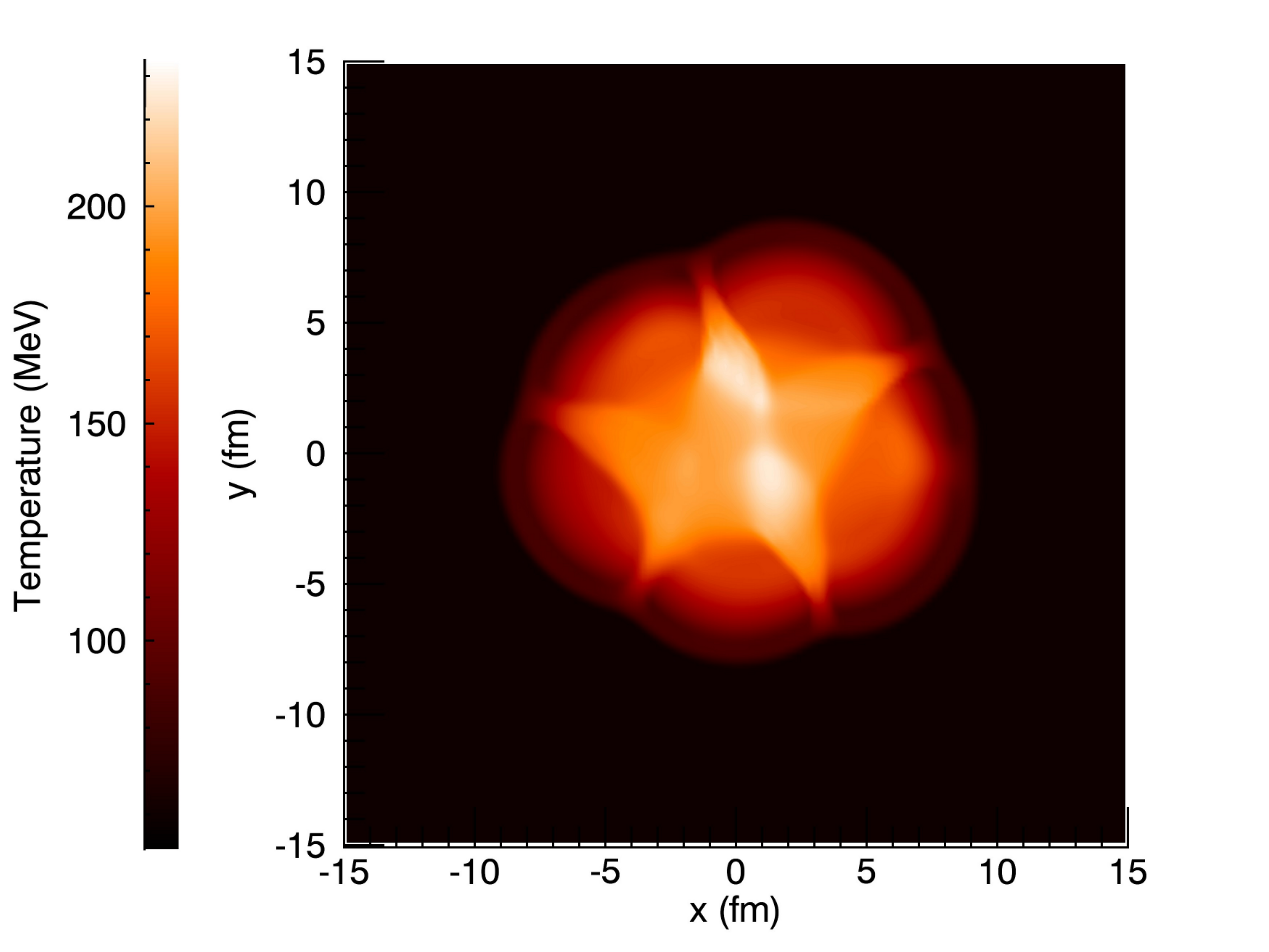}
\includegraphics[clip,width=.30\textwidth]{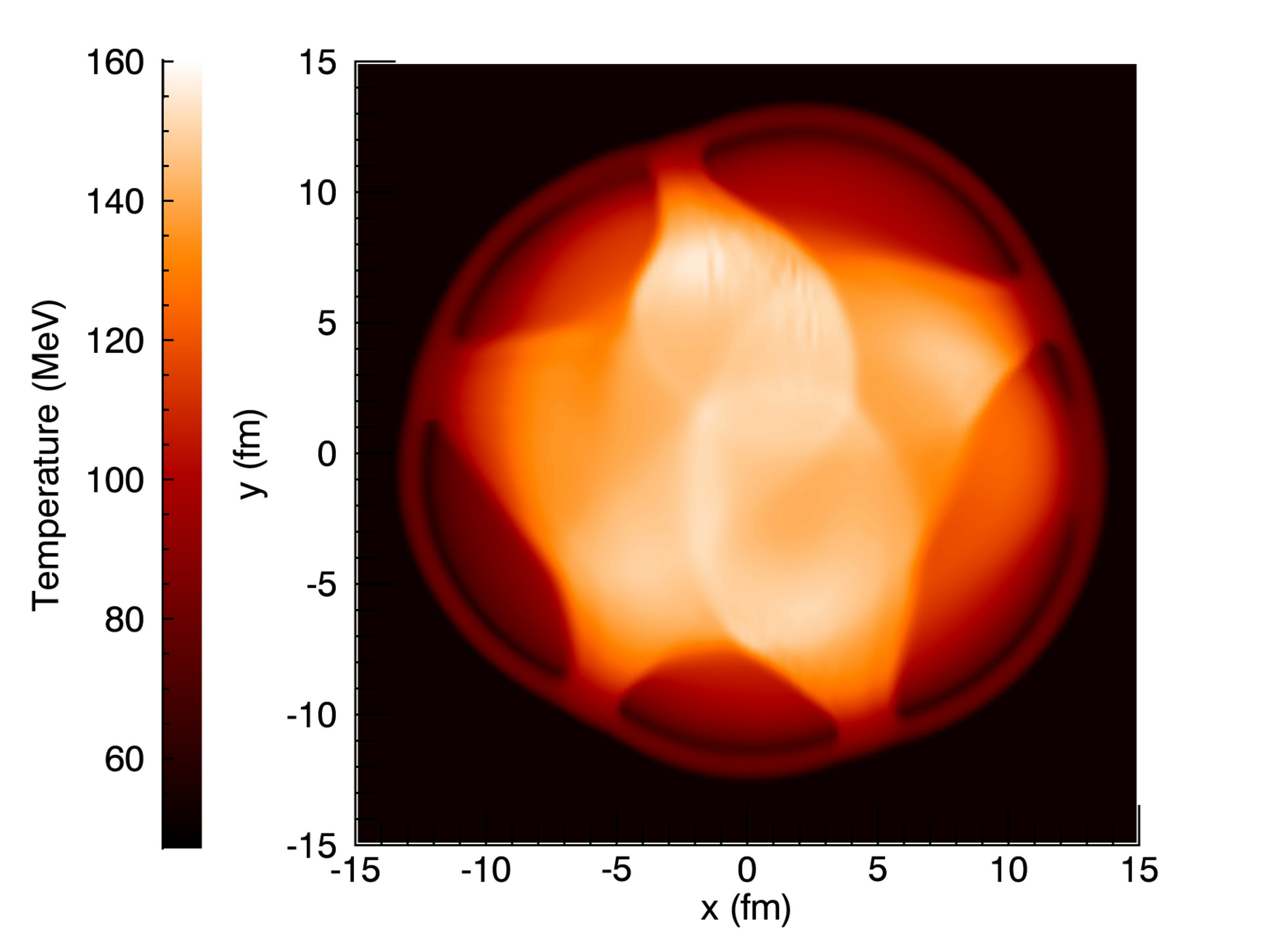}\hspace{-10mm}
\includegraphics[clip,width=.30\textwidth]{contour_t1_vis.pdf}
\includegraphics[clip,width=.30\textwidth]{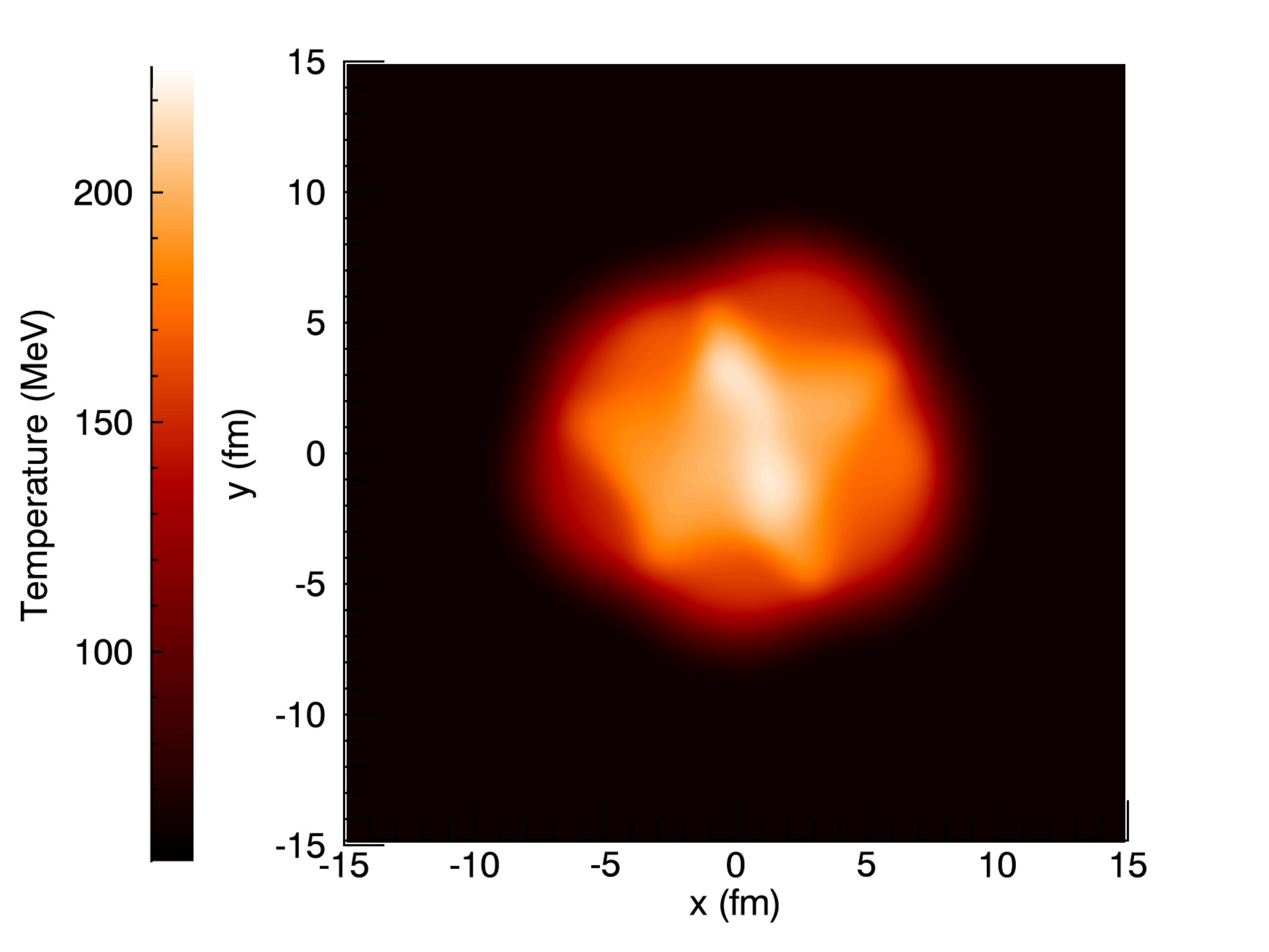}
\includegraphics[clip,width=.30\textwidth]{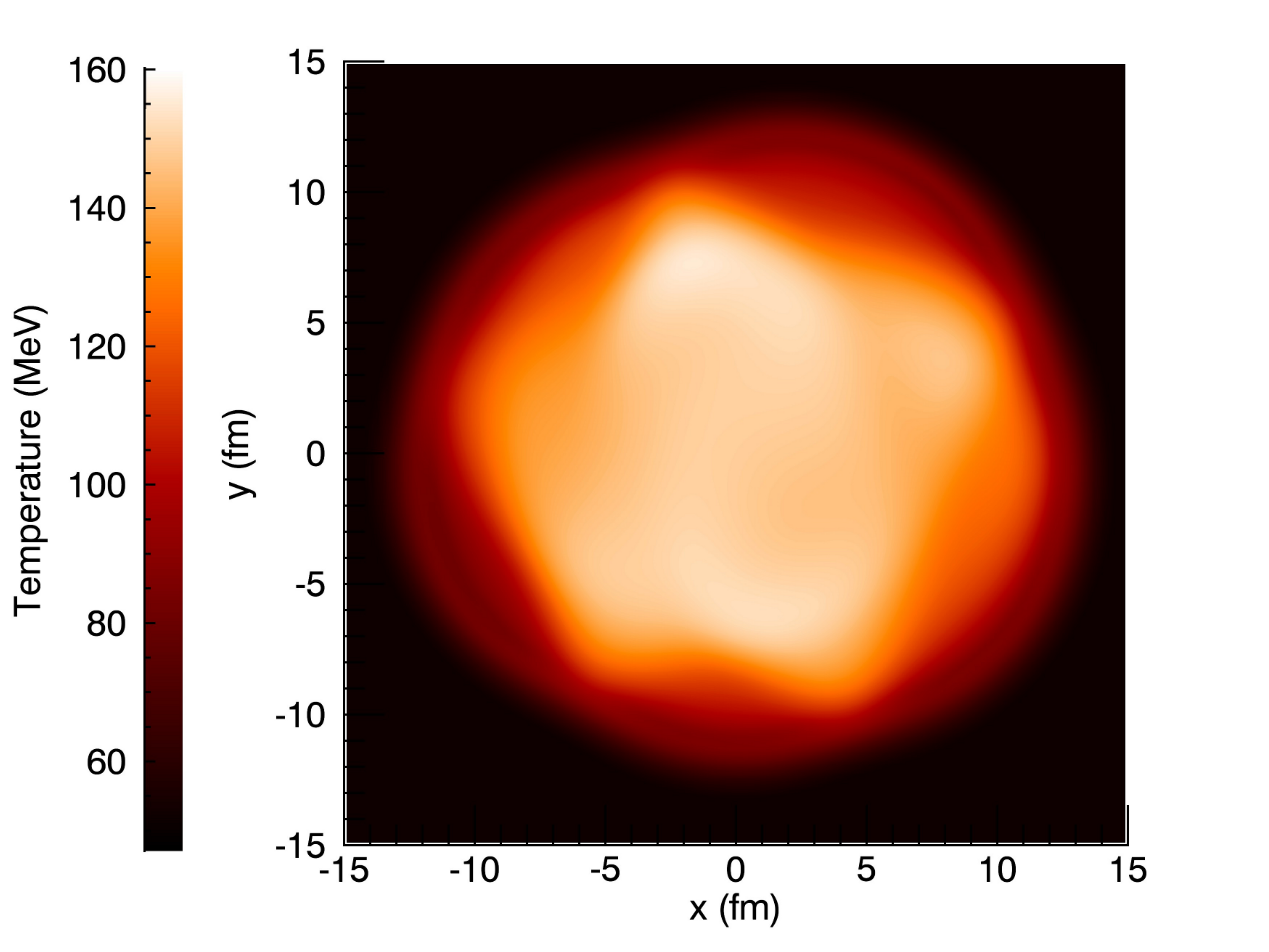}
\caption{Temperature scans at various times -- at $\tau=1$, $5$, and $10$ fm/c -- obtained 
from inviscid (upper panel) and viscous (lower panel) ECHO-QGP simulations with Glauber-MC initial conditions. 
The differences between the two cases are clearly visible. The effect of shear viscosity can be
 seen in the smoothening the profiles.}
\label{fig:tempglamc}
\end{center}
\end{figure*}

In the present section, we demonstrate the capability of running \2d ideal and viscous RHD
simulations with ECHO-QGP in the case of fluctuating Glauber-MC initial conditions. 
The local temperature profile is set at the initial time $\tau=1$~fm/c for one 
particular nuclear configuration generated through the Glauber-MC routine implemented in 
ECHO-QGP (we assume Au-Au collisions with $\sigma=0.6$~fm,
$K=19$~GeV/fm$^2$, and $\alpha=0.2$); 
then the subsequent evolution is followed both in the ideal and in the viscous case. 
In Fig.~\ref{fig:tempglamc} the initial
and later stages of the evolution at  $\tau= 5$ and $10$~fm/c are shown,
where the upper row refers to the ideal run and the lower one to the viscous run.
Here we assume a square numerical box ranging from $-15$ to $15$~fm and 
made up by 151 grid points in both directions. 
The choice for the EOS is EOS-LS, while in the viscous run we set $\eta/s=0.08$.

Clearly, the dynamical effects of shear viscosity are 
reflected in the smoother spatial profiles: the surfaces of discontinuity arising from the
transverse expansion of the initial peaks of energy (shock fronts) are clearly
visible only in the inviscid case. These results demonstrate the capability of 
ECHO-QGP to handle also complex initial conditions with events displaying sizable fluctuations.
The full analysis including the study of higher-order flow harmonics, 
of the impact on the freeze-out stage and of the final particle spectra is beyond 
the scope of the present investigation and is left for future work. 

\section{Conclusions and outlook}

In this paper we have presented ECHO-QGP, a code for {$(3\!+\!1)$-D}
relativistic viscous hydrodynamics specially designed for the physics of heavy-ion collisions.
The code has been built on top of the \emph{Eulerian Conservative High Order} 
code for General Relativistic
Magneto-HydroDynamics (GRMHD) \cite{luca_1}, originally developed and widely used
for high-energy astrophysical applications. ECHO-QGP shares with the original code
the conservative (shock-capturing) approach {-- needed to treat shocks and other
hydrodynamical discontinuities that invariably arise due to the intrinsic nonlinear 
nature of the equations --} and the high accuracy methods for time integration, and spatial
interpolation and reconstruction routines, needed to capture small-scale fluid features
and turbulence. With respect to the {original version} of the code,
where only the ideal case was treated, here second-order dissipative effects have
been included through the evolution of the Israel-Stewart equations for the bulk and shear
stress tensor components, both coupled to the other hydrodynamical equations.
In order for the code to be {suited to QGP applications}, four major
improvements have been implemented in the code:
\begin{enumerate}
\item other than the simple Minkowski metric, in any dimensionality, Bjorken coordinates
have been included with possibility to evolve in $\tau$ any kind of situation, from uniform
states up to \2d (boost invariance along $\eta_s$) and full \3d configurations;
\item any analytical or tabulated equation of state can be used, even at non-zero
baryonic chemical potential; {here in particular we have tested the ideal
ultra-relativistic EOS $P=e/3$ (EOS-I), a tabulated one arising from weak-coupling QCD 
calculations and often adopted in the literature (EOS-LS) and a couple of tabulated 
hybrid EOS's obtained} by matching those for a hadron resonance gas 
(in full, EOS-CE, or partial, EOS-PCE, chemical equilibrium at low temperatures) 
with lattice-QCD results;
\item all kinds of initializing conditions are possible. Among them, RHIC experiments
 are simulated by initializing the hydrodynamical quantities with smooth energy
density (or entropy density) profiles based on the optical Glauber model,
with both participants and binary collisions contributions,
and with different choices for the impact parameter $b$.
Also fluctuating initial conditions in \2d and \3d
can be initialized with a Monte Carlo Glauber routine implemented in the code;
\item a new freeze-out routines is developed and implemented in the ECHO-QGP package.
The procedure has been tested against other available freeze-out routines 
in the literature, both in \2d and in \3d. The comparisons appear consistent
and well convincing, in spite of {the} simpler approach followed.
\end{enumerate}

The code has been extensively tested against analytical solutions, wherever available, or against other 
authors' results, both in Cartesian and Bjorken coordinates, and both {for ideal and for} viscous hydrodynamics. 
Our results always show smooth and accurate profiles, due to the conservative
and high-order properties of the numerical method employed. In particular, thanks to the use of high-order
methods, a limited number of grid points is typically enough to obtain sharp profiles.

The freeze-out routine implemented in the code has been first tested against another existing one
(AZHYDRO) and, once interfaced to the output of ECHO-QGP, it produced particle 
spectra with the expected behavior: stronger radial flow for larger mass hadrons, 
mass ordering of the $v_2$ at low $p_T$, taming of the rise of $v_2(p_T)$ in the presence of viscosity 
(notice that viscous corrections have been so far implemented only in the hydrodynamic 
evolution and not in the Cooper-Frye decoupling algorithm). 

ECHO-QGP turned out to be able to address very granular fluctuating initial conditions. 
This will be fundamental in view of performing event-by-event hydrodynamic simulations, 
mandatory in order to interpret non-trivial experimental measurements like the 
non-vanishing $v_2$ in central events and the appearance of odd flow-harmonics. 
The study of higher harmonics and of event-by-event flow measurements will provide 
a rich information on the initial state and on the transport coefficients of the medium. 
Equally interesting will be the application of ECHO-QGP, with its Glauber-MC 
initialization, to the case of high-multiplicity p-A collisions, where recent 
theoretical~\cite{BozekpA} and experimental analysis~\cite{ATLASpA,CMSpA} 
suggest the possibility of formation of a medium with a collective behavior.

As a further item to address, we would like to include in ECHO 
the possibility of dealing with a finite-density EOS (with possibly 
a first-order phase transition), so to have a tool able to provide 
prediction of interest for the heavy-ion program foreseen at FAIR.
Finally, it is our intention to recover in ECHO-QGP the possibility of evolving also 
electromagnetic fields, either assuming the validity of the ideal MHD 
approximation~\cite{luca_1}, or including magnetic dissipation with the resistive 
term in the Ohm law~\cite{bucciantini2}. There are several motivations for
studying the effects of strong magnetic fields in ultrarelativistic nucleus-nucleus collisions,
such as the recently proposed Chiral Magnetic Effect \cite{kharzeev_CMFI} 
which is supposed to produce an observable separation of positive and negative 
charges with respect with the reaction plane. Such a tool would be unique among 
the codes for QGP studies, and it would represent a very promising 
cross-fertilization opportunity between the astrophysical and high-energy 
physics communities.

\begin{acknowledgements}

This work has been supported by the Italian Ministry of Education and
Research grant PRIN 2009 \emph{Il Quark-Gluon Plasma e le collisioni nucleari 
di alta energia} and by the INFN project RM31. 
F.~B.~acknowledges the support of HIC for FAIR during his sabbatical leave in
Frankfurt. We are thankful to J.~Rizzo, P.~Cea, and L.~Cosmai for help in the initial
stage of this project, and to R. Tripiccione for many interesting informative 
discussions about the freeze-out procedure. 
G.~P.~acknowledges financial support from the Italian Ministry of Research 
through the program \emph{Rita Levi Montalcini}.

\end{acknowledgements}


\appendix

\section{Fluid description in Minkowski and Bjorken coordinates}
\label{app:appendix_a}

In this appendix we summarize the essential formulas establishing the link between the fluid description 
in Minkowski and Bjorken coordinates. While what contained in the text is already sufficiently self-consistent, 
here we wish to establish a mapping with the notation adopted by other authors~\cite{heinz04,hira01} and widely
employed in phenomenological studies, such as blast-wave fits. For the sake of clarity in 
this appendix four-vectors in Bjorken coordinates will be denoted by a ``prime'' and components 
of the three-velocity by a ``tilde''; in the text such a distinction will be neglected.

\subsection{Minkowski coordinates}

Here we specify the notation employed for the different rapidities entering into our RHD setup.
\begin{itemize}
\item Fluid rapidity:
\beq
Y\equiv\frac{1}{2}\ln\frac{1+v^z}{1-v^z}\quad\longrightarrow\quad v^z=\tanh Y
\eeq
with $v_z$ the longitudinal component of the fluid velocity $\v$;
\item Space-time rapidity:
\beq
\eta_s\equiv\frac{1}{2}\ln\frac{t+z}{t-z}.
\eeq
\item Particle rapidity (of the emitted hadron):
\beq
y\equiv\frac{1}{2}\ln\frac{E+p_z}{E-p_z};
\eeq
\end{itemize}

The particle four-momentum is conveniently expressed in terms of its transverse mass $m_\perp$ and rapidity:
\beq
p^\mu\equiv(m_\perp\cosh y,\p_\perp,m_\perp\sinh y)
\eeq

The fluid four-velocity, defined as
\beq
u^\mu\equiv \gamma\,(1,\v)=\gamma\,(1,\v_\perp,v_z) ,\quad{\rm with}\quad \gamma\equiv 1/\sqrt{1-\v^2},
\eeq
can be recasted in terms of the fluid-rapidity as
\beq
u^\mu=\frac{1}{\sqrt{1-\cosh^2Y\,\v_\perp^2}}(\cosh Y,\cosh Y\,\v_\perp,\sinh Y).
\eeq
This suggest to define the ``transverse velocity''
\beq
\u_\perp\equiv\cosh Y\,\v_\perp,
\eeq
so that
\beq
u^\mu\equiv\gamma_\perp\,(\cosh Y,\u_\perp,\sinh Y) ,\quad{\rm with}\quad \gamma_\perp\equiv 1/\sqrt{1-\u_\perp^2}.
\eeq
The scalar product between the particle momentum and the fluid velocity, entering into the Cooper-Frye decoupling prescription, reads
 (with metric [-,+,+,+])
\beq
-p\cdot u=\gamma_\perp[m_\perp\cosh(y-Y)-\p_\perp\!\cdot\!\u_\perp],\label{eq:pdotu}
\eeq
expressing the fact that particles tend to be emitted with rapidity close to the one of the fluid-cell.

In the general case the velocity field of the fluid depends on all the four space-time coordinates: one has $\u_\perp\!\equiv\!\u_\perp(\tau,\r_\perp,\eta_s)$ and $Y\!\equiv\! Y(\tau,\r_\perp,\eta_s)$.

In the case of longitudinal boost-invariance of the fluid profile one has \emph{always}
\beq
v_z=\frac{z}{t}\quad\longrightarrow\quad Y\equiv\eta_s,
\eeq
so that the fluid velocity reduces to
\beq
u^\mu=\gamma_\perp\,(\cosh\eta_s,\u_\perp,\sinh\eta_s),
\eeq
which only depends on $\u_\perp(\tau,\r_\perp)$.

\subsection{Bjorken coordinates}

One can go from the Minkowsy coordinates
\beq
u^\mu\equiv(u^0,u^1,u^2,u^3)
\eeq
to the Bjorken (sometimes also known as Milne) coordinates
\beq
u'^m\equiv[u'^\tau,u'^x,u'^y,u'^\eta]\label{eq:uBj}
\eeq
through the transformation
\beq
u'^m\equiv\frac{\partial x'^m}{\partial x^\nu}u^\nu.
\eeq
One has:
\begin{subequations}
\begin{align}
u'^\tau=\frac{d\tau}{dt}u^0+\frac{d\tau}{dz}u^3=\gamma_\perp\cosh(Y-\eta_s)\\
u'^\eta=\frac{d\eta_s}{dt}u^0+\frac{d\eta_s}{dz}u^3=\gamma_\perp\frac{1}{\tau}\sinh(Y-\eta_s).
\end{align}
\end{subequations}
Hence
\beq
u'^m=\gamma_\perp\left[\cosh(Y-\eta_s),\u_\perp,\frac{1}{\tau}\sinh(Y-\eta_s)\right]\label{eq:uBj}
\eeq
The four-momentum in Milne coordinates analogously reads:
\beq
p'^m=\left[m_\perp\cosh(y-\eta_s),\p_\perp,\frac{1}{\tau}m_\perp\sinh(y-\eta_s)\right].
\eeq
Its contraction with $u'^m$ and the hypersurface element in Eqs.~(\ref{eq:uBj}) and 
(\ref{eq:dsigma}) allows one to recover the freeze-out formula employed in the text in 
Eq.~(\ref{cooperfrye}) and expressed in terms of the output variables of ECHO-QGP. 
Notice that its contraction with the expression of the fluid four-velocity in Eq.~(\ref{eq:uBj}) 
provides the result in Eq.~(\ref{eq:pdotu}).

After a further change of variables, introducing the definitions
\beq
\widetilde{v}^\eta\equiv\frac{\tanh(Y-\eta_s)}{\tau}\quad{\rm and}\quad\widetilde{v}^{x/y}\equiv\frac{u_\perp^{x/y}}{\cosh(Y-\eta_s)},
\eeq
one has:
\beq
u'^m=\widetilde{\gamma}\left[1,\widetilde{v}^x,\widetilde{v}^y,\widetilde{v}^\eta\right], 
\eeq
where
\beq
\widetilde{\gamma}\equiv\gamma_\perp\,\cosh(Y-\eta_s)=\frac{1}{\sqrt{1-(\widetilde{v}^x)^2-(\widetilde{v}^y)^2-\tau^2(\widetilde{v}^\eta)^2 }}
\eeq
in agreement with the definition $\widetilde{\gamma}\!\equiv\!(1-g_{ij}\tilde{v}^i\tilde{v}^j)^{-1/2}$ quoted in the text.

In the case of a longitudinal boost-invariant expansion one has $Y\equiv\eta_s$, so that:
\beq
u'^m=\gamma_\perp[1,\u_\perp,0].
\eeq

\section{Source terms}

\label{app:appendix_b}

In the present appendix, we write down explicitly the source terms needed for the
evolution of the set of balance laws as in Eq.~(\ref{echoeq2}). For the momentum
and energy equations we have, respectively
\begin{align}
\mathbf{S}(S_i) & = |g|^{1/2} [ \tfrac{1}{2} T^{\mu\nu} \partial_i g_{\mu\nu} ], \\
\mathbf{S}(E) & = |g|^{1/2} [ - \tfrac{1}{2} T^{\mu\nu} \partial_0 g_{\mu\nu} ],
\end{align}
whereas for the evolution of the bulk viscous pressure $\Pi$ and for 
the spatial components of the viscous stress tensor $\pi^{ij}$ more terms
are required. We recall here the general definitions for the expansion scalar, 
shear tensor, and vorticity, which are
\begin{align}
 \theta & = d_\mu u^\mu = \partial_\mu u^\mu + \Gamma^\mu_{\mu\nu}u^\nu, \\
\sigma^{\mu\nu} & = \tfrac{1}{2}(d^\mu u^\nu + d^\nu u^\mu) + 
\tfrac{1}{2}(u^\mu Du^\nu + u^\nu Du^\mu) \nonumber \\
& - \tfrac{1}{3}(g^{\mu\nu} + u^\mu u^\nu)\theta, \\
\omega^{\mu\nu} & =  \tfrac{1}{2}(d^\mu u^\nu - d^\nu u^\mu) + 
\tfrac{1}{2}(u^\mu Du^\nu - u^\nu Du^\mu),
\end{align}
with
\begin{align}
d^\mu u^\nu & = g^{\mu\alpha} (\partial_\alpha u^\nu + \Gamma^\nu_{\alpha\beta} u^\beta), \\
Du^\nu & = u^\alpha  (\partial_\alpha u^\nu + \Gamma^\nu_{\alpha\beta} u^\beta).
\end{align}
Moreover, the $\mathcal{I}^{\mu\nu}$ terms are provided by
\begin{align}
\mathcal{I}^{\mu\nu}_0 & = - u^\alpha 
(\Gamma^\mu_{\alpha\beta}\pi^{\nu\beta} + \Gamma^\nu_{\alpha\beta}\pi^{\mu\beta}), \\
\mathcal{I}^{\mu\nu}_1 & = g_{\alpha\beta}(\pi^{\mu\alpha}u^\nu + \pi^{\nu\alpha}u^\mu) Du^\beta, \\ 
\mathcal{I}^{\mu\nu}_2 & = - \lambda\ g_{\alpha\beta}(\pi^{\mu\alpha}\omega^{\nu\beta} + 
\pi^{\nu\alpha}\omega^{\mu\beta} ).
\end{align}
In the remainder we shall specify to either Minkowski or Bjorken coordinates.

\subsection{Minkowski coordinates}

The simplest case is that of Minkowskian Cartesian coordinates $(t,x,y,z)$, with metric 
$g_{\mu\nu}=g^{\mu\nu}=\mathrm{diag}(-1,1, 1, 1)$ ($|g|^{1/2}=1$), and vanishing 
Christoffel symbols. Thus, no source terms are needed for the evolution of the 
energy-momentum tensor. Covariant derivatives simply become
\begin{align}
d^t u^\nu & = - \partial_t u^\nu, \\
d^i u^\nu & = \partial_i u^\nu, \\
D u^\nu & = u^t \partial_t u^\nu + u^i \partial_i u^\nu,
\end{align}
so for instance the expansion scalar is
\be
\theta = \partial_t u^t + \partial_i u^i,
\ee
with $i=x,y,z$. Notice that $\mathcal{I}_0^{\mu\nu}\equiv 0$, while for
$\mathcal{I}_1^{\mu\nu}$ and $\mathcal{I}_2^{\mu\nu}$ the standard definitions
apply.

\subsection{Bjorken coordinates}

Bjorken coordinates $(\tau, x, y, \eta_s)$ have still a diagonal metric 
$g_{\mu\nu}=\mathrm{diag}(-1, 1, 1, \tau^2)$, and $g^{\mu\nu}=\mathrm{diag}(-1, 1, 1, 1/\tau^2)$
($|g|^{1/2}=\tau$), but here $\partial_\tau g_{\eta\eta} = 2\tau \neq 0$ leading to the 
non-vanishing Christoffel symbols $\Gamma^{\tau}_{\eta\eta}=\tau$ and 
$\Gamma^{\eta}_{\eta \tau}=1/\tau$. Then, while
the source term for the momentum equation is still zero, that for the energy equation becomes
\be
\mathbf{S}(E) =  -\tau^2 T^{\eta\eta}.
\ee
Non-Minkowskian covariant derivatives are
\begin{align}
d^\tau u^\eta & = - (\partial_\tau u^\eta + u^\eta/\tau), \\
d^\eta u^\tau & = (\partial_\eta u^\tau + \tau u^\eta)/\tau^2, \\
d^\eta u^\eta & = (\partial_\eta u^\eta + u^\tau/\tau)/\tau^2, \\
D u^\tau & = u^\tau \partial_\tau u^\tau + u^i\partial_i u^\tau + \tau u^\eta u^\eta, \\
D u^\eta & = u^\tau \partial_\tau u^\eta + u^i\partial_i u^\eta + 2u^\tau u^\eta/\tau,
\end{align}
with $i=x,y,\eta$, and the expansion scalar is now
\be
\theta = \partial_\tau u^\tau + \partial_i u^i + u^\tau/\tau.
\ee
In Bjorken coordinates the non-vanishing $\mathcal{I}_0^{\mu\nu}\equiv 0$
terms are
\begin{align}
\mathcal{I}_0^{x\eta} & = -(u^\tau \pi^{x\eta} + u^\eta\pi^{\tau x})/\tau, \\
\mathcal{I}_0^{y\eta} & = -(u^\tau \pi^{y\eta} + u^\eta\pi^{\tau y})/\tau, \\
\mathcal{I}_0^{\eta\eta} & = -2(u^\tau \pi^{\eta\eta} + u^\eta\pi^{\tau\eta})/\tau,
\end{align}
while $\mathcal{I}_1^{\mu\nu}$ and $\mathcal{I}_2^{\mu\nu}$ are defined in the usual way.

\end{document}